\documentclass[fleqn,usenatbib]{mnras}

\usepackage{newtxtext,newtxmath}
\usepackage{bm}

\usepackage[T1]{fontenc}

\usepackage{enumitem}
\setlist[itemize,1]{
                  align=left, 
                  itemindent=!
                  }

\DeclareRobustCommand{\VAN}[3]{#2}
\let\VANthebibliography\thebibliography
\def\thebibliography{\DeclareRobustCommand{\VAN}[3]{##3}\VANthebibliography}

\usepackage[noabbrev]{cleveref}
\usepackage{graphicx}	% Including figure files
\usepackage{amsmath}	% Advanced maths commands
\usepackage{cases}      % Numbered cases

\newcommand{\asterequation}[1]{\refstepcounter{equation}\tag{\theequation \,*}\label{#1}}

\title[Ionization chemistry in the inner disc]{Ionization chemistry in the inner disc: a combined treatment of ionic and thermionic emission and arbitrary grain size distributions}

\author[M. Williams and S. Mohanty]{
Morgan Williams$^{1}$\thanks{E-mail: m.williams22@imperial.ac.uk (MW)} and
Subhanjoy Mohanty$^{1}$
\\
$^{1}$Astrophysics Group, Imperial College London, Blackett Laboratory, Prince Consort Road, London SW7 2AZ, UK
}

\date{Accepted XXX. Received YYY; in original form ZZZ}

\pubyear{2024}

\begin{document}
\label{firstpage}
\pagerange{\pageref{firstpage}--\pageref{lastpage}}
\maketitle

\begin{abstract}
  In the inner regions of protoplanetary discs, ionization chemistry controls the fluid viscosity, and is thus key to understanding various accretion, outflow and planet formation processes. The ionization is driven by thermal and non-thermal processes in the gas-phase, as well as by dust–gas interactions that lead to grain charging and ionic and thermionic emission from grain surfaces. The latter dust–gas interactions are moreover a strong function of the grain size distribution. However, analyses of chemical networks that include ionic/thermionic emission have so far only considered grains of a single size (or only approximately treated the effects of a size distribution), while analyses that include a distribution of grain sizes have ignored ionic/thermionic emission. Here, we: (1) investigate a general chemical network, widely applicable in inner disc regions, that includes gas-phase reactions, ionic and thermionic emission, {\it and} an arbitrary grain size distribution; (2) present a numerical method to solve this network in equilibrium; and (3) elucidate a general method to estimate the chemical time-scale. We show that: (a) approximating a grain size distribution by an ``effective dust-to-gas ratio'' (as done in previous work) can predict significantly inaccurate grain charges; and (b) grain charging significantly alters grain collisional time-scales in the inner disc. For conditions generally found in the inner disc, this work facilitates: (i) calculation of fluid resistivities and viscosity; and (ii) inclusion of the effect of grain charging on grain fragmentation and coagulation (a critical effect that is often ignored).
\end{abstract}


%%%%%%%%%%%%%%%%%%%%%%%%%%%%%%%%%%%%%%%%%%%%%%%%%%

%%%%%%%%%%%%%%%%% BODY OF PAPER %%%%%%%%%%%%%%%%%%

\section{Introduction}

Magnetic effects are thought to play a crucial role in driving accretion and outflow in protoplanetary discs. These effects include, e.g., the magnetorational instability~\citep[MRI;][]
{Balbus_1991} and magneto-hydrodynamical (MHD) winds~\citep{Blandford_Payne_1982}. To be efficient, these phenomena require sufficient coupling between the magnetic field and the fluid.

The degree of field-fluid coupling is determined by the resistivities of the fluid~\citep[e.g.][]{Gammie_1996,Jin_1996, Fleming_2000, Sano_Stone_2002b, Bai_Stone_2011}. The resistivities in turn are set by the disc ionization chemistry, driven by various thermal and non-thermal processes. In models of protoplanetary discs, the chemistry is usually investigated via a specified chemical network. Over the past several decades, networks of increasing complexity have been developed to better represent reality. Initial networks omitted dust grains, assuming that grains would settle to the mid-plane and hence be irrelevant for studying the layered accretion model~\citep[e.g.,][]{Gammie_1996}. More complete subsequent networks accounted for grains with gas–grain and grain–grain interactions, and also included a range of non-thermal ionization sources and intricate gas phase reactions, including charge transfer between species~\citep[e.g.,][]{Ilgner_Nelson_2006, Bai_Goodman_2009}.

Due to computational constraints, time-dependent chemical reaction networks, such as those of~\cite{Ilgner_Nelson_2006}, usually involve a single grain size with a limited set of charge states { or assume a steady-state distribution of possible charges for each grain size bin~\citep{Balduin_2023}}. Conversely, time-independent ({kinetic} equilibrium) chemical networks typically do include a distribution of grain sizes with a full range of possible charge states. {Such networks have been used for calculations specialized to the cool outer disc~\citep{Okuzumi_2009, Fuji_2011,Dzyurkevich_2013, Mori_2016, Ivlev_2016, Marchand_2021}}, with no thermal ionization.~\cite{Marchand_2022_2} later performed similar calculations including the effects of thermal ionization.

Meanwhile, our understanding of grain { effects on the ionization} has also evolved~\citep[e.g.][]{Desch_Turner_2015, Thi_2019}. Earlier works~\citep[e.g.,][]{Sano_2000, Ilgner_Nelson_2006, Wardle_2007, Salmeron_Wardle_2008, Bai_Goodman_2009, Mohanty_2013} assumed that grains act only to reduce the ionization state of the gas, by adsorbing free charges from the gas and allowing recombinations on their surfaces. As~\cite{Desch_Turner_2015} showed, this is not always true for conditions prevalent in the inner disc ($\gtrsim$ 800 K): grains can also {\it increase} the number density of free charges in the gas phase, due to thermionic and ion emission (emission of electrons and ions respectively from grains, following grain surface-catalysed ionization of alkali atoms condensed onto grains).

However, the equilibrium chemical network~\cite{Desch_Turner_2015} solved only included a single grain size, with a single value of charge.~\cite{Jankovic_2021} extended this to a distribution of grain sizes in a simplified fashion, by combining a single grain size with an ``effective'' dust-to-gas ratio~\citep[motivated by earlier work by][]{Bai_Goodman_2009}. As we show here (Appendix \ref{sec:appendeffectivedg}), however, while this technique gives acceptable results for gas-phase charge densities (and thus resistivities), the grain charges it predicts are inaccurate. This is important because grain charges are vital for correctly computing grain–grain collision rates, and thus fragmentation and coagulation rates~\citep{Okuzumi_2009, Okuzumi_2011,Okuzumi_2011b, Akimkin_2020, Akimkin_2023}, which determine the grain size distribution itself in dynamical calculations. This size distribution in turn controls key aspects of disc physics: not only the (gas-phase) charge densities and thus resistivities discussed here, but also the disc opacity~\citep[e.g.,][]{Draine_2016} and the coupled dynamics of dust and gas.  

To summarize: currently, disc equilibrium chemical networks that rigorously treat a distribution of grain sizes do not yet include ionic and thermionic emission from grains~\citep[e.g.,][]{Marchand_2022_2}, while equilibrium networks that do include the latter effects either do not include a grain size distribution or only treat it approximately~\citep[e.g.,][]{Desch_Turner_2015,Jankovic_2021}.\footnote{{ As an aside, the time-dependent network (albeit with a steady-state charge distribution in each grain size bin) of~\cite{Balduin_2023}, that includes a distribution of grain sizes, and accounts for thermionic emission and charge exchange between ions/neutrals and grains (including for alkalis), does not account for condensation , assuming ions and neutral to be freely available in the gas-phase. As such, it also does not account for ionic emission.}}

Our goal here is to redress this lack for inner disc equilibrium chemical networks. Specifically, we present a numerical method to {\it exactly} solve (to numerical precision) a generalized equilibrium chemical network that is widely applicable in inner disc conditions, which includes: {\it (i)} thermal and non-thermal ionization processes; {\it (ii)} ionic and thermionic emission; and {\it (iii)} an arbitrary distribution of grain sizes, with size-dependent grain charges.  

Our chemical network (see Section \ref{sec:chemnet}) is the same as in~\cite{Desch_Turner_2015}, except generalized to include a distribution of grain sizes and charges. Our solution technique (Subsection \ref{subsec:reaction-network}) is similar to that of~\cite{Marchand_2022_2}, generalized to include ionic and thermionic emission.  

In section Section \ref{sec:chemnet} below, we introduce our chemical network. In Section Section \ref{sec:results}, we discuss the results of our simulations. Specifically, we: validate the results of simulation against earlier work in relevant limits (Subsection \ref{subsec:tests}); discuss the behaviour of the full network (Subsection \ref{subsec:network-behaviour}); investigate the distribution of charge as a function of grain size ({Subsection \ref{subsec:dist_charge});  analyse the effect of MRN grain-size distributions with various slopes (Subsection \ref{subsec:effect-of-distribution}); analyse an illustrative arbitrary (non-MRN) grain size distribution (Subsection \ref{subsec:arbitrary}); determine the applicability of chemical equilibrium (Subsection \ref{subsec:validity-equilibrium-results}); { and compute resistivities using our network (Subsection \ref{subsec:resistivies})}. Finally, our conclusions are presented in Section \ref{sec:conclusions}.

% . This relies on the assumption that charge transfer between grains can be neglected, valid when grains have large like charges due to Coulomb repulsion. We illustrate this technique as a modification to the equilibrium reaction network of~\cite{Desch_Turner_2015} but could be incorporated easily into more complex time-dependent networks. This enables the determination of the ionization state and the charge distribution on dust grains for later use in fragmentation-coagulation routines.

\section{Chemical Network}
\label{sec:chemnet}

The network we have used is an adaptation of the network employed by~\cite{Desch_Turner_2015}, consisting of ODEs for:
{
\begin{itemize}
    \item electrons ($n_e$);
    \item a molecular ion species ($n_{\text{m}^+}$), produced through non-thermal ionization alone;
    \item a metal ion species ($n_{\text{M}^+}$), produced by charge transfer to a neutral metal atom from the molecular ion;
    \item a neutral alkali species ($n_{\text{alk}^0}$) which is converted to the alkali ion ($n_{\text{alk}^+}$) by thermal processes only;
    \item a phase of the alkali species (both neutral and ionic) condensed onto the grain surface, with a separate ODE for each grain bin (i.e., $n_{\text{alk cond}}^i$ for the $i$th grain size bin); and
    \item the charge on each grain size bin ($Z^i$).
\end{itemize}
}
We also have the equation of charge conservation. Differences with respect to the~\cite{Desch_Turner_2015} network are the separate ODEs for the condensed alkali and grain charge for each grain size bin, and the inclusion of the intermediate molecular ion species.
 
\subsection{Gas phase chemistry}
\label{subsec:gas-phase}
\subsubsection{Non-thermal ionization and downstream processes}
Non-thermal ionization of $\mathrm{H_2}$ occurs at a volumetric rate $\zeta n_{\mathrm{H_2}}$. The non-thermal ionization rate of $\mathrm{H_2}$ includes terms due to ionization by radionuclides, cosmic rays and high energy stellar irradiation. The ionization rate of $\mathrm{H_2}$ due to radionuclides in discs is $\zeta \approx 7.6 \times 10^{-19}$~\citep{Umebayashi_Nakano_2009}, the largest contributor being the decay of $^{26}\mathrm{Al}$. This study will focus on conditions applicable to the inner-disc mid-plane; therefore, we assume that any cosmic rays or high energy stellar irradiation are shielded by the large column depths. We thus use $\zeta = 7.6 \times 10^{-19}$ in all our simulations.

In our model, through charge transfer to molecules, non-thermal ionization of $\mathrm{H_2}$ results in an identical volumetric production rate of molecular ions $\text{m}^+$. For definiteness, we use the molecular ion $\mathrm{HCO}^+$ in all subsequent simulations. We assume that the abundance of CO, from which this ion is formed, is sufficiently high to allow rapid production of $\mathrm{HCO}^+$ through charge transfer. Our simulations show low abundances of HCO$^+$, indicating minimal depletion of CO. Coupled with the high abundance of CO, this supports our assumption.

The cations immediately produced by non-thermal ionization of $\mathrm{H}_2$ are $\mathrm{H}_2^+$ or $\mathrm{H}^+$. Before reaching neutral metal atoms $\text{M}^0$\,, e.g., Mg, Na, Ca, Fe~\citep{Oppenheimer_Dalgarno_1974}, the positive charge on these cations must be transferred, via reactions with neighbouring neutral molecular species, to form molecular ions, e.g., $\text{H}_3^+$, $\text{H}_2\text{O}^+$, $\text{HCO}^+$ ~\citep[e.g.][]{Oppenheimer_Dalgarno_1974}; these molecular ions may then transfer their charge to a neutral metal atom, forming $\text{M}^+$. However, the molecular ions in the chain may instead recombine, radiatively or dissociatively, meaning that the complete charge transfer from $\mathrm{H}_2^+$ or $\mathrm{H}^+$ to produce a metal ion $\text{M}^+$ is not 100\% efficient. The metal ion that we use in this work is Mg$^+$, due to the high abundance of Mg, $x_{\text{Mg}} = 3.67 \times 10^{-5}$~\citep{Asplund_2009}.~\footnote{\cite{Asplund_2009} in fact provides logarithmic abundances in the form $X_i = \log_{10} (n_i/n_{\text{H}}) + 12$, where $X_{\text{H}}$ is defined to be 12. Thus, to compute $x_i = n_i / \left(\sum_j n_{j}\right)$ from these values, one must compute $x_i = 10^{X_i} / \left(\sum_j 10^{X_j}\right)$. As done in~\cite{Jankovic_2021}, to convert from this abundance to number density of a species $i$, we use $n_{i}^{\text{tot}}=\frac{2x_{i}}{2-x_{\text{H}}} \frac{\rho}{\mu m_{\text{H}}}$, where $x_{\text{H}} = 9.21 \times 10^{-1}$~\citep{Asplund_2009}, $\rho$ is the gas density and $\mu$ is the mean molecular weight, taken to be 2.34. The number density of H$_2$ is $n_{\text{H}_2}=\frac{x_{\text{H}}}{2-x_{\text{H}}} \frac{\rho}{\mu m_{\text{H}}}$.}

Charge may transfer from a molecular ion m$^+$ to a metal atom M$^0$ at a rate
\begin{equation}
\mathcal{R_{\mathrm{ct}}} = \beta n_{\mathrm{m}^+} n_{\mathrm{M}^{0}} = \beta n_{\mathrm{m}^+} (n_{\mathrm{M}^{\text{tot}}} -n_{\mathrm{M}^+}) \,\,\,\,\, ,
\end{equation}
where $\beta = 3 \times 10^{-9} \, \text{cm}^3 \, \text{s}^{-1}$~\citep{Oppenheimer_Dalgarno_1974} and $n_{\mathrm{M}^{\text{tot}}}$ is the total number density of M atoms and ions. The UMIST~\citep{UMIST_2022} value for this rate constant for Mg$^+$ and HCO$^+$ agrees with the~\cite{Oppenheimer_Dalgarno_1974} value to within the defined accuracy. Alternatively, the molecular ion ($\text{m}^+$) may recombine dissociatively at a rate:
\begin{equation}
\mathcal{R_{\mathrm{diss-rec}}} = \alpha n_{\mathrm{m}^+} n_{\mathrm{e}} \,\,\,\,\, ,
\end{equation}
where $\alpha = 3 \times 10^{-6} T^{-1/2} \mathrm{cm^3 s^{-1}}$~\citep{Oppenheimer_Dalgarno_1974}. The UMIST~\citep{UMIST_2022} value for this rate constant for HCO$^+$ is valid only over the range 10-300\,K. Therefore, we use the~\cite{Oppenheimer_Dalgarno_1974} value. The molecular ion may also collide with dust grains, where it is assumed that the molecular ions gain an electron from the grains due to the ionization potential of the ion being larger than the work function of the grains. This means that electrons associated with the grains fall into the deeper potential well of the molecular ion, neutralizing it, before the neutral molecule is eventually re-emitted to the gas phase.

Once produced, the metal ions M$^+$ recombine radiatively at a rate given by:

\begin{equation}
   \mathcal{R_{\mathrm{gas, \; 2-rec, \; M^+}}} = \gamma n_{\mathrm{e}} n_{\mathrm{M^+}} \,\,\,\,\, ,
   \label{2-rec-m}
\end{equation}
where $\gamma = 3 \times 10^{-11} T^{-1/2} \mathrm{cm^3 s^{-1}}$~\citep{Oppenheimer_Dalgarno_1974}. Similarly to the case for dissociative recombination, the UMIST~\citep{UMIST_2022} value for this rate constant for Mg$^+$ is defined over a subset of our desired temperature range (10-1000\,K). For this reason, and to mirror the choice of~\citep{Desch_Turner_2015}, allowing for direct comparison with their work, we again use the~\cite{Oppenheimer_Dalgarno_1974} value.

Just like the molecular ions, metal ions also collide with dust grains and are assumed to be neutralized and re-emitted as neutral atoms.

\subsubsection{Thermal Ionization and Downstream Processes}

The alkali ion species $\mathrm{alk^+}$ is produced by thermal processes alone; this includes collisional ionization in the gas phase and ion emission from the dust grains (see~\ref{subsec:grain-phase}). Collisional ionization has a rate

\begin{equation}
    \mathcal{R_{\mathrm{gas, \; coll-ion}}} = k_2 n_{\mathrm{H_2}} n_{\mathrm{alk^0}} \,\,\,\,\, ,
    \label{coll-ion}
\end{equation}
where $k_2$ is the rate constant. Following~\cite{Desch_Turner_2015}, we take the experimentally computed value~\citep{Ashton_Hayhurst_1973} $k_2 = 9.9 \pm 2.7 \times 10^{-9} T^{1/2} \exp(-\frac{\text{IP}}{kT}) \; \mathrm{cm^3 s^{-1}}$ with $T$ in units of K (while noting the discrepancy with the theoretical value from~\cite{Pneumann_Mitchell_1965}, for which no calculation or references are provided). See~\cite{Marchand_2022_2} for a discussion of the implications of using the experimental value in place of the theoretical value. However, note that the value chosen is relatively unimportant as grain-phase ionization and emission dominate the rate of thermal production of alkali ions and electrons in the gas-phase~\citep[][as well as our results, shown further along]{Desch_Turner_2015}.

The specific alkali adopted in this work is potassium. The reasons for this, in terms of the ionization fractions produced by different alkalis versus the threshold ionization fractions required for MRI activity (which is widely assumed to be the source of viscosity in the inner disc), are provided in Appendix~\ref{sec:append-mri-conds}. However, for completeness, we also show in Appendix~\ref{sec:sodium} the effect of including sodium alongside potassium in our network. While sodium is harder to ionize than potassium, it is over an order of magnitude more abundant. Hence sodium ionization dominates at the highest temperatures (although efficient MRI is possibly already assured by potassium ionization alone). 

The alkali species may recombine radiatively; the equation for this process is identical to equation~\ref{2-rec-m}, except with the substitution of alk$^+$ for M$^+$. 
\begin{equation}
   \mathcal{R_{\mathrm{gas, \; 2-rec, \; alk^+}}} = \gamma n_{\mathrm{e}} n_{\mathrm{alk^+}} \,\,\,\,\, .
   \label{2-rec-alk}
\end{equation}
For the same reasons as earlier, we use the value of $\gamma$ from \citet{Oppenheimer_Dalgarno_1974}. Alternatively to the two-body process, the alkali may recombine through a three-body process, at a rate
\begin{equation}
    \mathcal{R_{\mathrm{gas, \; 3-rec, alk^+}}} = k_{-2} n_{\mathrm{H_2}} n_e n_{\mathrm{alk^+}} \,\,\,\,\, ,
\end{equation}
where $k_{-2} = 4.4 \pm 1.1 \times 10^{-24} \, T^{-1} \, \mathrm{cm^6 \, s^{-1}}$~\citep{Ashton_Hayhurst_1973} with $T$ in units of K.\footnote{The rate constant is denoted $k_{-2}$ as this process is the inverse of collisional ionization (equation~\ref{coll-ion}), for which the rate constant is denoted $k_{2}$.} Three-body recombination dominates over the two-body radiative recombination  at lower temperature and higher density but we include both processes in our calculations. 

The neutral alkali species, $\text{alk}^0$, and alkali ion species, ${\text{alk}^+}$, may be adsorbed by grains; adsorption by grains in the $i$th grain size bin gives a number density $n_{\text{alk cond}}^i$. Whether the condensed species is re-emitted to the gas phase as $\text{alk}^0$ or $\text{alk}^+$ depends on the effective work function of the grains and the temperature (see~\ref{subsec:grain-phase}).

\subsection{Grain phase chemistry}
\label{subsec:grain-phase}

Our network includes spherical dust grains, with a distribution of $N$ sizes, with a single value of charge for each grain size. The spherical approximation is necessary for dust-phase chemistry as this requires the computation of an electrostatic potential of the grain, for which only the sphere has an analytic form.\footnote{However, we note experimental investigations indicate a fractal dimension $D \sim 2$ to be more realistic~\citep[e.g.][]{Wurm_Blum_1998} and that the fractal dimension changes during coagulation~\citep{Okuzumi_2009}. We eschew this complication here.
} The number density of grains of size $a^{i}$ is $n_{\mathrm{gr}}^{i}$. The charge on grains of size $a^{i}$ is $Z^{i}$. The total number density of grains is $n_{\mathrm{gr}}^{\text{tot}} = \sum_{i=1}^{N_{\text{gr}}} n_{\mathrm{gr}}^{i}$ (in the remainder of this work we omit the limits on sums over the grain size distribution for brevity). We will consider both arbitrary distributions of grain sizes, as well as standard MRN distributions of grain sizes~\citep{MRN_1977}. 

For an MRN distribution, the number density of grains in range $[a, a + da]$ is $(dn/da) da = A a^{-q} da$. The normalization constant $A$ is determined by the integral $\int_{a_{\text{min}}}^{a_{\text{max}}} m (a)(dn/da) da = f_{\text{dg}} \rho$, where $a_{\text{min}}$ and $a_{\text{max}}$ are the minimum and maximum grain size respectively, $m(a)$ is the mass of the grains ($=\frac{4 \pi}{3} \rho_{\text{gr}} a^3$, where $\rho_{\text{gr}}$ is the bulk density of the grains), $f_{\text{dg}}$ is the dust-to-gas ratio and $\rho$ is the gas density. Performing the above integral, we obtain:
\begin{equation}
  A =
    \begin{cases}
       \frac{f_{\text{dg}} \rho}{\rho_{\text{gr}}} \left(\frac{1}{4\pi / 3}\right) (4-q) \left( \frac{1}{a_{\text{max}}^{4-q}} - \frac{1}{a_{\text{max}}^{4-q}} \right) & \text{if } q \neq 4\\
      \frac{f_{\text{dg}} \rho}{\rho_{\text{gr}}} \left(\frac{1}{4\pi / 3}\right) \left( \frac{1}{\ln(a_{\text{max}}/a_{\text{min}})} \right) & \text{if } q = 4 \,\,\,\,\, .
    \end{cases}       
\end{equation}
In this work, we assume a minimum grain size $a_{\text{min}} = 10^{-5} \, \text{cm}$, a maximum grain size $a_{\text{max}} = 10^{-1} \, \text{cm}$ and a fiducial $q = 3.5 \,$. To fix the normalization $A$, we use: a dust-to-gas ratio $f_{\text{dg}} = 10^{-2}$; a bulk density for the grains of $\rho_{\text{gr}} = 3.3 \mathrm{\, g \, cm^{-3}}$, corresponding to silicates (though this could be extended to account for compositional changes or porosity); and density $\rho = \mu m_{\text{H}} \left( \frac{2-x_\text{H}}{x_{\text{H}}} \right) n_{\text{H}_2}$ ($\rho \approx \mu m_{\text{H}} n_{\text{H}_2}$), where  $n_{\text{H}_2} = 10^{14}$\,g\,cm$^{-3}$. 

Dust grains adsorb gas phase species. The adsorption of species x by grains of size $a^i$ and charge $Z^i$ occurs with rate
\begin{equation}
    \mathcal{R}_{\mathrm{x}, \, \mathrm{ads}}^{i} (Z^i) = n_{\text{x}} n_{\mathrm{gr}}^i \pi ({a^{i}})^2 \left( \frac{8 k T}{\pi m_\text{x}} \right)^{1/2} S_\text{x} \Tilde{J}_{\text{x}} (Z^i e / q_x) = n_{\text{x}} \nu_{\text{x}}^{i} \,\,\,\,\, ,
\end{equation}
where $n_{\text{x}}$ is the number density of species x in the gas phase, $m_{\text{x}}$ is the mass of species x, $S_{\text{x}}$ is the sticking coefficient of species x, $\Tilde{J_\text{x}} (Z^i e / q_x)$ is the focusing factor for species x, of charge $q_x$, providing the enhancement to the geometrical collision cross-section~\citep{DS_87} and $e$ is the elementary charge. We take the sticking coefficient of electrons $S_e$ to be 0.6 and the sticking coefficient of ions and neutrals to be unity~\citep{Desch_Turner_2015, Jankovic_2021}. An exact calculation of the electron sticking coefficient was performed by~\cite{Bai_2011} and is broadly consistent with our choice.

The adsorption of $\mathrm{alk^0}$ and $\mathrm{alk^+}$ by grains with size $a^i$ gives a number density of condensed $\text{alk}$, $n_{\text{alk \, cond}}^i$. The condensed $\text{alk}$ vibrate on the grain lattice; the frequency of this vibration is taken to be the same value as~\cite{Desch_Turner_2015} used for the vibration of K on Pt of $\nu = 3.7 \times 10^{13} \;  \text{s}^{-1}$~\citep{Hagstrom_2000}. The vibration cycle has a probability $\exp(-\frac{E_a}{kT})$ of resulting in evaporation of alk~\citep{Desch_Turner_2015}, where $E_a$ is the activation energy. This results in a rate of evaporation of alk from grains of size $a^i$:

\begin{equation}
    \mathcal{R}_{\text{alk} \; \text{evap}}^i = n_{\mathrm{alk \; cond}}^i \nu \exp \left( -\frac{E_a}{k T} \right) = n_{\mathrm{alk \; cond}}^i \nu_{\text{evap}} \,\,\,\,\, .
    \label{eqn:evap}
\end{equation}
We follow~\cite{Desch_Turner_2015} in setting the activation energy to 3.25\,eV, and refer the reader to~\cite{Desch_Turner_2015} for a discussion of how this value was calculated.

The fraction of the alkali species $\text{alk}$ emitted from the grains as ions ($\mathrm{alk^+}$) is given by the Saha–Langmuir equation~\citep{Desch_Turner_2015} as
\begin{equation}
    f_+^i = \frac{1}{1 + \frac{g_0}{g_+} \exp \left( \frac{\text{IP}-W_{\text{eff}}^i}{kT}\right)} \,\,\,\,\,,
    \label{eqn:f+i}
\end{equation}
where $W_{\mathrm{eff}}$ is the effective work function of the grains. The effective work function is defined as $W_{\mathrm{eff}} = W - e\phi $, where $W$ is the work function and $\phi$ is the electrostatic potential of the grain. For spherical grains of size $a$, $\phi = -Ze/a$. We follow~\cite{Desch_Turner_2015} and take $W = 5.0 \; \text{eV}$.

This gives a rate of emission of $\mathrm{alk^+}$ from grains of size $a^i$ (termed ion emission) of
\begin{equation}
    \mathcal{R}^i_{\mathrm{alk^+ \; evap}} = f_+^i n_{\mathrm{alk \; cond}}^i \nu_{\text{evap}} = \nu_{\text{alk}^+}^i n_{\mathrm{alk \; cond}}^i\,\,\,\,\, .
\end{equation}
Therefore, the rate of emission of neutral atoms (alk$^0$) from grains of size $a^i$ is
{
\begin{equation}
    \mathcal{R}^i_{\mathrm{alk^0 \; evap}} = (1-f_+^i) n_{\mathrm{alk \; cond}}^i \nu_{\text{evap}} = \nu_{\text{alk}^0}^i n_{\mathrm{alk \; cond}}^i \,\,\,\,\, .
\end{equation}
}
The ability of grains to convert $\text{alk}^0$ to $\text{alk}^+$ through ion emission is an important contributor to the ionization fraction of the disc at high temperatures~\citep{Desch_Turner_2015}.

As~\cite{Desch_Turner_2015} also noted, at high temperatures grains are able to emit electrons from their surface due to thermionic emission. For grains of size $a^i$ and charge $Z^i$, electrons are emitted at a rate
\begin{equation}
    \mathcal{R}_{\mathrm{therm}}^{i} (Z^i) = n_{\mathrm{gr}}^i 4 \pi ({a^{i}})^2 \lambda_R \frac{4 \pi m_e (k T)^2}{h^3} \exp \left(-\frac{W_{\text{eff}}^i}{k T} \right) \,\,\,\,\, ,
    \label{thermionic}
\end{equation}
where $\lambda_{R}$ is the Richardson constant, experimentally $\approx 1/2$~\citep{Corwell_1965}. We take its value to be exactly $1/2$ in our calculations. At high temperatures, equation~\ref{thermionic} shows that the grains will also positively contribute to the ionization state through emission of electrons.

 \subsection{Reaction network}
 \label{subsec:reaction-network}

 The above processes give rise to the following equilibrium reaction network (where starred equations will be invoked further below)
\begin{align}
&\frac{d n_{\text{gr}}^i Z^{i}}{d t} = 
\mathcal{R}_{\mathrm{alk^+}, \; \mathrm{ads}}^{i} (Z^i) +
\mathcal{R}_{\mathrm{m^+}, \; \mathrm{ads}}^{i} (Z^i) +
\mathcal{R}_{\mathrm{M^+}, \; \mathrm{ads}}^{i} (Z^i)  - \mathcal{R}_{\mathrm{e}, \; \mathrm{ads}}^{i} (Z^i) 
\notag \\
&
+ \mathcal{R}_{ \mathrm{therm}}^{i} (Z^i) - \mathcal{R}^i_{\mathrm{alk^+ \; evap}}(Z^i) = 0
\asterequation{Zi} 
\\
&
\frac{d n_{\mathrm{alk}^+}}{dt} = -\mathcal{R_{\mathrm{gas, \; 3-rec, \; alk^+}}} - \mathcal{R_{\mathrm{gas, \; 2-rec, \; alk^+}}} + \mathcal{R_{\mathrm{gas, \; coll-ion}}} 
\notag \\
&
+ \sum_i \mathcal{R}^i_{\mathrm{alk^+ \; evap}}(Z^i)
- \sum_{i} \mathcal{R}_{\mathrm{alk^+}, \; \mathrm{ads}}^{i} (Z^i) = 0
\asterequation{nkp}
\\
&
\frac{d n_{\mathrm{e}}}{d t} = +\zeta 
n_{\mathrm{H}_{2}}- \mathcal{R_{\mathrm{gas, \; 2-rec, \; M^+}}} - \sum_{i} \mathcal{R}_{\mathrm{e}, \; \mathrm{ads}}^{i} (Z^i)  -\mathcal{R_{\mathrm{diss-rec}}}
\notag \\
& 
- \mathcal{R_{\mathrm{gas, \; 2-rec, \; alk^+}}} -\mathcal{R_{\mathrm{gas, \; 3-rec, \; alk^+}}}+\mathcal{R_{\mathrm{gas, \; coll-ion}}}
\notag \\
&
+ \sum_{i} \mathcal{R}_{ \mathrm{therm}}^{i} (Z^i) = 0
\asterequation{ne}
\\
&\frac{d n_{\mathrm{m^+}}}{d t} = +\zeta n_{\mathrm{H}_{2}} - \mathcal{R}_{\mathrm{ct}} -\mathcal{R_{\mathrm{diss-rec}}}- \sum_{i} \mathcal{R}_{\mathrm{m^+}, \; \mathrm{ads}}^{i} (Z^i) = 0 
\asterequation{nmol}
\\
&\frac{d n_{\mathrm{M^+}}}{d t} = +\mathcal{R}_{\mathrm{ct}}-\mathcal{R_{\mathrm{gas, \; 2-rec, \; M^+}}}- \sum_{i} \mathcal{R}_{\mathrm{M^+}, \; \mathrm{ads}}^{i} (Z^i) = 0 
\label{ni}
\\
&
\frac{d n_{\mathrm{alk^0}}}{d t} = \mathcal{R_{\mathrm{gas, \; 2-rec, \; alk^+}}} + \mathcal{R_{\mathrm{gas, \; 3-rec, \; alk^+}}} - \mathcal{R_{\mathrm{gas, \; coll-ion}}} 
\notag \\
&
+ \sum_i \mathcal{R}^i_{\mathrm{alk^0 \; evap}}(Z^i)
- \sum_{i} \mathcal{R}_{\mathrm{alk^0}, \; \mathrm{ads}}^{i} (Z^i) = 0
\label{nk0}
\\
&
\frac{d n_{\mathrm{alk \; cond}}^{i}}{d t} = \mathcal{R}_{\mathrm{alk^+}, \; \mathrm{ads}}^{i} (Z^i) + \mathcal{R}_{\mathrm{alk^0}, \; \mathrm{ads}}^{i} (Z^i) - \mathcal{R}_{\text{alk} \; \text{evap}}^i = 0
\,\,\,\, ,
\label{nkcond}
\end{align}
where all rates dependent on grain charge are explicitly indicated with a suffix ($Z^i$).
In addition to the above equations, we have an equation for charge conservation
\begin{equation}
    \sum_{i} Z^i n_{\text{gr}}^i + n_{\mathrm{alk}^+} + n_{\mathrm{m^+}}  + n_{\mathrm{M^+}} - n_{\mathrm{e}} = 0 \,\,\,\,\, .
\label{charge_conservation}
\end{equation}
Equation \eqref{nkcond} is linear in $n_{\text{alk cond}}^i$ and provides the equilibrium value of $n_{\text{alk cond}}^i$ directly
      \begin{equation}
        n_{\text{alk cond}}^i = \frac{1}{\nu_{\text{evap}}} (\nu_{\mathrm{alk^+}}^i n_{\mathrm{alk^+}} + \nu_{\mathrm{alk^0}}^i n_{\mathrm{alk^0}}) \,\,\,\,\,.
      \end{equation}
In combination with $n_{\text{alk tot}} = n_{\mathrm{alk^+}} + n_{\mathrm{alk^0}} + \sum_{i} n_{\text{alk cond}}^i$, this provides the equilibrium value of $n_{\mathrm{alk^0}}$ in terms of $n_{\mathrm{alk^+}}$

\begin{equation}
  n_{\mathrm{alk^0}} = \frac{n_{\text{alk tot.}} - \left(1 + \frac{\sum_{i}\nu_{\mathrm{alk^+}}^i}{\nu_{\text{evap}}}\right) n_{\mathrm{alk^+}}}{1 + \frac{\sum_{i}\nu_{\mathrm{alk^0}}^i}{\nu_{\text{evap}}}} \,\,\,\,\, .
\end{equation}

Therefore, we solve the subset of~\crefrange{Zi}{nmol} for $\left[ Z^1, \ldots, Z^{N_{\text{gr}}}, n_{\mathrm{alk}^+}, n_{\mathrm{e}}, n_{\mathrm{m}^+} \right]$, with $n_{\mathrm{m}^+}$ given by the charge conservation equation.
\begin{equation}
    n_{\mathrm{M^+}}  = n_{\mathrm{e}} - \sum_{i} Z^i n_{\text{gr}}^i - n_{\mathrm{alk}^+} - n_{\mathrm{m}^+}
\end{equation}

\subsubsection{Solution techniques}
  We have used two techniques to solve the above system of equations: (1) a globally convergent, relatively slow non-linear successive over-relaxation (nSOR) method~\citep[see e.g.,][]{Ortega_Rheinboldt_2000}; and (2) a locally convergent, faster multidimensional Newton-like method. The first method is employed for finding initial guesses, while the second is employed for evolutionary calculations where parameter values, e.g. $T$, $P$, $f_{\text{dg}}$, are changed by small amounts. We can also evolve between two sets of significantly differing parameter values by incrementally evolving the parameters with the Newton-like method to prevent divergence from the root while maintaining speed.

  Our system of equations can be represented as $\bm{F}(E_1, ..., E_{N}):\mathbb{R}^{N} \to \mathbb{R}^{N}$, where the equations $E_i$ correspond to~\crefrange{Zi}{nmol} and $N$ (as throughout) is the number of equations (i.e.,
 $N = N_{\text{gr}} + 3$). Each of these equations depends on the set of parameters, $x_i \, : i \in [1, N]$, i.e., on the values Z for each grain size in the distribution, $n_{\mathrm{alk}^+}$, $n_{\mathrm{e}}$, $n_{\mathrm{m}^+}$.
  
  The $i$th step of the $k$th iteration of the nSOR method works by finding the root of the $i$th equation with respect to $x_i$. The root of this equation is denoted $\hat{x}_i$.
  \begin{equation}
    E_i(x_1^{(k)}, ..., x_{i-1}^{(k)}, \hat{x}_{i}, .., x_{N}^{(k-1)}) = 0 \,\,\,\, .
    \label{eqn:sor}
  \end{equation}
  Note that the variables $x_j \, : j \in [1, i-1]$ take the value assigned to them at the $k$th iteration, while the variables $x_j \, : j \in [i+1, N]$ take their values from those at the $k-1$th iteration.
  
  The variable $x_i$ is updated at the $k$th iteration to

  \begin{equation}
      x_{i}^{(k)} = x_{i}^{(k-1)} + \omega^{(k)} (\hat{x}_{i} - x_{i}^{(k-1)}) \,\,\,\,\, ,
  \end{equation}
where $\omega^{(k)}$ is the relaxation parameter for the $k$th iteration with value in range $(0,2)$. For a fixed relaxation parameter of unity, successive over-relaxation is equivalent to the Gauss–Seidel method. The term over-relaxation reflects the possibility of having $\omega^{(k)}$ exceed unity. Typically the first few iterations require a value of $\omega^{(k)}$ which is less than unity, while later iterations can be allowed to approach the solution faster by a choice of $\omega^{(k)}$ in range $(1,2)$. Of order 100-1000 iterations are typically required to achieve an initial guess that enables the Newton-like method to converge.

The Newton-like method chosen is the modified Powell's Hybrid method~\citep{Powell_1970} first implemented in \texttt{MINPACK}~\citep{more1984minpack} and subsequently in \texttt{GSL}~\citep{gough2009gnu}. We use the latter in this work.
This is a trust-region algorithm, which first computes the Newton step $\bm{\delta x_N}$ by inverting ${\bm{\mathsf{J}}} \bm{\delta x_N} = - \bm{F}$ (where ${\bm{\mathsf{J}}}$ is the Jacobian matrix of the vector $\bm{F}$). The components of the Jacobian matrix are computed analytically. If the step is within the trust-region $\lvert \bm{\delta x} \rvert < \delta$ (the default value $\delta = 100$), it is accepted. Otherwise, the step taken is a linear combination of the Newton step $\bm{\delta x_N} = - {\bm{\mathsf{J}}}^{-1} \bm{F}$ and the gradient direction of $\lvert \bm{F} \rvert^2$.
\begin{equation}
    \bm{\delta x} = - \alpha {\bm{\mathsf{J}}}^{-1} \bm{F} - \beta \bm{\nabla} \lvert \bm{F} \rvert^{2} \,\,\,\,\, ,
\end{equation}
where $\alpha$ and $\beta$ are constants chosen to minimize $\lvert \bm{F} \rvert^{2}$, while keeping $\delta x$ within the trust-region. We set a criterion of $\lvert \bm{F} \rvert < 10^{-7}$ for convergence. Of order 10 iterations are typically required to achieve convergence.
\subsection{Validity of chemical equilibrium}
\label{subsec:chemical-equilibrium}

To assume chemical equilibrium in the context of a protoplanetary disc, the chemical time-scale must be shorter than the dynamical time-scale, thermal time-scale, and the shortest time-scale for collisions between grains. The first two conditions ensure constant $n_{\text{H}_2}$ and $T$, and also ensure that dynamics do not change the dust distribution. The third condition ensures that the underlying grain size distribution does not evolve collisionally on the time-scales of chemical reactions and that charge exchange between grains is negligible. Note that, if viscous heating provides the dominant heating rate (as we expect in the inner disc), the thermal time-scale ($\sim t_{\text{dyn}}/\alpha$) is (much) longer than the dynamical time-scale $t_{\text{dyn}}$ (since $\alpha$$<$1). Therefore, the thermal time-scale does not further constrain the chemical time-scale and is omitted from the analysis. { Below, we show how we calculate the chemical and grain collision time-scales.}

\subsubsection{Chemical time-scale}

To determine the chemical time-scale, we consider the time-scales for small perturbations away from the equilibrium solution to decay. The full system,~\crefrange{Zi}{nkcond}, without assuming equilibrium may be written in vector form
\begin{equation}
    \bm{\dot{x}} = \bm{F(\bm{x})} \,\,\,\,\, .
    \label{x-equation}
\end{equation}
Away from equilibrium, we have two constraints: charge conservation (equation \ref{charge_conservation}); and the constraint that alk number densities ($n_{\text{alk}^0}$, $n_{\text{alk}^+}$, $n_{\text{alk cond}}^{i}$) must sum to $n_{\text{alk}}^{\text{tot}}$. To avoid linear dependence due to these constraints, we remove equations (\ref{ni}) and (\ref{nk0}) from the system.

The system may be linearized about equilibrium i.e. $\bm{x} = \bm{x_0} + \bm{\Delta x}$, $\bm{\dot{x}} = \bm{\dot{\Delta x}}$ to give
\begin{equation}
    \bm{\dot{\Delta x}} = \bm{F}(\bm{x_0} + \bm{\Delta x}) \approx \bm{F}(\bm{x_0}) + {\bm{\mathsf{J}}}(\bm{x_0}) \bm{\Delta x} \,\,\,\,\, ,  
    \label{eqn:lambda}
\end{equation}
where ${\bm{\mathsf{J}}}$ is the Jacobian matrix.

Considering a perturbation about equilibrium of the form $\bm{\Delta x} \propto e^{\lambda t}$, an eigen equation must be solved for the values of $\lambda$
\begin{equation}
({\bm{\mathsf{J}}}-\lambda {\bm{\mathsf{I}}}) \bm{\Delta x} = \bm{0} \,\,\,\,\, .
\end{equation}
For a stable equilibrium, it is required that the real part of $\lambda_i$ satisfies $\Re (\lambda_i) < 0 \, \forall \, i$. Provided this is satisfied, the chemical time-scale will be
\begin{equation}
t_{\text{chem}} = \max \left( \frac{1}{\lvert \Re (\lambda_i) \rvert} \right) \,\,\,\,\, .
\label{eqn:chemical-time}
\end{equation}
Concurrently, the dynamical time-scale, at semi-major axis $r$, is
\begin{equation}
    t_{\text{dyn}}(r) = \frac{1}{\Omega_K (r)} \,\,\,\,\, ,
    \label{eqn:dynamical-time}
\end{equation}
where $\Omega_K (r)$ is the Kepelerian frequency at $r$.

\subsubsection{Grain collision time-scale}

The shortest time-scale for one grain to collide with another is given by
{
\begin{equation}
    t_{\text{coll. dust}} = \min_{i,j} \left( \frac{1}{K_{ij} n_{\text{gr}}^j} \right) \,\,\,\,\, ,
    \label{eqn:dust-time}
\end{equation}
where $K_{ij}$ is termed the collision kernel. The full collision kernel is defined in~\cite{Okuzumi_2011} as an integral over the Maxwellian distribution of relative velocities $\mathbf{\Delta {\rm v}}_{ij}$, centred on $\mathbf{\Delta {\rm v}}^D_{ij}$, the sum of the systematic (non-Brownian) relative velocity sources (or drift velocities). This distribution may be denoted as $P(\lvert \mathbf{\Delta {\rm v}}_{ij} - \mathbf{\Delta{\rm v}}_{ij}^D \rvert)$\,. Hence, we have
\begin{equation}   
K_{ij} = \int P(\lvert \mathbf{\Delta {\rm v}}_{ij} - \mathbf{\Delta{\rm v}}_{ij}^D \rvert) \, \sigma_{ij} \, \lvert \mathbf{\Delta {\rm v}}_{ij} \rvert \, d \mathbf{\Delta {\rm v}}_{ij} \,\,\,\,\, ,    
\label{eqn:Kij}
\end{equation}
where $\sigma_{ij}$ is the collision cross-section including the {\it Coulomb} electrostatic interaction between the grains. Note that the velocities are defined at infinite separation.\footnote{The screening-length (e.g. Debye length), due to the re-distribution of electrons surrounding each charged grain, is shorter than the mean free path of grains for all but the highest densities, temperatures and dust-to-gas ratios, and lowest ionization fractions. In this region of parameter space, the changing relative velocity of grains $\mathbf{\Delta \rm{v}}_{ij}$ (due to the electrostatics), as they approach one another, will affect the collision rate appreciably. Considering this effect makes $\mathbf{\Delta \rm{v}}_{ij}^D$ a function of the impact parameter $b$ and separation of the grains $r$. Therefore, $K_{ij}$ should be computed as an integral over $b$ and $r$. We leave this complication to later work, if required.}

\cite{Okuzumi_2011} provide limiting forms for $K_{ij}$ (now referring to absolute values of the velocities without the vector notation) in the case of like charged particles interacting through a Coulomb potential:
\begin{subnumcases}{K_{ij} \approx}
\pi (a_i + a_j)^2 \times \notag
\\
\Delta {\rm v}_{ij}^{\text{Br}} \exp\left(-\frac{U(a_i + a_j)}{k T}\right),& if $\Delta {\rm v}_{ij}^{D} \ll \Delta {\rm v}_{ij}^{\text{Br}}$, \label{eqn:case1} \\
\pi (a_i + a_j)^2 \times \notag
\\ 
\Delta {\rm v}_{ij}^{D} \left(1-\frac{U(a_i + a_j)}{\text{KE}_{ij}^{D}}\right),& if $\Delta {\rm v}_{ij}^{D} \gg \Delta {\rm v}_{ij}^{\text{Br}}$. \label{eqn:case2} 
\end{subnumcases}
where $U(a_i + a_j)$ is the electrostatic potential at the point of contact, $\Delta {\rm v}_{ij}^{\text{Br}} = \left( \frac{8 k T (m_i + m_j)}{\pi m_i m_j}\right)^{\frac{1}{2}}$ is the relative velocity due to Brownian motion and $\text{KE}_{ij}^D = \frac{1}{2} \frac{m_i m_j}{m_i + m_j} \left(\Delta {\rm v}_{ij}^{D}\right)^{2}$ is the kinetic energy due to drift at infinite separation. Note that, formally equation (\ref{eqn:case2}) is only applicable in the limit $\text{KE}_{ij}^D \gg U(a_i + a_j)$, i.e., weak Coulomb effects. However, we, like~\cite{Akimkin_2023}, use this limiting form for the entire range $\text{KE}_{ij}^D > U(a_i + a_j)$,  for the purpose of computational efficiency. Also following~\cite{Akimkin_2023}, we set $K_{ij} = 0$, if $\text{KE}_{ij}^D < U(a_i + a_j)$. In reality, a non-zero $K_{ij}$ persists, because there will always be a contribution to $K_{ij}$ due to the Maxwellian distribution of velocities, even when $\Delta {\rm v}_{ij}^{D} \gg \Delta {\rm v}_{ij}^{\text{Br}}$.

\cite{Akimkin_2023} also derive a form for $\sigma_{ij}$ including the effect of a dipole between the grains, which requires root-finding for each pairwise interaction; for computational efficiency we avoid this complication.}

Sources of {(non-Brownian)} relative velocities  {$\Delta {\rm v}_{ij}^{D}$} between grains include: turbulent eddies~\citep{OrmelCuzzi_2007}, settling~\citep{Weidenschilling_1980}, radial and azimuthal drift~\citep{Weidenschilling_1977} and ambipolar drift~\citep{Guillet_2020}. We consider each of these in turn.\\

{\noindent \it Turbulence}\\

We take the relative velocities due to turbulence $\Delta
{\rm{v}}_{ij}^{\text{turb}}$ to be given by the limiting solutions provided in equations (26), (28) and (29) of~\cite{OrmelCuzzi_2007}.\\

{\noindent \it Settling}\\

Differential settling of dust grains gives a contribution
\begin{equation}
     \Delta {\rm{v}}_{ij}^{\text{settle}} = \lvert \Delta \text{St}_{ij} \rvert \, \Omega_{K} z \,\,\,\,\, ,
\end{equation}
where $z$ is the vertical displacement above the mid-plane and $\lvert \Delta \text{St}_{ij} \rvert$ is the absolute value of the difference in the dimensionless stopping times (Stokes numbers) of particles $i$ and $j$. To determine the Stokes number, we must know which drag regime we are in: Stokes ($a\gtrsim l_{\text{mfp}}$, where $a$ is the particle size and $l_{\text{mfp}}$ is the mean free path of gas) or Epstein ($a\lesssim l_{\text{mfp}}$). In terms of the disc variables, the mean free path of gas is:
\begin{equation} 
    l_{\text{mfp}} = \frac{1}{n \sigma} \approx 5 \left( \frac{n_{\text{H}_2}}{10^{14} \, \text{cm}^{-3}}\right)^{-1} \, \text{cm} \,\,\,\,\, ,
\end{equation}
where $n$ is the gas number density, and $\sigma \approx 2 \times 10^{-15} \, \text{cm}^{-2}$ is the cross-section for collisions between gas particles, assumed to be $\text{H}_2$ molecules. The approximate mean-free path we have derived exceeds the maximum grain size $a_{\text{max}} = 0.1$\,cm that we use; therefore, we expect to be within the Epstein regime at 1\,au.

In the Epstein regime, the dimensionless stopping time of a particle is 
\begin{equation}
    \text{St}_{i} = \frac{\rho_{\text{gr}} a^i}{\rho c_s} \Omega_K \,\,\,\,\,.
\end{equation}
If we are not in the Epstein regime, either due to the presence of large particles or an increase in density (at smaller semi-major axis), we merely have to compute a form of the dimensionless stopping time valid for the Stokes regime.\\

{\noindent \it Radial and azimuthal drift}\\

The terms due to radial and azimuthal drift depend on the underlying disc structure; therefore, their inclusion is left to future work where a disc model is present. However, in the inner disc, where the dimensionless stopping times are small, we expect these contributions to be small.\\

{\noindent \it Ambipolar drift}\\

{The final driver of relative velocities that we consider is ambipolar drift~\citep{Guillet_2020}. In the context of protoplanetary discs, while polycyclic aromatic hydrocarbons (PAHs) may be well–coupled to the field~\citep{Mohanty_2013}, grains are not; therefore, the relative velocity due to ambipolar drift is negligible. To see this, consider the Hall parameter for grains, $\beta_{\text{gr}}$ (ratio of the gyrofrequency of grains to their collision frequency with neutrals, which measures the degree to which grains are coupled to the field).

\begin{multline}
    \beta_{\text{gr}} = \frac{\lvert{Z}\rvert e B}{m_{\text{gr}} c}\frac{1}{\gamma_{\text{gr}} \rho_{n}} \approx 6 \times 10^{-6} \, \lvert Z \rvert \left(\frac{B}{10 \,\, \text{G}} \right) \left( \frac{a}{10^{-5} \, \text{cm}} \right)^{-2} \\ \times
    \left(\frac{T}{800 \, \text{K}}\right)^{-\frac{1}{2}} \left(\frac{n_{\text{H}_2}}{10^{14} \, \text{cm}^{-3}} \right)^{-1} \,\,\,\, ,
    \label{eqn:grain_hall}
\end{multline}
where $\gamma_{\text{gr}} = \langle \sigma {\rm v} \rangle_{\text{gr}} / (m_{\text{gr}} + m_{n}) \approx \langle \sigma {\rm v} \rangle_{\text{gr}} / m_{\text{gr}} $ is the drag coefficient for the grains due to collisions with the neutrals, $\langle \sigma {\rm v} \rangle_{\text{gr}}$ being the rate constant for momentum transfer between grains and neutrals, $m_{\text{gr}}$ is the mass of a given grain and $m_{\text{n}}$ the mass of the neutrals. The density of the neutrals is $\rho_n \approx \mu m_{\text{H}} n_{\text{H}_2}$. We follow \cite{Mohanty_2013} in using  $\langle \sigma {\rm v} \rangle_{\text{gr}} = \pi a^2 \left( \frac{128 k T}{9 \pi m_n}\right)^{\frac{1}{2}} \,$cm$^{3}$\,s$^{-1}$\,, from ~\cite{Wardle_Ng_2013}. Note that $\beta_{\text{gr}}$ is independent of the mass of the grain $m_{\text{gr}}$, and depends only on the grain cross-sectional area $a^2$.

We see from~\cref{eqn:grain_hall} that $\beta_{\text{gr}} \ll 1$, and thus the grains are completely decoupled. From the weak dependence of $\beta_{\text{gr}}$ on disc properties, we do not expect this to change throughout the inner disc. Therefore, relative velocity due to ambipolar drift may be neglected.}\\

{\noindent \it Combining sources}\\

Relative velocities between grains are typically combined in quadrature~\citep[e.g.,][]{Robinson_2024}
\begin{equation}
    \Delta {\rm{v}}_{ij}^{D} = \sqrt{\lvert \boldsymbol{\Delta} \boldsymbol{{\rm v}}_{ij}^{\text{lam}} \rvert ^2 + (\Delta {\rm v}_{ij}^{\text{turb}})^2} \,\,\,\,\, ,
    \label{eqn:quad-velocities}
\end{equation}
where $\boldsymbol{\Delta{\rm v}}_{ij}^{\text{lam}}$ is the drift velocity between two grains due to laminar sources, while $\Delta {\rm v}_{ij}^{\text{turb}}$ is the drift velocity due to turbulence. Note that we have not included a source due to Brownian motion in the above equation, as this is already accounted for by the integration over a Maxwellian in~\cref{eqn:Kij}. Of the sources that we have listed, the only laminar source of relative velocities that we consider is due to settling. 

In this work, we shall compute the time-scales for collisions between grains due to Brownian motion, turbulence and settling separately. For Brownian motion, we use the first limiting solution in~\cref{eqn:Kij} to compute $t_{\text{coll. dust}}^{\text{Br}}$. For turbulence and settling, we use the second limiting solution of~\cref{eqn:Kij}. Instead of summing velocities in quadrature, i.e. 
~\cref{eqn:quad-velocities}, $\Delta {\rm{v}}_{ij}^{D} = \Delta {\rm v}_{ij}^{\text{turb}}$ is used to compute $t_{\text{coll. dust}}^{\text{turb}}$ and  $\Delta{\rm{v}}_{ij}^{D} = \Delta {\rm v}_{ij}^{\text{settle}}$ used to compute $t_{\text{coll. dust}}^{\text{settle}}$. If $t_{\text{coll. dust}}^{\text{turb}}$ or $t_{\text{coll. dust}}^{\text{settle}}$ is the shortest time-scale, the overall shortest time-scale for collisions between grains $t_{\text{coll. dust}}$ can then be, at most, a factor of $\sqrt{2}$ shorter $\left(\text{if } t_{\text{coll. dust}}^{\text{turb}} = t_{\text{coll. dust}}^{\text{settle}}\right)$. Otherwise if $t_{\text{coll. dust}}^{\text{Br}}$ is shorter than the other dust collision time-scales $t_{\text{coll. dust}} = t_{\text{coll. dust}}^{\text{Br}}$.

\section{Results}
\label{sec:results}

\subsection{Tests}
\label{subsec:tests}

\subsubsection{Desch \& Turner (2015)}

Here we: (i) verify that the results of our network, with a distribution of grain sizes, agree with those of the single grain size network of~\cite{Desch_Turner_2015}, in the limit of the distribution of grain sizes tending to a single size; and (ii) investigate the change in the ionization state caused by broadening the grain size distribution. 

To test that our network is consistent with that of~\cite{Desch_Turner_2015}, we first simplify the physics of our network to more closely match theirs. Specifically, we remove the intermediate molecular ion species m$^+$ \,from our network and assume a 100$\%$ efficiency of charge transfer from $\text{H}_2^+$ \,to $\text{M}^0$. This entails replacing the charge transfer term in \cref{ni} with $ \zeta n_{\mathrm{H_2}}$, and removing the dissociative recombination term from \cref{ne}.

With this simplified physics in our network, we compute the ionization fractions in the disc as a function of $a_{\text{max}}$. For comparison with~\cite{Desch_Turner_2015}, we use the same disc parameters they do: an MRN distribution with  $q=3.5$ and $a_{\text{min}} = 10^{-4} \text{ cm}$, dust-to-gas ratio $f_{\text{dg}} = 0.01$, $T = 1000$ K and $n_{\text{H}_2} = 10^{14} \text{ cm}^{-3}$. The value of $a_{\text{min}} = 10^{-4} \text{ cm}$ (which is different from our fiducial value of $10^{-5} \text{ cm}${)} is chosen to match the single grain size used by~\cite{Desch_Turner_2015}. We have also used $\zeta = 1.4 \times 10^{-22} \, \text{s}^{-1}$, established by~\cite{Jankovic_2021} to be the actual non-thermal ionization rate that~\cite{Desch_Turner_2015} had used in their simulations. 

With the physics and parameter values matching those used by~\cite{Desch_Turner_2015}, our results are plotted in Fig.~\ref{fig:test-dt}. It is clear that our network converges to the single grain size network of~\cite{Desch_Turner_2015} in the limit that our distribution collapses to a single grain size i.e., as $a_{\text{max}} \to a_{\text{min}}$. 

Conversely, as $a_{\text{max}}$ increases, i.e., as the distribution broadens, our network moves away from the single grain size results. As $a_{\text{max}}$ increases at fixed dust-to-gas ratio (i.e., fixed total mass of grains), the total surface area of grains decreases.

First, as $a_{\text{max}}$ increases, the decreased total surface area causes a decrease in the total amount of charge condensed on the grains.
Second, the Mg$^+$ abundance increases with increased $a_{\text{max}}$, also due to the reduced total surface area for adsorption of Mg$^+$. Third, the electrons and K$^+$ behave differently from Mg$^+$. Mg$^+$ is produced exclusively in the gas-phase, but this is not true for electrons and K$^+$. As we shall see, the dominant mechanism for production of electrons and K$^+$ are on the surface of the grains. On the one hand, this means that, as we decrease the total surface area, we decrease the rate of K$^0$ adsorption, which is required to produce electrons and K$^+$ on the grains. On the other hand, there is a proportionate decrease in the rate of adsorption of electrons and K$^+$ with decreased total surface area of grains. Taken together, this means that the abundance of electrons and K$^+$ stay approximately constant as a function of $a_{\text{max}}$.
\begin{figure}
  \centering
  \includegraphics[width=\columnwidth]{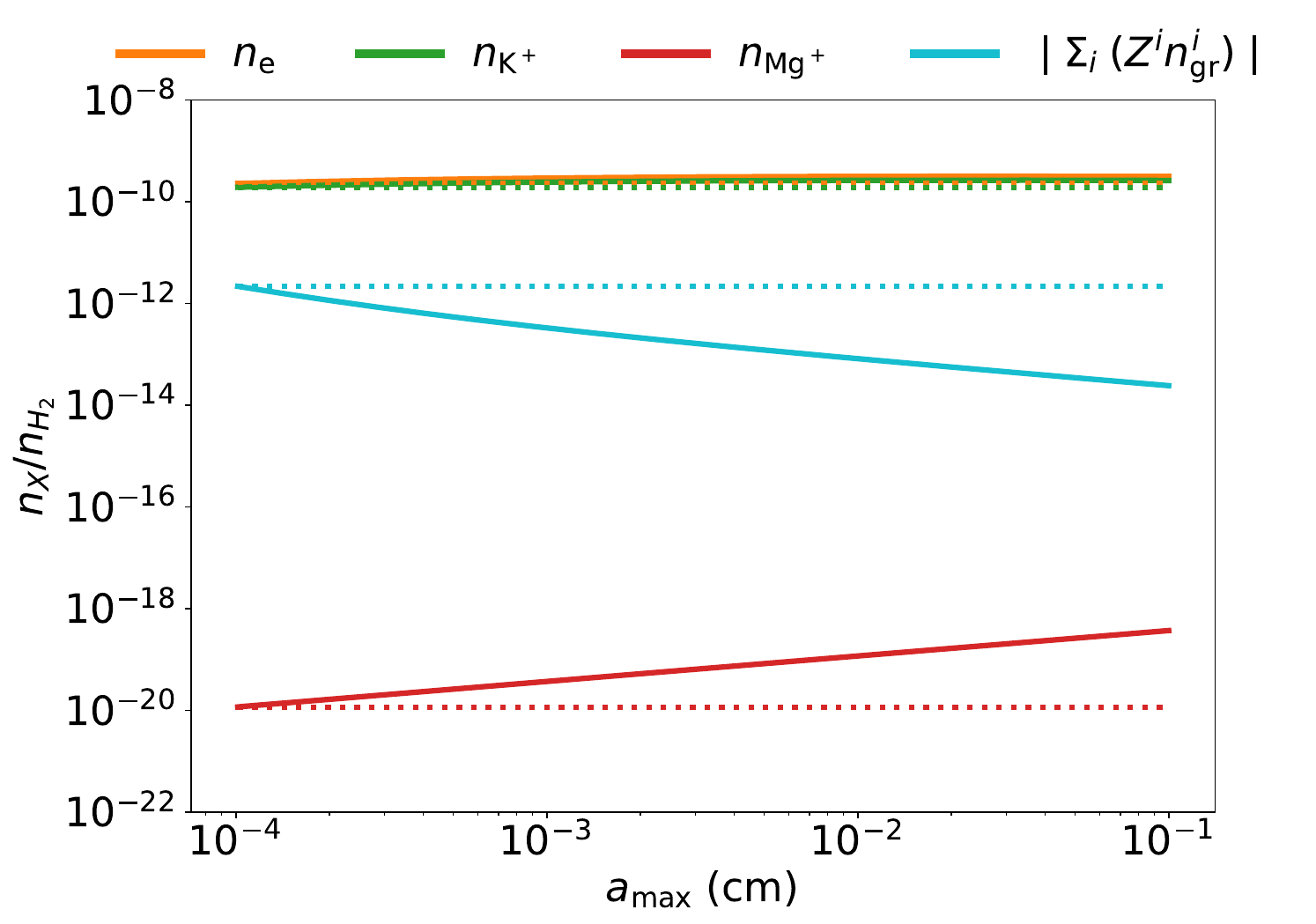}
  \caption{Solid lines show number densities of charged species as a function of the maximum grain size of the distribution of grains, $a_{\text{max}}$ with $a_{\text{min}} = 10^{-4} \text{ cm}$, $q=3.5$ and $f_{\text{dg}} = 0.01$. $\lvert \Sigma_i (Z^i n_{\text{gr}}^i) \rvert$ shows the total charge density on the grains. The temperature $T = 1000 \; \text{K}$ and $n_{\text{H}_2} = 10^{14} \; \mathrm{cm^{-3}}$. The dotted line shows the \protect\cite{Desch_Turner_2015} result for a single grain size with $a = a_{\text{min}}$. As expected, the abundances for the network with a  distribution of grains tends to this \protect\cite{Desch_Turner_2015} result as $a_{\text{max}} \to a_{\text{min}}$.}
  \label{fig:test-dt}
\end{figure}

\subsubsection{Marchand et al. (2022)}
\label{subsubsec:marchand}

\cite{Marchand_2022_2} solve an equilibrium chemical network with: (i) a slightly different solution technique to our own; and (ii) a distribution of charges for each grain size. We wish to verify that our network and solution technique can reproduce their results. To do this, we must first align our physics with ~\cite{Marchand_2022_2}. This includes excluding: thermionic and ion emission; the molecular ion, intermediate between $\text{H}_2^+$ and the metal ion ($\text{Mg}^+$). Conversely, we include non-thermal ionization of the alkali metal (which has little practical impact, and is thus not included elsewhere). We incorporate these choices in our network. 
On the other hand, we ignore charge transfer between the metal ($\text{Mg}$) and alkali ($\text{K}$) (which is a small effect), but do include three-body gas recombination for the alkali ions, in our calculation of their network.

The parameters we choose are an MRN distribution of dust grains with $q=3.5$ with $a_{\text{min}} = 10^{-5} \text{ cm}$, $a_{\text{max}} = 10^{-1} \text{ cm}$, $f_{\text{dg}} = 0.01$ and $n_{\text{H}_2} = 10^{14} \; \mathrm{cm^{-3}}$. 

The comparison is shown in Fig.~\ref{fig:test-marchand}. We see an excellent agreement between the two models, despite the difference in solution technique and absence of a distribution of grain charges for each grain size in our model. We discuss the reason for the last point in Subsection \ref{subsec:dist_charge}.
\begin{figure}
    \centering
    \includegraphics[width=\columnwidth]{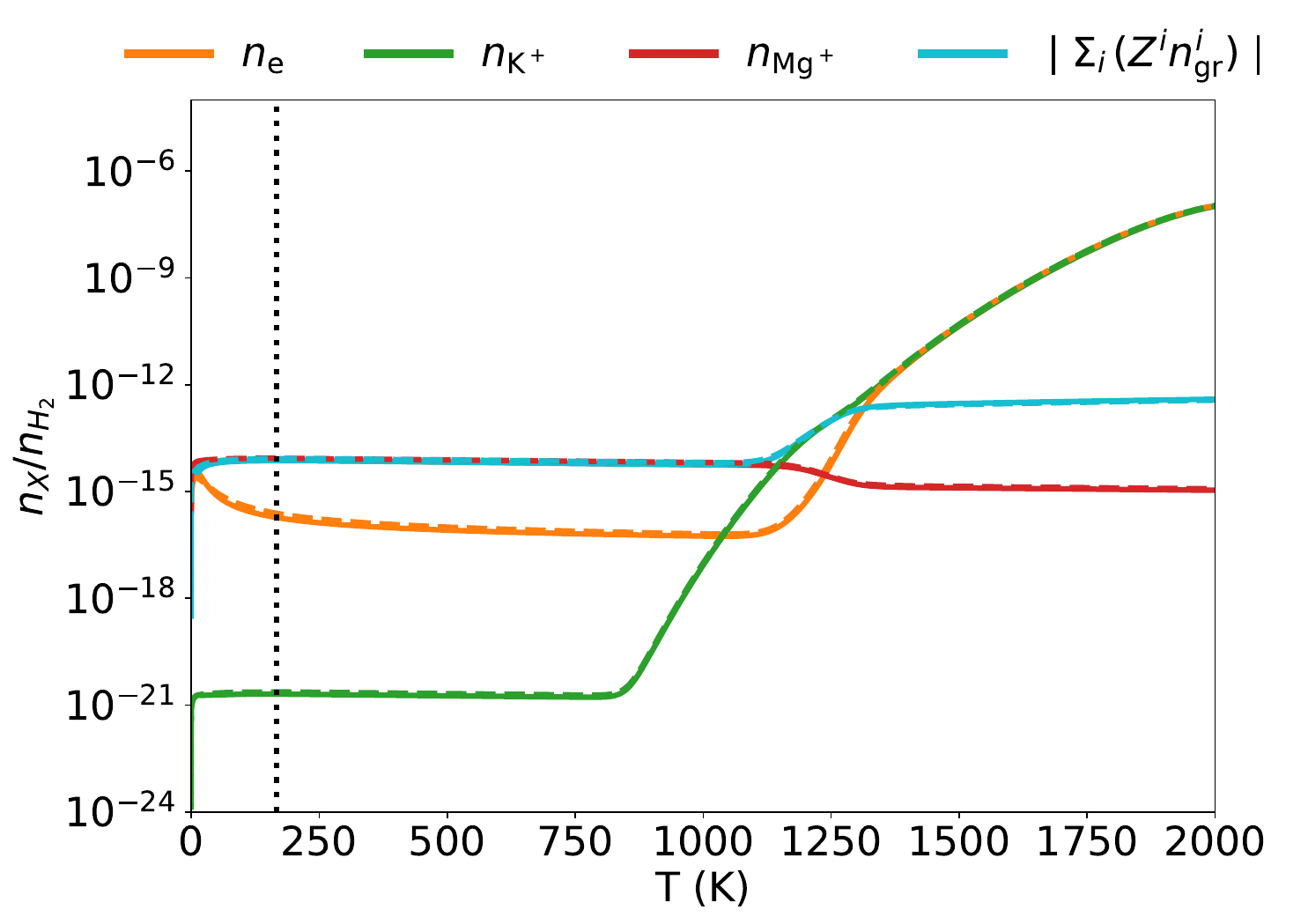}
    \caption{Ionization as a function of $T$ of our network (solid) and the \protect\cite{Marchand_2022_2} network (dashed) with modifications described in the text for $q=3.5$ and $a_{\text{min}} = 10^{-5} \text{ cm}$, $a_{\text{max}} = 10^{-1} \text{ cm}$, $f_{\text{dg}} = 0.01$ and $n_{\text{H}_2} = 10^{14} \; \mathrm{cm^{-3}}$. $\lvert \Sigma_i (Z^i n_{\text{gr}}^i) \rvert$ shows the total charge density on the grains. The black dashed vertical line indicates the temperature above which $\tau \equiv akT/e^2 > 1$ for all grains in the network {(see \ref{subsec:dist_charge} for the significance of $\tau > 1$)}. Good agreement between the models is obtained for all $T$, not simply those temperatures where $\tau > 1$ for all grains.}
    \label{fig:test-marchand}
\end{figure}
\subsubsection{Result of tests}
We have seen that our network reproduces the abundances obtained previously for a single grain size, and the abundances obtained with a similar network for a distribution of grain sizes but without ion and thermionic emission.

\subsection{Network behaviour}
\label{subsec:network-behaviour}
In this section, we seek to isolate the impact of grains in our network and explore this impact as a function of temperature.

In Fig.~\ref{fig:network-comparison}, we plot the equilibrium number densities of species as a function of temperature: {\it solid lines} show the case of our full network including grains (i.e., solving the full network described in Subsection \ref{subsec:reaction-network}), while {\it dashed lines} show the no-grain case (i.e., excluding the grain terms from our network). Both cases are for a single value of $n_{\text{H}_2} = 10^{14} \; \mathrm{cm^{-3}}$. We discuss the behaviour in the low and high temperature regimes below.

\subsubsection{Low temperature}

\begin{figure}
  \centering
    \includegraphics[width=\columnwidth]{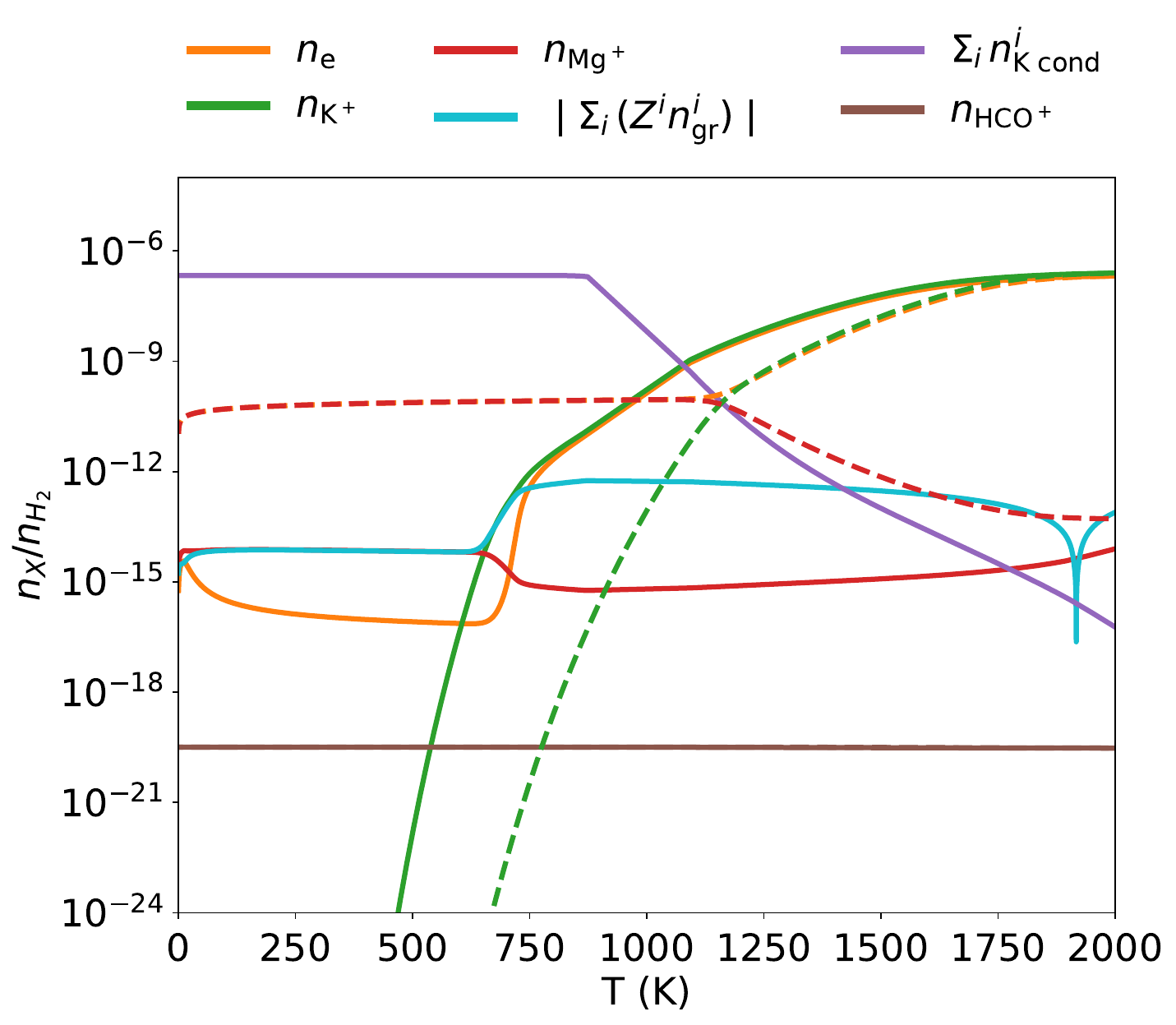}
  \caption{Solid lines show number densities as a function of temperature for an MRN distribution of grains with $a_{\text{min}} = 10^{-5} \text{ cm}$, $a_{\text{max}} = 10^{-1} \text{ cm}$, $q=3.5$ and $f_{\text{dg}} = 0.01$. Dashed lines show the grain-free case (for which there is no condensed potassium or grain charge). For the case with grains, $\lvert \Sigma_i (Z^i n_{\text{gr}}^i) \rvert$ shows the total charge density on the grains and $\lvert \Sigma_i (n_{\text{K cond}}^i) \rvert$ is the total number density of condensed potassium on the grains. For both cases, $n_{\text{H}_2} = 10^{14} \; \mathrm{cm^{-3}}$.}
  \label{fig:network-comparison}
\end{figure}

At low temperatures ($T\lesssim 500$K for the dusty case, $T\lesssim 1100$K for the dust-free case), we see that the dominant charged species in the gas-phase are electrons and Mg$^+$, our metal ion. This is a result of non-thermal ionization, producing electrons and molecular ions (here these are HCO$^+$ ions). These molecular ions then transfer their charge to neutral metal atoms (here Mg$^0$), making Mg$^+$.  

In the model with grains, we see that the abundances of electrons and Mg$^+$ are orders of magnitude lower than in the dust-free case, implying that adsorption onto grains is the dominant mechanism for removing these species from the gas-phase. 

Moreover, in the model with grains, at temperatures $\lesssim 600$ K, we see that the electron abundance decreases with increasing temperature. This is due to the increased thermal velocity of the electrons: as the velocity increases with temperature, electrons can overcome the electrostatic repulsion of the negatively charged grains, and are thus adsorbed onto the grains. { Note that Mg$^+$ is actually attracted to negatively charged grains; there is no repulsion to overcome, and its gas phase abundance remains approximately constant below $\lesssim 600$ K. Once the electrons cross the electrostatic repulsion barrier of the grains, their abundance in the gas-phase become orders of magnitudes below the abundance of Mg$^+$ on account of their larger thermal velocities.} 

Finally, in both the dusty and dust-free cases, and throughout the temperature range, the abundance of the molecular ion (HCO$^+$) is identical and constant. This implies that the HCO$^+$ abundance is not dependent on the other gas phase abundances. This can only be the case if charge transfer is the dominant pathway for removal of HCO$^+$. In this case, the HCO$^+$ abundance depends only on the number density of neutral metal atoms $n_{\text{Mg}^{0}}$ that charge is transferred to, and  $n_{\text{Mg}^{0}} \approx n_{\text{Mg}^{\text{tot}}} = \text{constant}$ due to the large total abundance of $\text{Mg}$.

\subsubsection{High temperature}
For both the network with grains and without, a sharp increase in the ionized potassium abundance as a function of temperature is observed due to thermal ionization sources. For the network with grains, this rise begins at lower temperature. This is due to ion emission from the grain surface. If the grains are not too negatively charged, the work function of the grains is a deeper potential well than the ionization potential of the alkali atom. This means all the neutral alkali on the grain surface lose their outermost electron to the grains. When evaporation of these K$^+$ from the surface of the grains begins to occur at sufficiently high temperatures ($kT \to E_a$), the gas phase number density of K$^+$ increases rapidly. Since the activation energy of the grains ($E_a$) is lower than the first ionization potential of potassium (for $kT\to$IP collisional ionization of potassium occurs in the gas phase), ion emission is able to increase the abundance of potassium ions in the gas phase considerably at a lower temperature than is possible without grains.

After the increase in the potassium ion abundance with $T$, for the model with grains, there is a small lag before the electron abundance also increases with temperature. As the gas phase potassium ion abundance increases with temperature due to ion emission, the grains become more negatively charged. Coupled with the increased thermal energy, this means electrons can more readily escape the grains through thermionic emission. The origin of the lag in temperature for this to occur is the fact that the grain charge must become more negative, and the thermal energy greater, until the effective work function of the grains can be overcome. As noted by~\cite{Desch_Turner_2015}, this is when $\lvert Ze^2/a \rvert \to W$. When the temperature becomes sufficiently high, both the K$^+$ ion and electron produced by ionization of K$^0$ on the surface of the grains are immediately released. Thermionic emission becomes increasingly efficient; hence, the grains becomes more positive. 

The average charge on grains as a function of temperature is plotted in Fig.~\ref{fig:z-T}, and compared to the average grain charge we would obtain without ionic and thermionic emission. { At low temperatures ($T \lesssim 600 \,$K), the average grain charge for the networks with and without emission is identical, since no emission process is important at these temperatures. Above this threshold, the average grain charge in the network with emission decreases rapidly to a minimum. This sharp change in average grain charge with temperature, in the network with emission, is due to the exponential onset of ion emission once $kT \to E_a$. At the minimum of the grain charge, ionization of K$^0$ to K$^+$ on the surface of the grains is impeded, and recombination of K$^+$ on the surface becomes more favourable. Emission of electrons through thermionic emission also increases with increasing temperature. Both of these factors conspire to increase the average grain charge with increasing temperature.

On the other hand, the average charge for the network without emission remains roughly constant until the temperature exceeds $\sim 1200 \,$K. Above this temperature, the average grain charge decreases, rapidly at first due to the onset of collisional ionization, and then more slowly as electrons gain higher thermal velocities.}

Note that we use silicate grains in our network, which should sublimate at $\gtrsim 1500$\,K. However, at these temperatures the abundance of gas-phase species converge to the same values for the dusty and grain-free cases, so correcting for the sublimation of grains would not alter the gas phase abundances significantly.

\begin{figure}
  \centering        
  \includegraphics[width=\columnwidth]{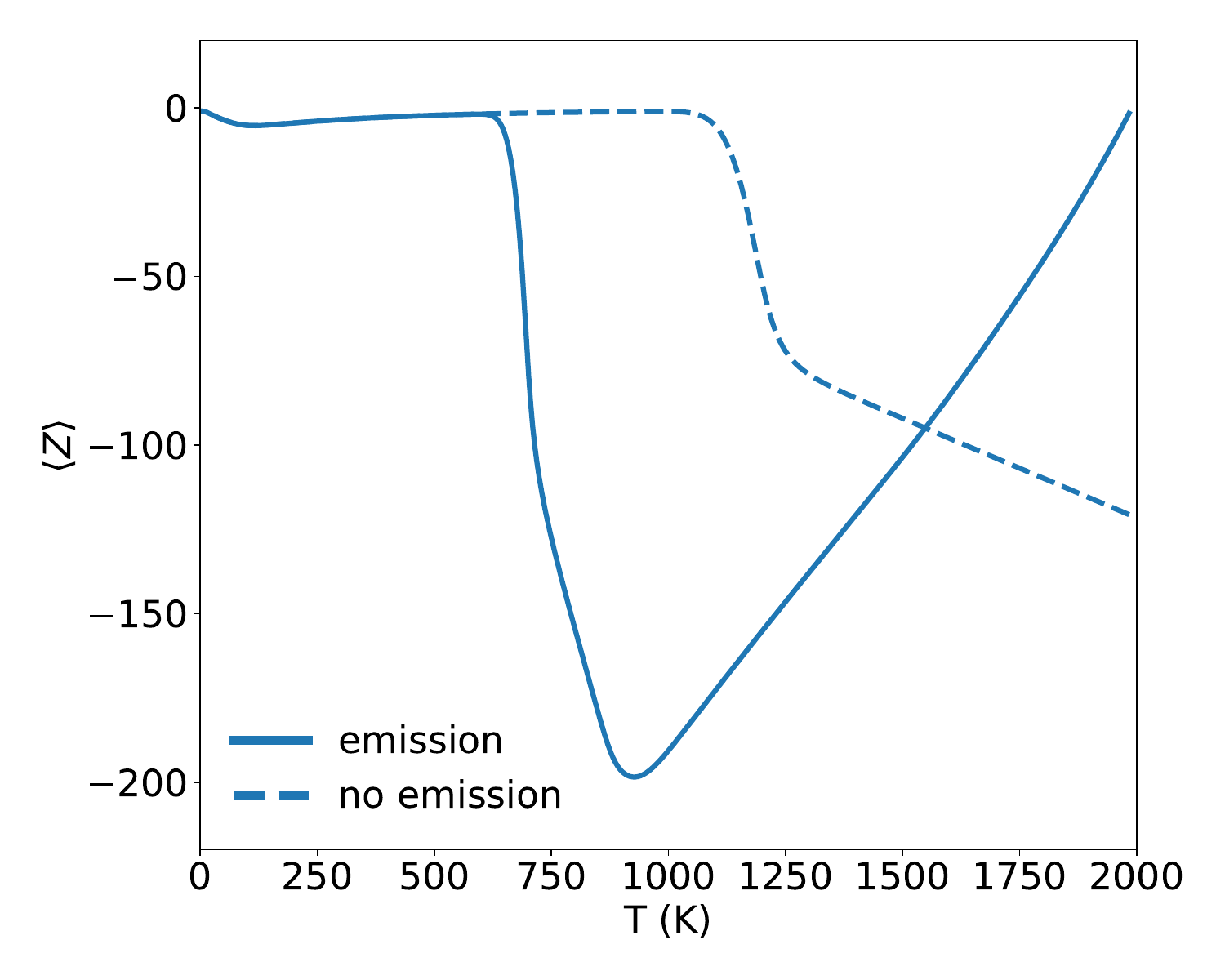}
  \caption{The solid line shows the average charge on the grains, $\langle Z \rangle = \sum_i Z^i n_{\text{gr}}^i / \sum_i n_{\text{gr}}^i$, as a function of temperature for an MRN distribution of grains with $a_{\text{min}} = 10^{-5} \text{ cm}$, $a_{\text{max}} = 10^{-1} \text{ cm}$, $q=3.5$ and $f_{\text{dg}} = 0.01$ at $n_{\text{H}_2} = 10^{14} \; \mathrm{cm^{-3}}$. The dashed line illustrates the case for the same population of grains and $n_{\text{H}_2}$, but without the inclusion of thermionic or ion emission.}
  \label{fig:z-T}
\end{figure}

\subsection{Distribution of charge on grains}
\label{subsec:dist_charge}
{
In this section, we discuss the grain charge as a function of grain size and temperature. We first illustrate our results using our fiducial MRN distribution (but, as we show in~{Subsection \ref{subsec:arbitrary}} and in Appendix~\ref{sec:append-ds87}, the shape of the distribution is immaterial). 

To facilitate this discussion, it is useful to define a ``reduced temperature'': $\tau(a,T) \equiv a k T / e^2$. In their seminal work, ~\cite{DS_87} found a linear relationship between the average grain charge on a grain of given size, and the grain size for $\tau\gtrsim 1$, a result which is widely used~\citep[e.g.,][]{Okuzumi_2009,Marchand_2021}. However, they did not include ionic and thermionic emission. We wish to test if a similar result still holds when these emissions are included (though we do not include a distribution of charges for a given grain size).  

Fig.~\ref{fig:z_a} shows the grain charge as a function of grain size for $T$ $=$ [10\,K, 100\,K, 1000\,K]. It is immediately apparent that a linear relationship with a common slope is present in the log–log plots for all values of the temperature, at least for larger grain sizes. The grain size beyond which this relationship emerges is around the grain size for which $\tau = 1$. This grain size decreases with increasing temperature.

The slope of this linear relationship in the log-log plot is unity, thus $Z \propto a$. Therefore, $Z \propto a$ not only for low temperature, in the limit where ion emission and thermionic emission are negligible, but also at high temperature, where both of these processes are active. The reason for the above is discussed in Appendix~\ref{sec:append-ds87}.

If the straight line in the log-log plot were extrapolated, the intercept of this extrapolated straight line determines the constant of proportionality between $Z$ and $a$. It is clearly a function of $T$. We term this constant of proportionality $\psi$; this is the same notation used by~\cite{DS_87}.

We also note that the magnitude of the charge on the grains increases with increasing temperature; this is the result of the combined gas and grain-phase processes.
}
\begin{figure}
      \centering
\includegraphics[width=\columnwidth]{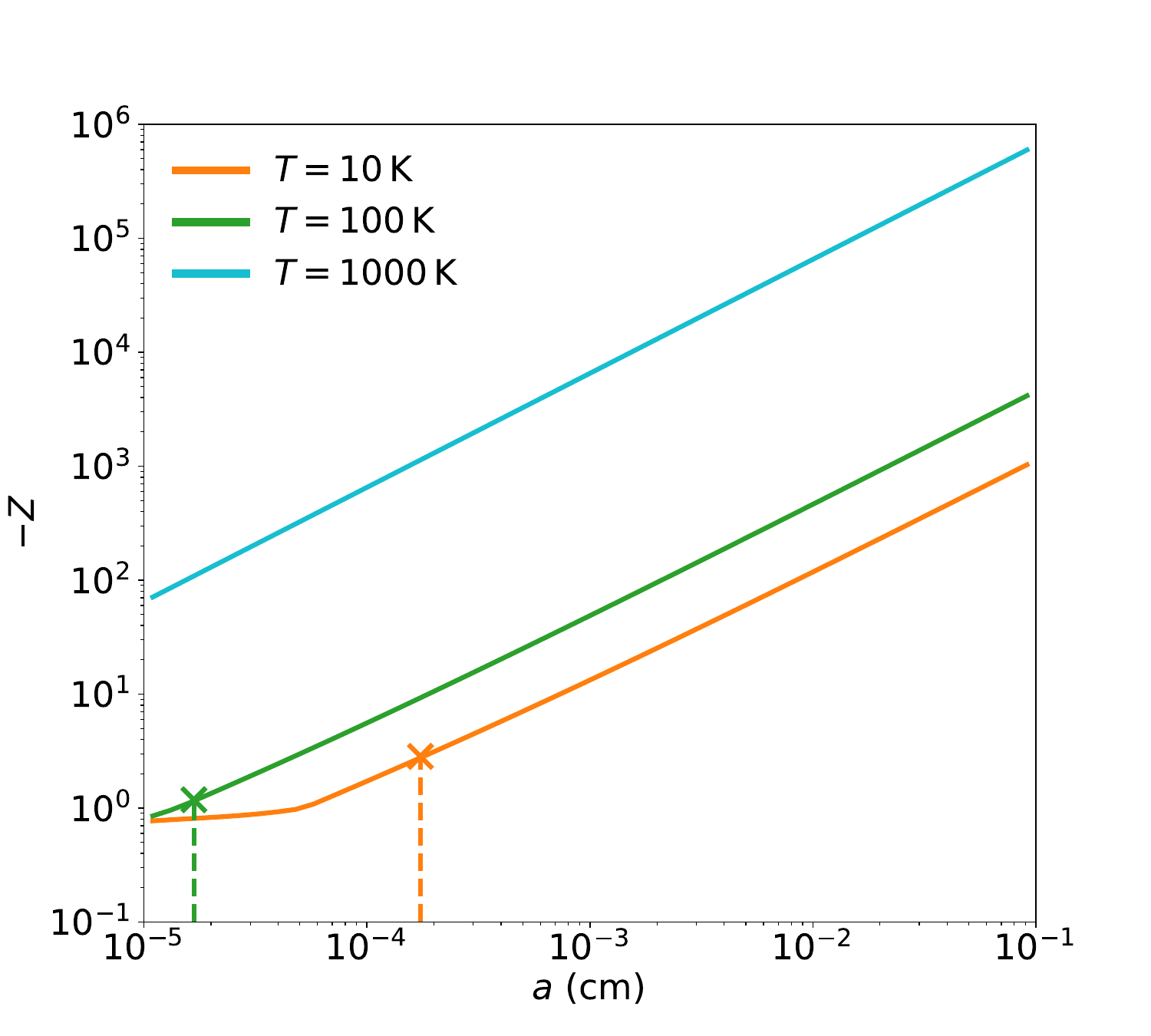}
      \caption{
        {
      Grain charge as a function of binned grain size (50 bins) in an MRN population of dust grains with $a_{\text{min}} = 10^{-5} \text{ cm}$, $a_{\text{max}} = 10^{-1} \text{ cm}$, $q=3.5$ and $f_{\text{dg}} = 0.01$. The grain size within the distribution for which $\tau = 1$ for each temperature is depicted with a vertical dashed line and a cross. At $T = 1000$\,K, all grains within the distribution have $\tau \gtrsim 1$. We see that the linear relationship for $\tau \gtrsim 1$ is observed for all temperatures.}  
      }
      \label{fig:z_a}
\end{figure}

\subsection{Effect of grain size distribution}
\label{subsec:effect-of-distribution}
Here we examine the effect of changing the input grain size distribution, keeping dust-to-gas ratio constant. The distributions used are MRN, with: the fiducial $q=3.5$; a distribution more strongly peaked at smaller grains, $q=5.0$; and a distribution which is flatter as a function of grain size, $q=2.0$.

The fiducial $q = 3.5$ has its mass dominated by large grains but its surface area dominated by small grains. The steeper value of $q = 5$ has both its surface area and mass dominated by small grains, giving a significantly larger total surface area for a given dust-to-gas ratio. Conversely, the $q = 2$ distribution has both its mass and surface area dominated by large grains. This leads to a significantly smaller total surface area.

Increased total surface area of the grains has been shown to reduce gas-phase abundances of charged species in the absence of ion and thermionic emission~\citep{Bai_Goodman_2009}; we wish to explore how different grain size distributions affect the abundances with these effects included.

\begin{figure*}
  \centering
  \includegraphics[width=\textwidth]{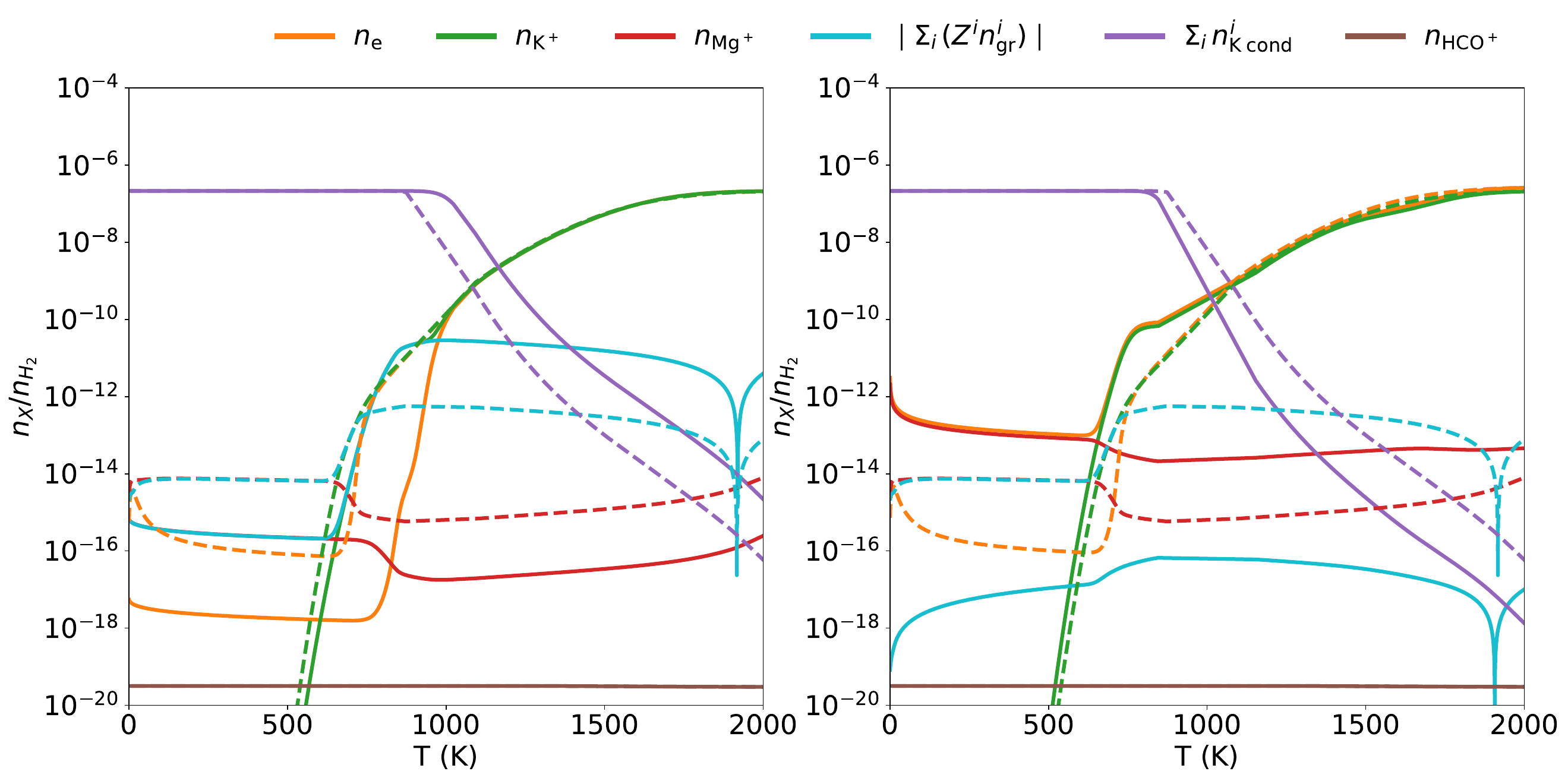}
  \caption{The ionization fractions of the reaction network are shown for MRN distributions with $q=5$ (left, solid) and $q=2$ (right, solid) and compared with the ionization state produced by an MRN distribution with the fiducial $q=3.5$ slope (left and right, dashed). The remaining parameters used are the fiducial $a_{\text{min}} = 10^{-5} \text{ cm}$, $a_{\text{max}} = 10^{-1} \text{ cm}$ and $f_{\text{dg}} = 0.01$ for fixed $n_{\text{H}_2} = 10^{14} \; \mathrm{cm^{-3}}$. $\lvert \Sigma_i (Z^i n_{\text{gr}}^i) \rvert$ shows the total charge density on the grains and $\lvert \Sigma_i (n_{\text{K cond}}^i) \rvert$ is the total number density of condensed potassium on the grains.}
  \label{fig:q-comparison}
\end{figure*}

In Fig.~\ref{fig:q-comparison}, the ionization fractions as a function of temperature obtained with $q =5$ and shallower $q =2$ are separately compared to the fiducial value of $q=3.5$. As expected, the $q = 5$ distribution, with the largest total surface area, has significantly lower number densities of gas phase species at low temperature. The $q = 2$ distribution, having the smallest total surface area, has significantly higher gas phase number densities. Lower total surface areas correspond to more negative values of $\psi$ at low temperature. A more negative $\psi$ implies a more negative charge on a given grain size bin.

Ion emission produces a steep rise in the potassium number density at $T\gtrsim 500\,K$, regardless of the grain size distribution. This is because, what limits ion emission at these temperatures is evaporation from the grains, i.e., not the process of producing potassium ions from potassium atoms on the surface of the grains. The evaporation temperature is evidently independent of the grain size distribution used; therefore, so is the threshold temperature for appreciable ion emission.

Lower total surface area (in this comparison, lower $q)$, means that ion emission produces a more negative value of $\psi$ for a given temperature, since each grain becomes more negatively charged. Since thermionic emission requires sufficiently negative $\psi$, distributions with lower $\psi$ achieve an appreciable increase in the electron number density due to thermionic emission at lower $T$. This sequence is shown in Fig.~\ref{fig:psi-temp}. 

Total surface area may also be altered by changing the total dust-to-gas ratio (i.e., total grain mass for a fixed gas mass). For example, thermionic emission would begin at a lower temperature if the grain surface area were decreased by decreasing the dust-to-gas ratio.

We note that this trade-off between total grain surface area and the dust-to-gas ratio  has been employed in previous work, to approximate the effects of a grain size-distribution with a single grain size and an ``effective dust-to-gas ratio'' \citep{Jankovic_2021, Bai_Goodman_2009}. As we show in Appendix \ref{sec:appendeffectivedg}, this method mostly yields reliable number densities for the dominant gas-phase charged species (and is thus useful for calculating fluid resistivities).  However, as we also show in Appendix \ref{sec:appendeffectivedg}, this technique predicts a total grain charge that can (in some regions of parameter space) diverge by an order of magnitude or more compared to the true value explicitly calculated in this paper. Moreover, since it considers only a single grain size by definition, this technique cannot capture the grain charge as a function of grain size, unlike the explicit method here (see Subsection \ref{subsec:dist_charge} for our grain charge distribution results). For both reasons, when considering phenomena that are directly and explicitly affected by the grain charge (e.g., most importantly, coagulation/fragmentation calculations), the ``effective dust-to-gas ratio'' method is inadequate, and the full explicit calculation shown in this paper is required.

\begin{figure*}
  \centering
  \includegraphics[width=\textwidth]{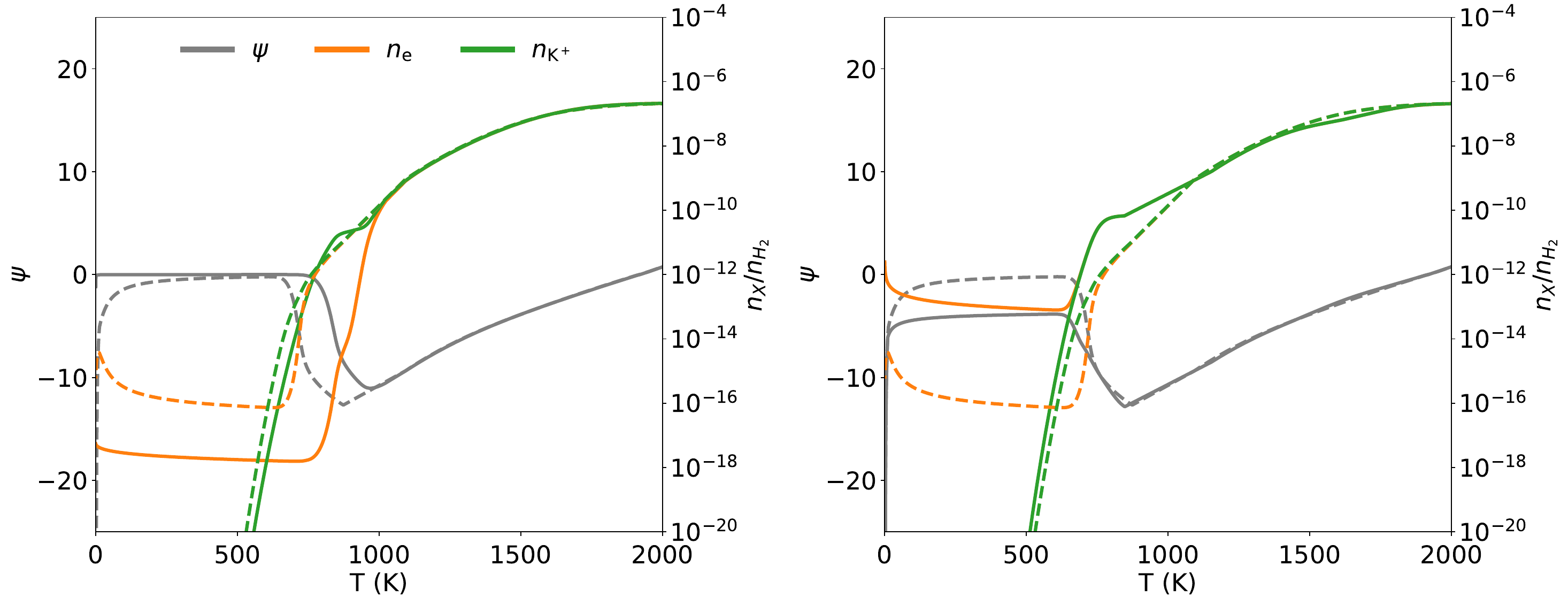}
  \caption{$\psi$ ($= Z/\tau$), number densities of electrons ($n_{\text{e}}$)  and potassium ions  ($n_{\text{K}^+}$) as a function of temperature, $T$, for the same MRN distributions as in Fig. \protect\ref{fig:q-comparison} i.e. $q = 5$ (left, solid) and $q = 2$ (right, solid) compared to the fiducial $q=3.5$ (dashed). As in the models in Fig. \protect\ref{fig:q-comparison}, the remaining parameters are the fiducial $a_{\text{min}} = 10^{-5} \text{ cm}$, $a_{\text{max}} = 10^{-1} \text{ cm}$ with $n_{\text{H}_2} = 10^{14} \; \mathrm{cm^{-3}}$.}
  \label{fig:psi-temp}
\end{figure*}

\subsection{An arbitrary grain size distribution}
\label{subsec:arbitrary}

Grain size distributions, resulting from fragmentation-coagulation and dynamics, are typically more complex than a simple MRN distribution. To integrate this network self-consistently within codes that include this dust physics, our network must be able to take arbitrary distributions as inputs. We show that we may compute the chemical abundances for such arbitrary distributions.

We have constructed a fictitious distribution which has the logarithm of the number density of grains { log–normally distributed with respect to the grain size} about a mean of $10^{-4}$\,cm, with a standard deviation $\sigma$ of 1 decade, i.e. $\log_{10} n(a) = A e^{{-}(\log_{10}(a) - (-4))^2/2}$. We have cut off this distribution below $-1 \sigma$ and above $+3 \sigma$, to match the values of $a_{\text{min}} = 10^{-5}$\,cm and $a_{\text{max}} = 10^{-1}$\,cm that we have used elsewhere in this work. The normalization $A$ was fixed so that the dust-to-gas ratio at $n_{\text{H}_2} = 10^{14} \, \mathrm{cm^{-3}}$ with $\mu=2.34$ is 0.01 . This distribution is plotted in Fig.~\ref{fig:dustpy-distribution}. 
    
We plot the abundances as a function of temperature, with $n_{\text{H}_2} = 10^{14} \, \mathrm{cm^{-3}}$, for this distribution of grains in Fig.~\ref{fig:ionization-state-dustpy-distribution} and{, to provide some physical intuition,} compare to the abundance we get from a network with a single size of $10^{-4}$\,cm (the peak of our fictitious distribution) and the same dust-to-gas ratio of 0.01. 

Below 1000\,K, when potassium is largely condensed on the grains, the gas-phase abundances of the distribution of grains are significantly larger. Due to the mass of the distribution of grains being dominated by the larger grains, the total surface area of grains for fixed dust-to-gas ratio is lower than for the network  with a single grain size of $10^{-4}$\,cm. 

Beyond 1000\,K, as in Fig.~\ref{fig:q-comparison}, the potassium and electron abundances of the distribution and single size networks converge. However, grain charge and Mg$^+$ abundances is significantly different in both cases.
} 
    
In Fig.~\ref{fig:z_a_lognorm}, we show the grain charge as a function of grain size for the arbitrary distribution. We plot this for the same values of $T = \left[ 10 \text{\,K}, 100 \text{\,K}, 1000 \text{\,K}\right]$ that we chose for the MRN distribution whose grain charge as a function of grain size we plotted in Fig.~\ref{fig:z_a}. This illustrates that the linear relationship between grain charge and grain size at fixed temperature is not dependent on the distribution being a power law.
    
\begin{figure}
    \centering
    \includegraphics[width=\columnwidth]{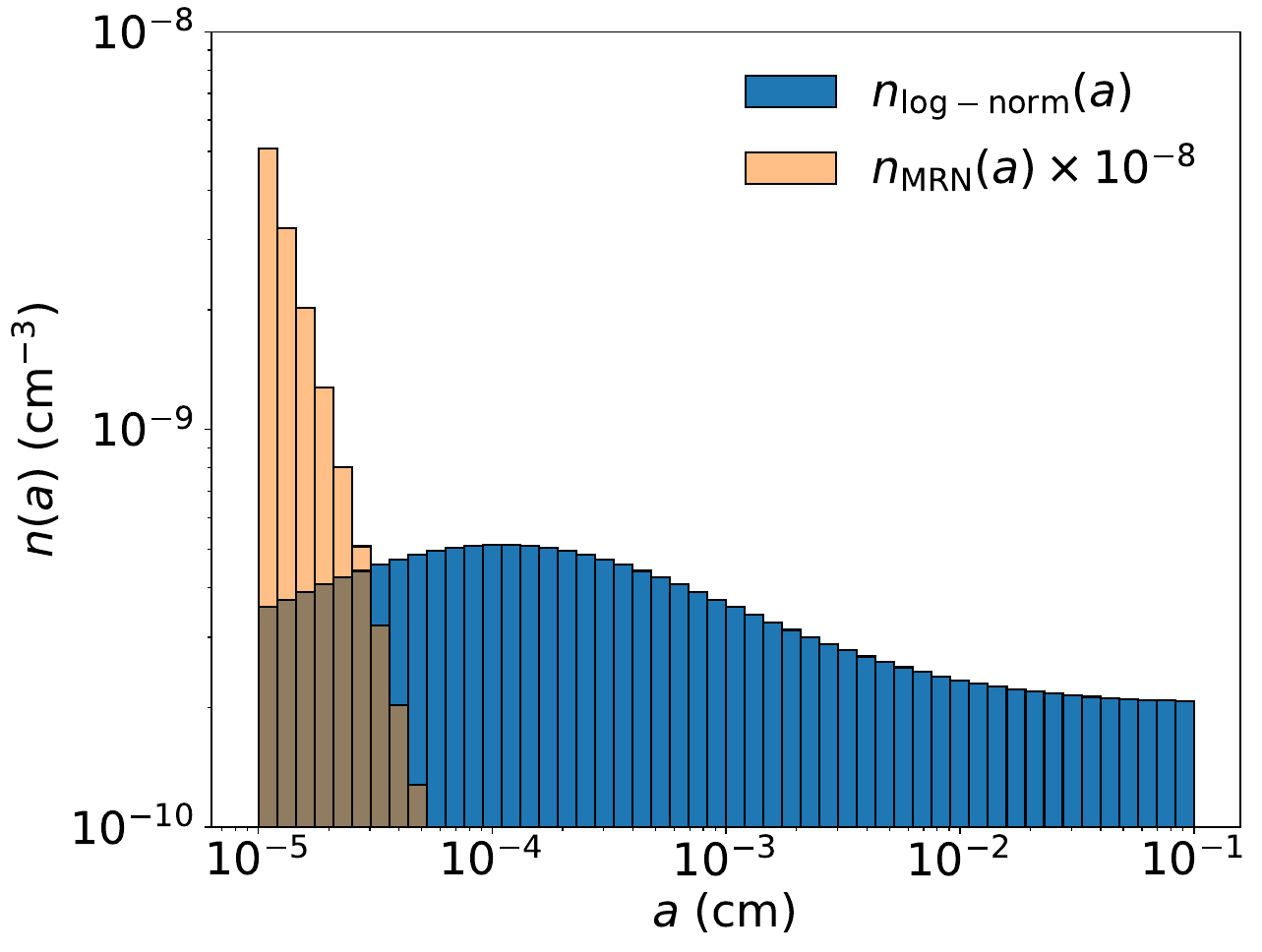}
    \caption{
    The blue histogram shows the number densities in each grain size bin of a fictitious dust size distribution with the logarithm of the number density of grains log-normally distributed about a mean of $10^{-4}$\,cm, with a standard deviation $\sigma$ of 1 decade. The distribution has been cut off below $-1\sigma$ and above $+3\sigma$, and normalized to produce a dust-to-gas ratio of 0.01 at $n_{\text{H}_2} = 10^{14} \, \mathrm{cm^{-3}}$. { For comparison, overplotted in orange is $10^{-8}$ times the number density of grains for an MRN distribution with the fiducial parameters $a_{\text{min}} = 10^{-5} \text{ cm}$, $a_{\text{max}} = 10^{-1} \text{ cm}$, $q=3.5$ and $f_{\text{dg}} = 0.01$ at the same density of $n_{\text{H}_2} = 10^{14} \, \mathrm{cm^{-3}}$.}}
    \label{fig:dustpy-distribution}
\end{figure}

\begin{figure}
    \centering
    \includegraphics[width=\columnwidth]{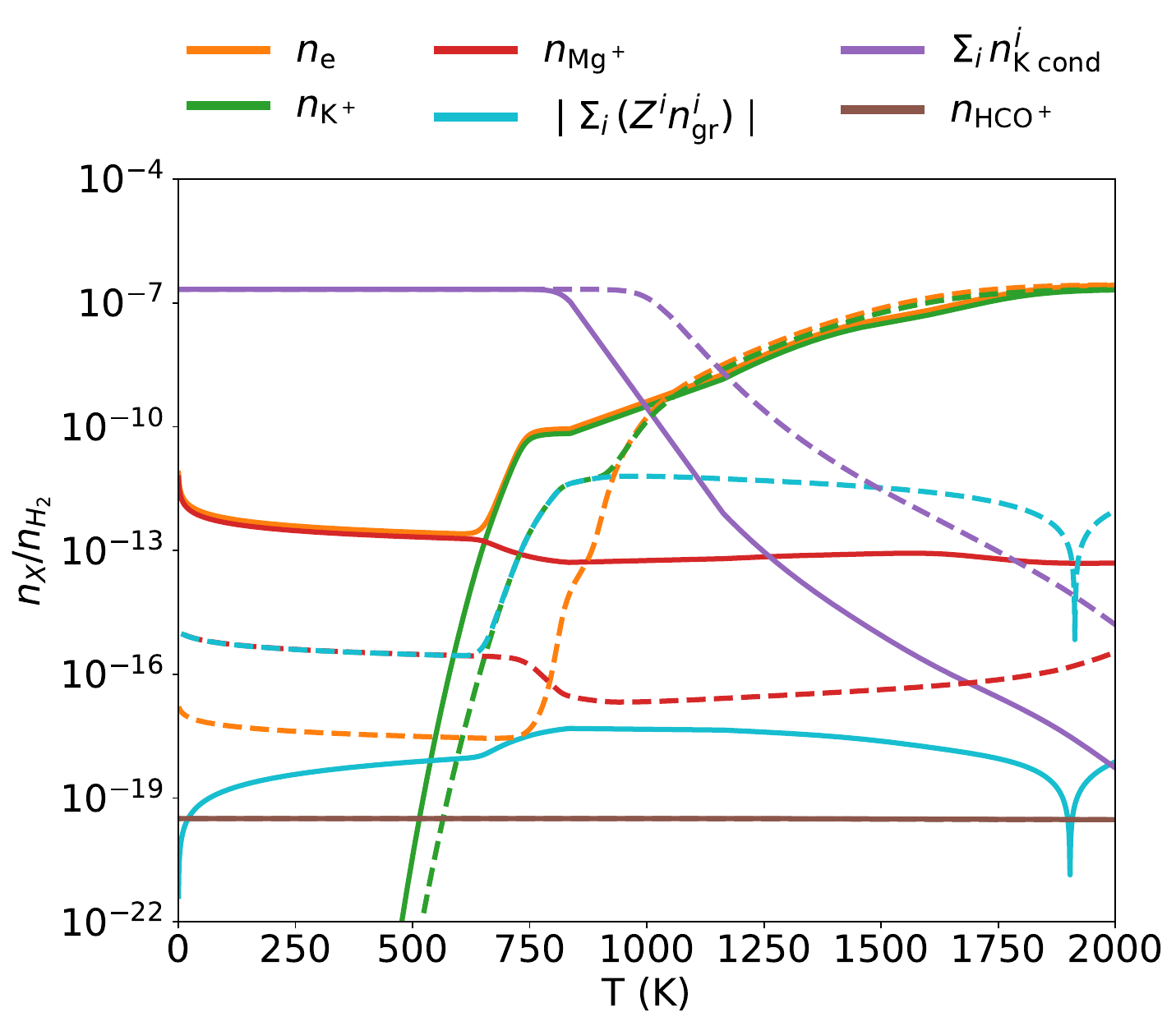}
    \caption{Abundances as a function of temperature $T$, for the fictitious distribution (solid) shown in Fig. \protect\ref{fig:dustpy-distribution} {, and grains with a single size of $10^{-4}\,\text{cm}$ (dashed)}. Both networks have $n_{\text{H}_2} = 10^{14} \; \mathrm{cm^{-3}}$ and $f_{\text{dg}} = 0.01$. $\lvert \Sigma_i (Z^i n_{\text{gr}}^i) \rvert$ shows the total charge density on the grains and $\lvert \Sigma_i (n_{\text{K cond}}^i) \rvert$ is the total number density of condensed potassium on the grains.} 
    \label{fig:ionization-state-dustpy-distribution}
\end{figure}
    
\begin{figure}
      \centering
\includegraphics[width=\columnwidth]{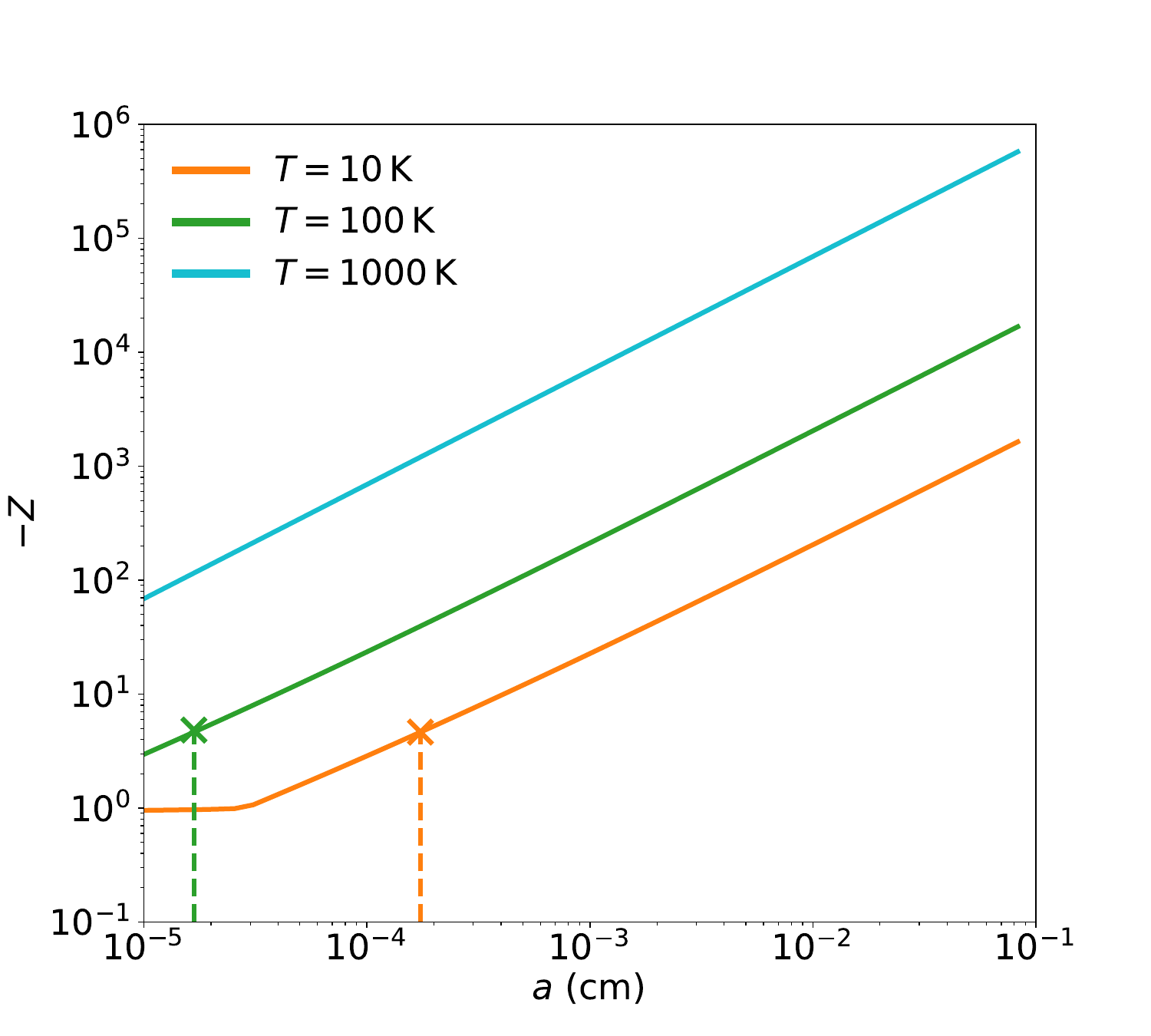}
      \caption{ 
      {
      The same plot of grain charge as a function of binned grain size (50 bins) as in Fig. \protect\ref{fig:z_a}, except for the fictitious distribution shown in Fig. \protect\ref{fig:dustpy-distribution} with $f_{\text{dg}} = 0.01$ at $n_{\text{H}_2} = 10^{14} \, \mathrm{cm^{-3}}$ and $\mu=2.34$. The grain size within the distribution for which $\tau = 1$ for each temperature is depicted with a vertical dashed line and a cross. At $T = 1000$\,K, all grains within the distribution have $\tau \gtrsim 1$. We see that the linear relationship for $\tau \gtrsim 1$ is observed for all temperatures.}
      }
      \label{fig:z_a_lognorm}
\end{figure}

\subsection{Validity of chemical equilibrium} 
\label{subsec:validity-equilibrium-results}

\begin{figure}
    \centering
    \includegraphics[width=\columnwidth]{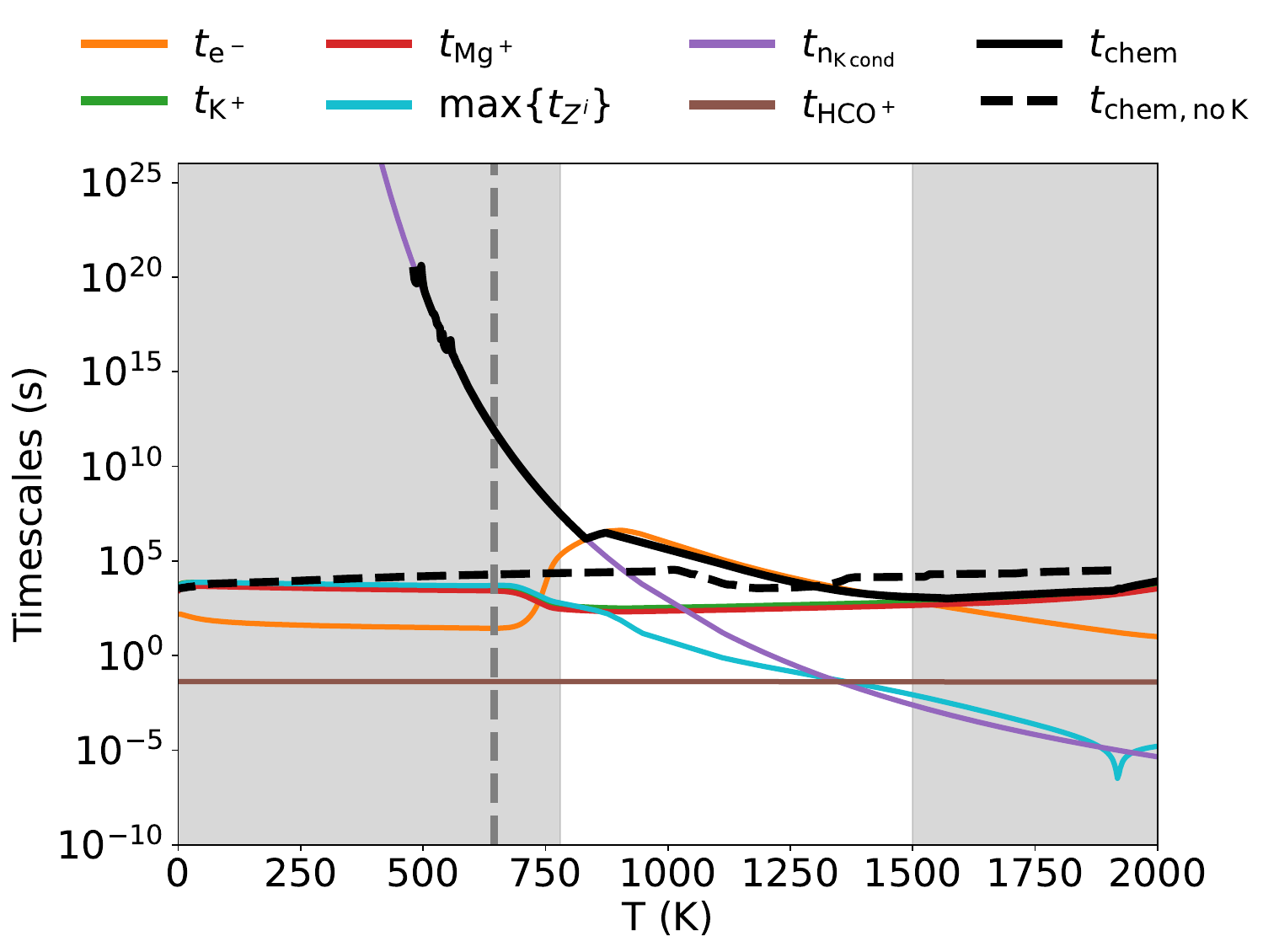}
    \caption{Comparison of chemical time-scales for $n_{\text{H}_2} = 10^{14} \text{cm}^{-3}$. The relevant chemical time-scale is $t_{\text{chem}}$ (the full system) to the right of the vertical dashed line, and $t_{\text{chem, no K}}$ (the system excluding potassium) to its left. The other time-scales plotted are the time-scales associated with individual species $t_{\text{x}}$. The longest of these individual time-scales maps on to the chemical time-scale $t_{\text{chem}}$.}
    \label{fig:chemical-timescales-midplane-1au}
\end{figure}

\begin{figure}
    \centering
    \includegraphics[width=\columnwidth]{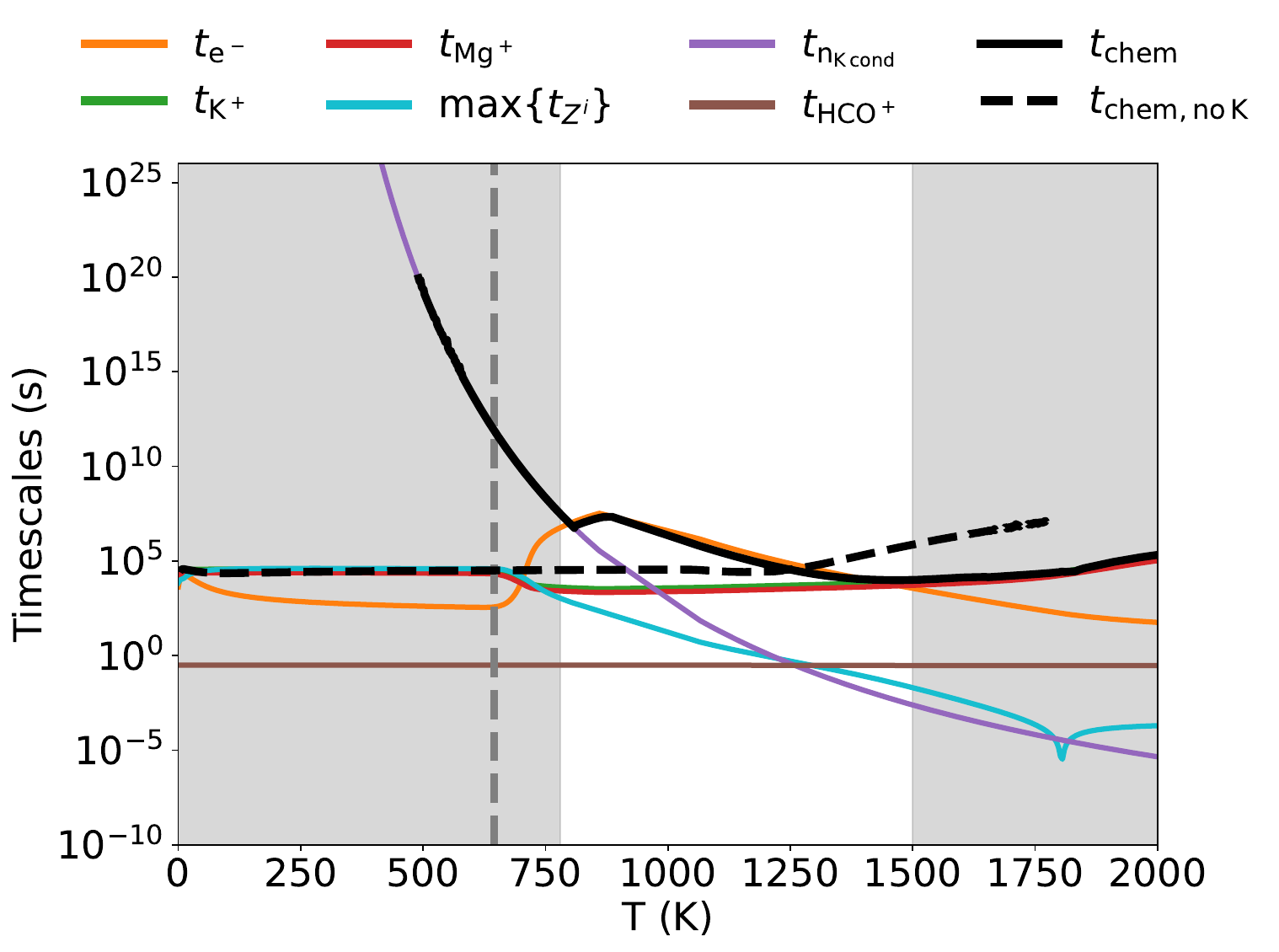}
    \caption{The same plot as Fig. \protect\ref{fig:chemical-timescales-midplane-1au}, but for $n_{\text{H}_2} = 10^{14}{\text{exp(-2)}} \text{ cm}^{-3}$, equivalent to two scale heights above the mid-plane value at approximately 1 au. Very similar behaviour is observed.}
    \label{fig:chemical-timescales-2scaleheights-1au}
\end{figure}

\begin{figure}
    \centering
    \includegraphics[width=\columnwidth]{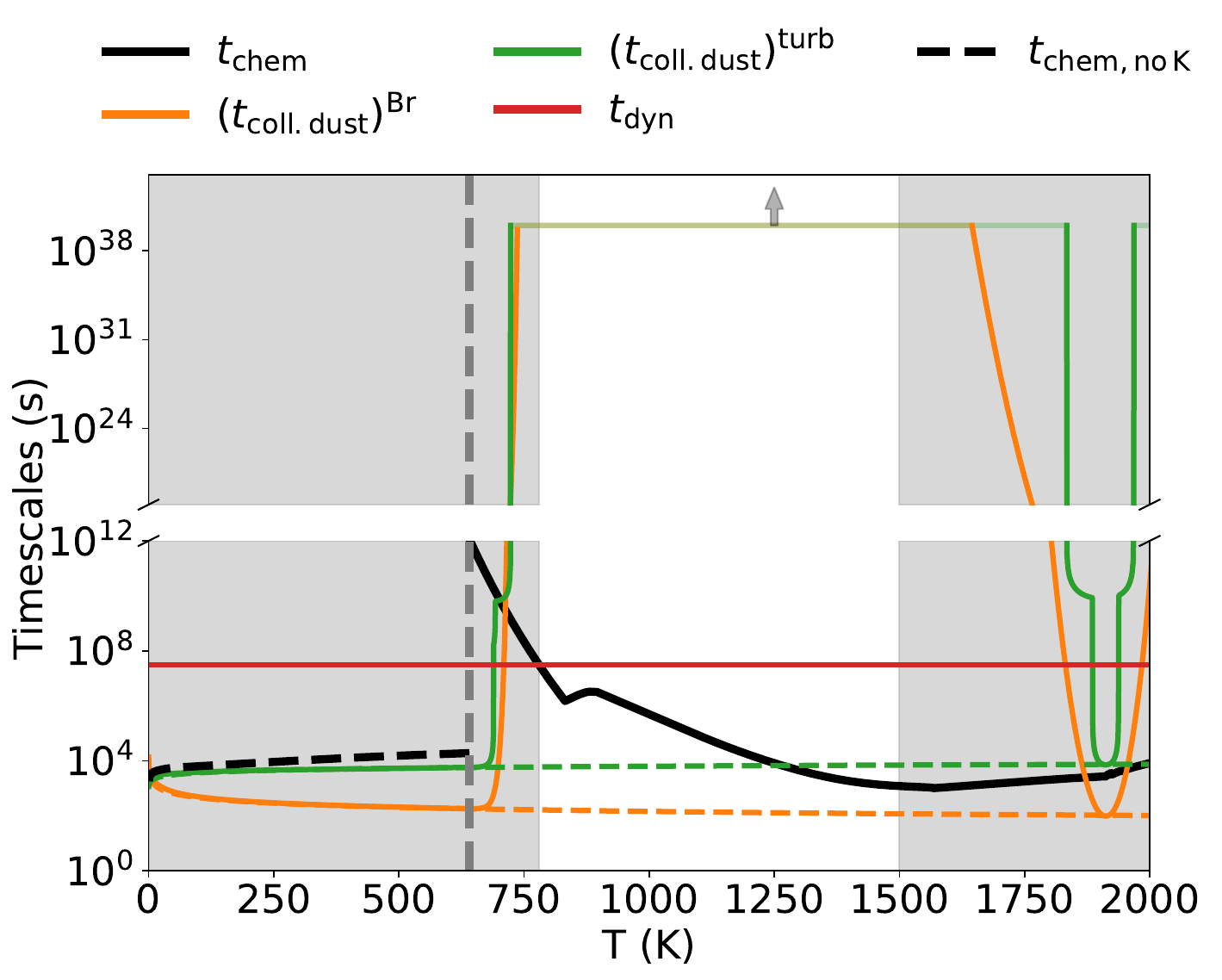}
    \caption{Comparison of time-scales for $n_{\text{H}_2} = 10^{14} \text{cm}^{-3}$. The relevant chemical time-scale is $t_{\text{chem}}$ (the full system) to the right of the vertical dashed line, and $t_{\text{chem, no K}}$ (the system excluding potassium) to its left. Dust collision time-scales for Brownian motion and turbulence are shown with the effect of grain charge included (solid) and excluded (dashed). Where grain charge is included, the grain collision time-scales in the region of interest (unshaded) become many orders of magnitude greater than all other system time-scales. In these regions, the chemical time-scale is also shorter than the other relevant time-scales, indicating that chemical equilibrium is valid. {(Note that our grain collision time-scales, with charge included, reach a plotted plateau of $10^{40} \,$s, shown in a lighter colour. In reality, the time-scales are even higher in this region, as indicated by the upward arrow. We have simply truncated our calculations here.)}}
    \label{fig:timescales-midplane-1au}
\end{figure}

Following the method in~\ref{subsec:chemical-equilibrium}, we compute the relevant time-scales (i.e., \crefrange{eqn:chemical-time}{eqn:dust-time}) for our network as a function of temperature for conditions applicable to the disc mid-plane at 1 au ($n_{\text{H}_2} = 10^{14} \, \mathrm{cm^{-3}}$) and two scale heights above our mid-plane value at 1 au ($n_{\text{H}_2} = 10^{14} \exp(-2) \; \mathrm{cm^{-3}}$). We also compute the time-scale associated with each species within our network. The time-scale for species x is defined as $t_{\text{x}} = \text{x} / \mathcal{R}_{\pm \text{x}}$, where $\mathcal{R}_{\pm \text{x}}$ is either the sum of all rates of production or the sum of all rates of destruction of species $\text{x}$; either of these choices are equivalent in equilibrium. Furthermore, we compute the chemical time-scale that we would obtain if potassium were omitted altogether $t_{\text{chem, no K}}$.

We show these chemical time-scales as a function of temperature for mid-plane conditions at 1 au in Fig.~\ref{fig:chemical-timescales-midplane-1au}. We have shaded regions not relevant for our study in grey, these are: (i) where the chemical time-scale is shorter than the dynamical time-scale (incidentally, this threshold produces a minimum temperature $T \sim 800$\,K, similar to that which we expect for the mid-plane at 1 au~\citep{Jankovic_2021}); and $T\gtrsim1500\,K$, greater than the sublimation temperature of silicate grains.  

We see that the chemical time-scale $t_{\text{chem}}$ of the whole system follows the longest of any of the individual species time-scales $t_{\text{x}}$. At the largest temperatures, the longest time-scales for any individual species are those associated with Mg$^+$ and K$^+$. Since these species are produced through non-thermal and thermal ionization respectively, but have similar time-scales, the limiting time-scale must be due to a common destruction rate. This destruction rate corresponds to adsorption onto grains, which is indeed the largest rate of destruction for both these species at these temperatures. For $900\,\text{K} \lesssim T \lesssim 1500\,\text{K}$, the electron time-scale is the longest. In this region, it is the rate of thermionic emission which is the dominant mechanism for production of electrons, while the dominant method of destruction is due to collisions with the negatively charged grains. Below 900\,K, non-thermal ionization is the dominant production mechanism for electrons, while collisions with grains remains the dominant destruction mechanism. 

However, the limiting time-scale for the system as a whole below 900\,K is not due to the electrons, but the condensed potassium. The evaporation time-scale becomes exponentially longer with decreasing temperature. At small enough temperatures, this sends the maximum $\Re(\lambda_i)$, found by solving \cref{eqn:lambda}, towards zero; eventually its value oscillates between positive (indicating an unstable equilibrium) and negative values. We clip the chemical time-scale where this occurs as this oscillation is purely due to finite numerical precision. If we consider a network without potassium, the chemical time-scale does not increase exponentially as $T \to 0$\,K as seen in Fig.~\ref{fig:chemical-timescales-midplane-1au}. As $T$ decreases from 900\,K, the thermal ionization rates becomes smaller than the non-thermal ionization rate, and the abundances of all species other than potassium are identical in the case with potassium or without. Here, for the purposes of determining the resistivities or grain charges, the inclusion of potassium is of no consequence. The temperature below which potassium does not appreciably alter the abundances of other species is shown with a vertical dashed line in Fig.~\ref{fig:chemical-timescales-midplane-1au}. Hence, the chemical time-scale of interest is that of this reduced system with no potassium, shown in the black dashed line in Fig.~\ref{fig:chemical-timescales-midplane-1au}. This time-scale is considerably reduced compared to the full system time-scale and matches the time-scales associated with potassium ions, magnesium ions and grains in the full system. This implies that the limiting time-scale is due to collisions of ions with grains.

The complementary plot, corresponding to two scale heights above this mid-plane density is shown in Fig.~\ref{fig:chemical-timescales-2scaleheights-1au}. Similar behaviour can be seen. However, time-scales are typically somewhat longer due to the lower densities. 

We now compare the chemical time-scale to the other system time-scales, namely the dynamical time-scale and the dust collision time-scales. This is plotted for a density applicable to the mid-plane at 1 au ($n_{\text{H}_2} = 10^{14} \, \mathrm{cm^{-3}}$) in Fig.~\ref{fig:timescales-midplane-1au}. For mid-plane conditions, the chemical time-scale, with the inclusion of the effect of charge on the dust collision rates, is shorter than all other system time-scales throughout the majority of the temperature range plotted (and throughout the whole unshaded range in which we are interested). This is with the exception of two regions: (i) close to 2000 K, where the grain charge transitions through zero (negative at low temperature to positive at high temperature due to the effect of thermionic emission); and (ii) $\lesssim750$\,K, where the chemical time-scale increases due to the rapidly decreasing rate of evaporation of potassium from the grains with decreasing temperature. Furthermore, below 750\,K the grain collision time-scales decrease with decreasing temperature as the grains are less negatively charged (see Fig.~\ref{fig:z-T}). At temperatures leftwards of the vertical dashed line, the chemical time-scale with no potassium becomes the chemical time-scale of interest. In this region, this chemical time-scale is shorter than the dynamical time-scale but still longer or comparable to the dust collision time-scales. Nevertheless, these regions are not germane for the reasons stated above.

In sum, since the chemical time-scale is the shortest of all time-scales in a temperature range and density applicable to the mid-plane at 1au~\citep[e.g.,][]{Jankovic_2021}, chemical equilibrium is a good approximation here.

Apart from in the regions near 2000 K and $\lesssim750$\,K, the grain–grain collision time-scales are so long, due to the negative charge on the grains, that a size distribution may be `frozen in' after entering the inner disc.

This paints a rosier picture for the survival of large grains in the inner disc than that presented by~\cite{Akimkin_2023}. Their charging model, which does not account for thermionic and ion emission, produces lower magnitude grain charges. 

This lower charging means that only a subset of all grain collisions are forbidden by the electrostatic barrier in their model. These are collisions between smaller grains. Small grains may still undergo collisions in their model, but must collide with larger grains. Since these collisions occur with relatively large relative velocity in their calculations for higher turbulence (as found in the inner disc~\citep{Mohanty_2018,Jankovic_2021}), a collision with a grown grain will typically occur with an impact velocity greater than the fragmentation velocity (for a fragmentation velocity of $1 \, \text{m s}^{-1}$).  

This means that their grain size distribution evolves towards a distribution very sharply peaked at the smallest grain sizes (see their Fig. 4 showing the evolving distribution for typical inner disc parameters). 

In our model of the charging, all collisions between grains at the mid-plane are orders of magnitude longer than the dynamical time, indicating that the distribution does not evolve collisionally. If grains can become large outside the inner disc, this grain charging can conserve their sizes. This difference has profound implications for the opacity and thus the disc structure.

\begin{figure}
    \centering
    \includegraphics[width=\columnwidth]{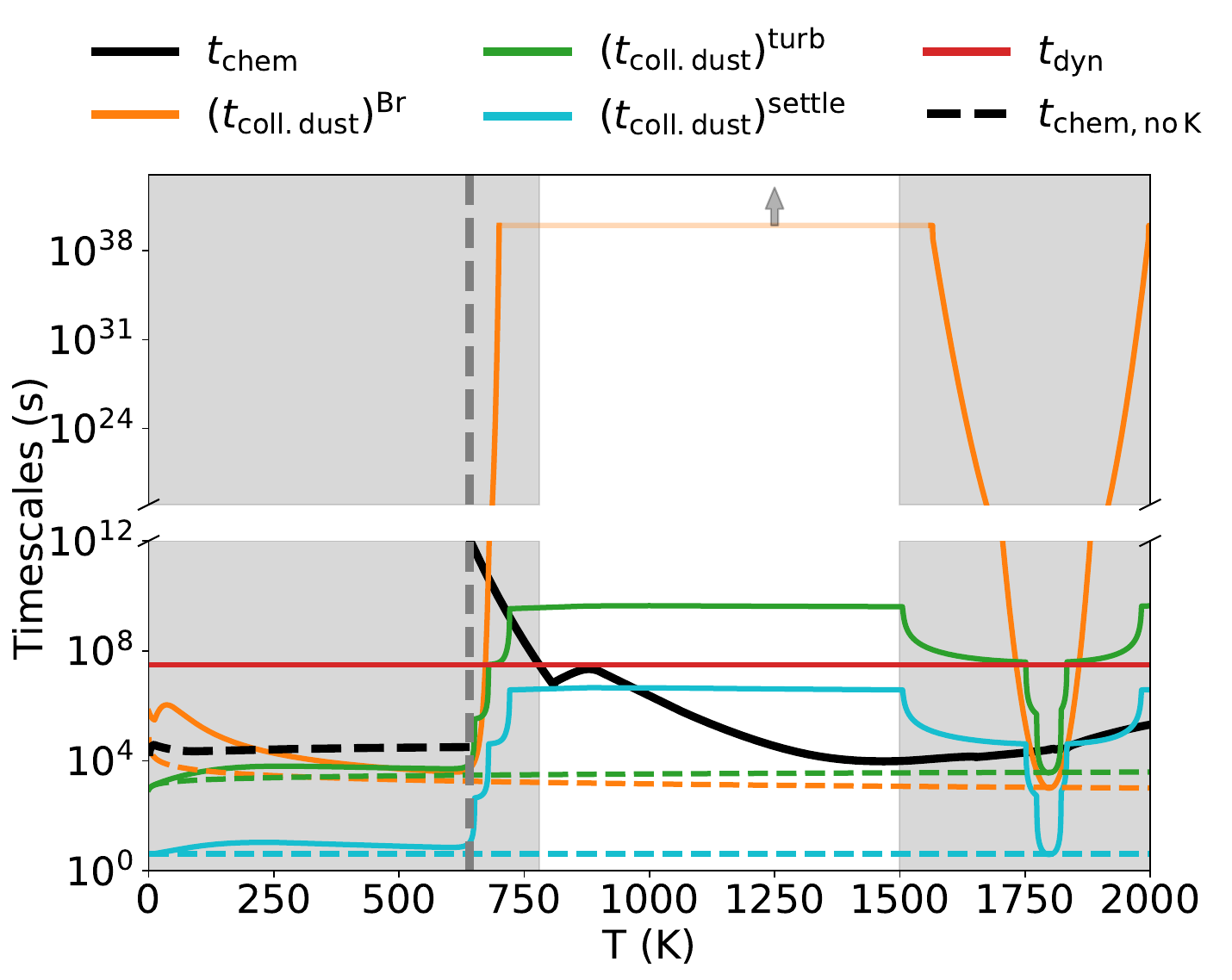}
    \caption{Comparison of time-scales for $n_{\text{H}_2} = 10^{14} / e^2 \text{ cm}^{-3}$, applicable to two scale-heights above the mid-plane value of the number density used in Fig.~\protect~\ref{fig:timescales-midplane-1au}. The time-scale for collisions between grains due to settling is now considered and found to be the shortest of the grain–grain collision time-scales. {(Note that our grain collision time-scales, with charge included, reach a plotted plateau of $10^{40} \,$s, shown in a lighter colour. In reality, the time-scales are even higher in this region, as indicated by the upward arrow. We have simply truncated our calculations here.)}}
    \label{fig:timescales-2scale-heights-1au}
\end{figure}

The time-scales of our network, two-scale heights above the value we adopt for the mid-plane at 1 au, are shown in Fig.~\ref{fig:timescales-2scale-heights-1au}. Due to the lower density of the gas, the chemical time-scale becomes longer but remains below the dynamical time-scale in the region of interest. This implies that chemical equilibrium is expected to be a good approximation throughout the vertical extent of the disc near the inner edge of the dead-zone. The shortest dust collision time-scale is now due to settling, which in some temperature ranges may becomes shorter than the chemical time-scale. Nevertheless, this is far above the mid-plane; the dust-to-gas ratio should be lower here than the value used in this calculation $f_{\text{dg}} = 0.01$ due to the effect of dust settling. This means that collisions between grains should be more infrequent than indicated here. We conclude that chemical equilibrium is also a good approximation here.

\subsection{Resistivities}
\label{subsec:resistivies}

Obtaining resistivities for non-ideal MHD calculations, that incorporate the effect of a distribution of grains, with the inclusion of ion and thermionic emission, is one of the primary motivations for this work. As such, we provide illustrative calculations of these resistivities below. The resistivities are computed following the method of~\cite{Wardle_2007}. We have assumed a constant r.m.s magnetic field $B$ of $10\,$\,G, informed by the values computed by~\cite{Jankovic_2021}. 

\begin{figure*}
    \centering
    \includegraphics[width=\textwidth]{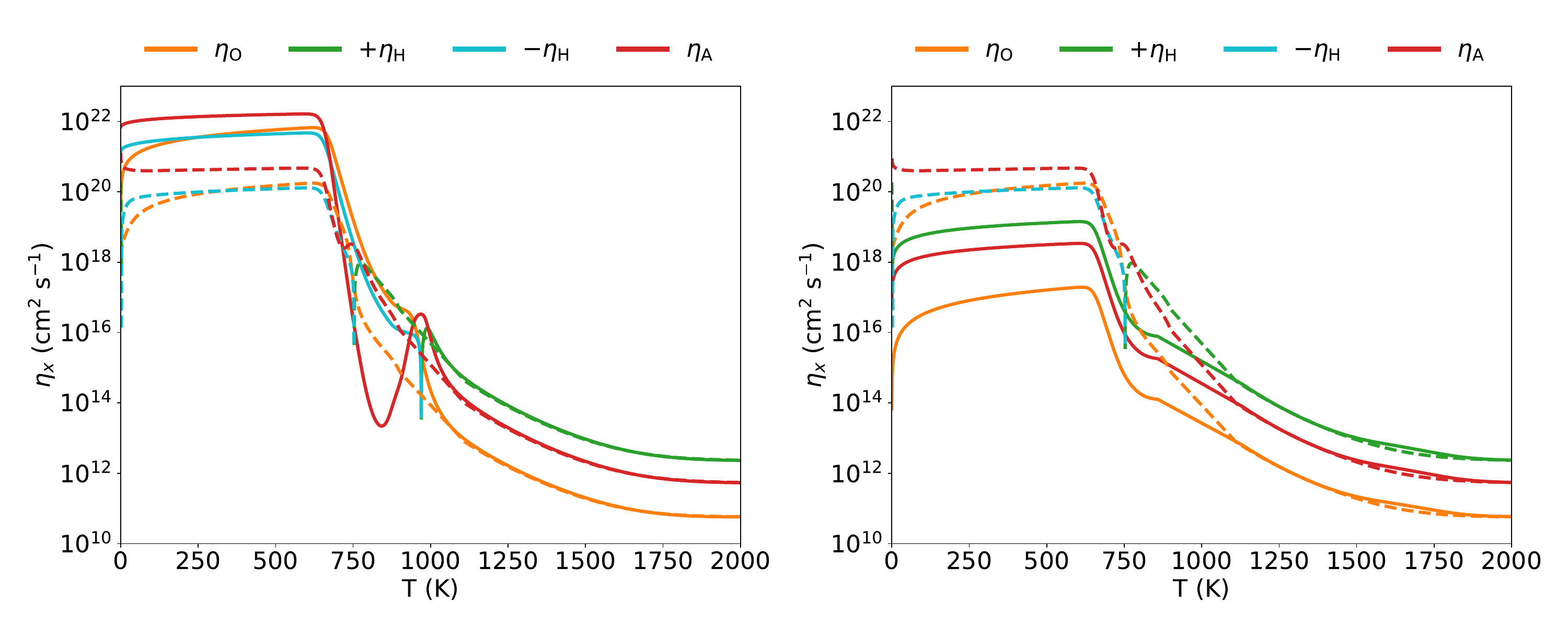}
    \caption{The Ohmic ($\eta_{\text{O}}$), Hall ($\eta_{\text{H}}$) and ambipolar ($\eta_{\text{A}}$) resistivities computed using ionization fractions computed by our reaction network for MRN distributions with $q=5$ (left, solid) and $q=2$ (right, solid) compared with the fiducial $q=3.5$ slope (left and right, dashed). The remaining parameters used are the fiducial $a_{\text{min}} = 10^{-5} \text{ cm}$, $a_{\text{max}} = 10^{-1} \text{ cm}$ and $f_{\text{dg}} = 0.01$ for fixed $n_{\text{H}_2} = 10^{14} \; \mathrm{cm^{-3}}$ and $B = 10\,$\,G. } 
    \label{fig:resistivities_midplane}
\end{figure*}

In Fig.~\ref{fig:resistivities_midplane}, we show the Ohmic, Hall and ambipolar resistivities as a function of temperature, at the fiducial density ($n_{\text{H}_2} = 10^{14} \, \mathrm{cm^{-3}}$). The grain distributions used are MRN, with: the fiducial $q=3.5$, the more sharply peaked $q=5.0$ and flatter $q=2.0$. We use the fiducial minimum and maximum grain size $a_{\text{min}} = 10^{-5}$\,cm and $a_{\text{max}} = 10^{-1}$\,cm with $f_{\text{dg}} = 10^{-2}$.

In both panels, we see that all resistivities are low in the high temperature limit due to relatively high ionization fractions. When electrons are the dominant negative charge carriers (throughout the entire temperature range for $q=2.0$, see Fig.~\ref{fig:q-comparison}), they are the best coupled species to the field, and the Hall resistivity is positive. At lower temperatures, the grains become the dominant negative charge carriers (for $q=3.5$ and $q=5.0$, see Fig.~\ref{fig:q-comparison}). Since the grains are more decoupled from the field than the ions, the Hall current changes direction relative to the case where electrons are the dominant negative charge carriers. This means that the Hall resistivity becomes negative. The magnitude of the Hall current is also weakened when the electrons become sub-dominant negative charge carriers, and the magnitude of the Hall resistivity decreases.

This transition from electrons being dominant negative charge carriers to grains also causes a decreases in the ambipolar diffusion, as the grains have a much weaker coupling to the field. Further decreases in temperature produce larger magnitude resistivities, which ultimately saturate as the dominant ionization fractions plateau (because the ionization is non-thermal, and temperature independent, and the recombination rate depends only weakly temperature).

\section{Conclusions}
\label{sec:conclusions}
In this paper, (1) we investigate a chemical network that applies to the inner regions of protoplanetary discs. The network includes electrons, molecular hydrogen, an alkali species, a non-alkali metal, a molecular ion, and grains with an arbitrary size distribution. We consider non-thermal ionization of hydrogen (by, e.g., XUV irradiation, cosmic rays and radioactivity), thermal ionization of alkalis, charge transfer between gas-phase species, gas–grain chemistry, ionic and thermionic emission from grain surfaces, and grain charging. (2) We present a numerical technique to solve this network {\it exactly} (to numerical precision), in equilibrium. The solution technique may, without loss of generality, be extended to multiple alkali, metal, and molecular species (as we demonstrate with two alkali species). (3) We also provide a general method to estimate the chemical time-scale for a network, and (4) supply analytic expressions for grain–grain collision time-scales for charged grains.  

We validate our equilibrium results against earlier work, in relevant limits. We further discuss the physical reasons for our results when all the above effects are included, in various parameter regimes. 

An important result of this work is that the grain charge is directly proportional to the grain size, above a minimum grain size. This result has previously been shown to hold in the absence of ionic and thermionic emission; we show here that it continues to hold when these effects are included, and explicate the reason for this behaviour in Appendix~\ref{sec:append-ds87}. 

We further show that approximating the effect of a range of grain charges via the ``effective dust-to-gas-ratio'' method, as done in earlier work, yields significantly erroneous grain charges. Finally, we show that grain charging has a very significant effect on grain collision time-scales (and grain collision velocities). 

While we have shown, for illustrative purposes, results at 1 au for a fiducial disc density, our network and results are valid throughout the ``inner disc'' regions, where thermal ionization of alkali species is efficient. As such, our chemical network, and the techniques provided for analysing it, are valuable for calculating the fluid resistivities, and thus the fluid viscosity, in the inner disc \citep[as calculated in, e.g.,][]{Mohanty_2018, Jankovic_2021}. Moreover, our grain charging results – both the amount of charging as a function of grain size, and the effect of grain charging on grain collision time-scales and velocities – are vital for discerning the fragmentation / coagulation outcomes of grain collisions, and thus for grain and disc evolution under inner disc conditions.     

\section*{Acknowledgements}
MW acknowledges the support of a Science and Technologies Facilities Council (STFC) PhD studentship.
MW and SM thank James Owen, Marija Jankovic, Neal Turner for extremely helpful discussions that helped sharpen the analysis of this paper. We also thank the anonymous referee for their careful reading and detailed comments, which were highly useful in improving the paper.

%%%%%%%%%%%%%%%%%%%%%%%%%%%%%%%%%%%%%%%%%%%%%%%%%%
\section*{Data Availability}
The data underlying this article will be shared on reasonable request
to the corresponding author.



\bibliographystyle{mnras}
\bibliography{example} % if your bibtex file is called example.bib

\begin{thebibliography}{}
\makeatletter
\relax
\def\mn@urlcharsother{\let\do\@makeother \do\$\do\&\do\#\do\^\do\_\do\%\do\~}
\def\mn@doi{\begingroup\mn@urlcharsother \@ifnextchar [ {\mn@doi@} {\mn@doi@[]}}
\def\mn@doi@[#1]#2{\def\@tempa{#1}\ifx\@tempa\@empty \href {http://dx.doi.org/#2} {doi:#2}\else \href {http://dx.doi.org/#2} {#1}\fi \endgroup}
\def\mn@eprint#1#2{\mn@eprint@#1:#2::\@nil}
\def\mn@eprint@arXiv#1{\href {http://arxiv.org/abs/#1} {{\tt arXiv:#1}}}
\def\mn@eprint@dblp#1{\href {http://dblp.uni-trier.de/rec/bibtex/#1.xml} {dblp:#1}}
\def\mn@eprint@#1:#2:#3:#4\@nil{\def\@tempa {#1}\def\@tempb {#2}\def\@tempc {#3}\ifx \@tempc \@empty \let \@tempc \@tempb \let \@tempb \@tempa \fi \ifx \@tempb \@empty \def\@tempb {arXiv}\fi \@ifundefined {mn@eprint@\@tempb}{\@tempb:\@tempc}{\expandafter \expandafter \csname mn@eprint@\@tempb\endcsname \expandafter{\@tempc}}}

\bibitem[\protect\citeauthoryear{{Akimkin}, {Ivlev}  \& {Caselli}}{{Akimkin} et~al.}{2020}]{Akimkin_2020}
{Akimkin} V.~V.,  {Ivlev} A.~V.,   {Caselli} P.,  2020, \mn@doi [\apj] {10.3847/1538-4357/ab6299}, \href {https://ui.adsabs.harvard.edu/abs/2020ApJ...889...64A} {889, 64}

\bibitem[\protect\citeauthoryear{{Akimkin}, {Ivlev}, {Caselli}, {Gong}  \& {Silsbee}}{{Akimkin} et~al.}{2023}]{Akimkin_2023}
{Akimkin} V.,  {Ivlev} A.~V.,  {Caselli} P.,  {Gong} M.,   {Silsbee} K.,  2023, \mn@doi [\apj] {10.3847/1538-4357/ace2c5}, \href {https://ui.adsabs.harvard.edu/abs/2023ApJ...953...72A} {953, 72}

\bibitem[\protect\citeauthoryear{Ashton \& Hayhurst}{Ashton \& Hayhurst}{1973}]{Ashton_Hayhurst_1973}
Ashton A.,  Hayhurst A.,  1973, \mn@doi [Combustion and Flame] {https://doi.org/10.1016/0010-2180(73)90008-4}, 21, 69

\bibitem[\protect\citeauthoryear{{Asplund}, {Grevesse}, {Sauval}  \& {Scott}}{{Asplund} et~al.}{2009}]{Asplund_2009}
{Asplund} M.,  {Grevesse} N.,  {Sauval} A.~J.,   {Scott} P.,  2009, \mn@doi [\araa] {10.1146/annurev.astro.46.060407.145222}, \href {https://ui.adsabs.harvard.edu/abs/2009ARA&A..47..481A} {47, 481}

\bibitem[\protect\citeauthoryear{{Bai}}{{Bai}}{2011}]{Bai_2011}
{Bai} X.-N.,  2011, \mn@doi [\apj] {10.1088/0004-637X/739/1/50}, \href {https://ui.adsabs.harvard.edu/abs/2011ApJ...739...50B} {739, 50}

\bibitem[\protect\citeauthoryear{{Bai} \& {Goodman}}{{Bai} \& {Goodman}}{2009}]{Bai_Goodman_2009}
{Bai} X.-N.,  {Goodman} J.,  2009, \mn@doi [\apj] {10.1088/0004-637X/701/1/737}, \href {https://ui.adsabs.harvard.edu/abs/2009ApJ...701..737B} {701, 737}

\bibitem[\protect\citeauthoryear{{Bai} \& {Stone}}{{Bai} \& {Stone}}{2011}]{Bai_Stone_2011}
{Bai} X.-N.,  {Stone} J.~M.,  2011, \mn@doi [\apj] {10.1088/0004-637X/736/2/144}, \href {https://ui.adsabs.harvard.edu/abs/2011ApJ...736..144B} {736, 144}

\bibitem[\protect\citeauthoryear{{Balbus} \& {Hawley}}{{Balbus} \& {Hawley}}{1991}]{Balbus_1991}
{Balbus} S.~A.,  {Hawley} J.~F.,  1991, \mn@doi [\apj] {10.1086/170270}, \href {https://ui.adsabs.harvard.edu/abs/1991ApJ...376..214B} {376, 214}

\bibitem[\protect\citeauthoryear{{Balduin}, {Woitke}, {J{\o}rgensen}, {Thi}  \& {Narita}}{{Balduin} et~al.}{2023}]{Balduin_2023}
{Balduin} T.,  {Woitke} P.,  {J{\o}rgensen} U.~G.,  {Thi} W.~F.,   {Narita} Y.,  2023, \mn@doi [\aap] {10.1051/0004-6361/202346442}, \href {https://ui.adsabs.harvard.edu/abs/2023A&A...678A.192B} {678, A192}

\bibitem[\protect\citeauthoryear{{Blandford} \& {Payne}}{{Blandford} \& {Payne}}{1982}]{Blandford_Payne_1982}
{Blandford} R.~D.,  {Payne} D.~G.,  1982, \mn@doi [\mnras] {10.1093/mnras/199.4.883}, \href {https://ui.adsabs.harvard.edu/abs/1982MNRAS.199..883B} {199, 883}

\bibitem[\protect\citeauthoryear{{Crowell}}{{Crowell}}{1965}]{Corwell_1965}
{Crowell} C.~R.,  1965, \mn@doi [Solid State Electronics] {10.1016/0038-1101(65)90116-4}, \href {https://ui.adsabs.harvard.edu/abs/1965SSEle...8..395C} {8, 395}

\bibitem[\protect\citeauthoryear{{Desch} \& {Turner}}{{Desch} \& {Turner}}{2015}]{Desch_Turner_2015}
{Desch} S.~J.,  {Turner} N.~J.,  2015, \mn@doi [\apj] {10.1088/0004-637X/811/2/156}, \href {https://ui.adsabs.harvard.edu/abs/2015ApJ...811..156D} {811, 156}

\bibitem[\protect\citeauthoryear{{Draine}}{{Draine}}{2016}]{Draine_2016}
{Draine} B.~T.,  2016, \mn@doi [\apj] {10.3847/0004-637X/831/1/109}, \href {https://ui.adsabs.harvard.edu/abs/2016ApJ...831..109D} {831, 109}

\bibitem[\protect\citeauthoryear{{Draine} \& {Sutin}}{{Draine} \& {Sutin}}{1987}]{DS_87}
{Draine} B.~T.,  {Sutin} B.,  1987, \mn@doi [\apj] {10.1086/165596}, \href {https://ui.adsabs.harvard.edu/abs/1987ApJ...320..803D} {320, 803}

\bibitem[\protect\citeauthoryear{{Dzyurkevich}, {Turner}, {Henning}  \& {Kley}}{{Dzyurkevich} et~al.}{2013}]{Dzyurkevich_2013}
{Dzyurkevich} N.,  {Turner} N.~J.,  {Henning} T.,   {Kley} W.,  2013, \mn@doi [\apj] {10.1088/0004-637X/765/2/114}, \href {https://ui.adsabs.harvard.edu/abs/2013ApJ...765..114D} {765, 114}

\bibitem[\protect\citeauthoryear{{Fleming}, {Stone}  \& {Hawley}}{{Fleming} et~al.}{2000}]{Fleming_2000}
{Fleming} T.~P.,  {Stone} J.~M.,   {Hawley} J.~F.,  2000, \mn@doi [\apj] {10.1086/308338}, \href {https://ui.adsabs.harvard.edu/abs/2000ApJ...530..464F} {530, 464}

\bibitem[\protect\citeauthoryear{{Fujii}, {Okuzumi}  \& {Inutsuka}}{{Fujii} et~al.}{2011}]{Fuji_2011}
{Fujii} Y.~I.,  {Okuzumi} S.,   {Inutsuka} S.-i.,  2011, \mn@doi [\apj] {10.1088/0004-637X/743/1/53}, \href {https://ui.adsabs.harvard.edu/abs/2011ApJ...743...53F} {743, 53}

\bibitem[\protect\citeauthoryear{{Galassi}, {Davies}, {Theiler}, {Gough}, {Jungman}, {Booth}  \& {Rossi}}{{Galassi} et~al.}{2009}]{gough2009gnu}
{Galassi} M.,  {Davies} J.,  {Theiler} J.,  {Gough} B.,  {Jungman} G.,  {Booth} M.,   {Rossi} F.,  2009, GNU scientific library reference manual.
Network Theory Ltd.

\bibitem[\protect\citeauthoryear{{Gammie}}{{Gammie}}{1996}]{Gammie_1996}
{Gammie} C.~F.,  1996, \mn@doi [\apj] {10.1086/176735}, \href {https://ui.adsabs.harvard.edu/abs/1996ApJ...457..355G} {457, 355}

\bibitem[\protect\citeauthoryear{{Guillet}, {Hennebelle}, {Pineau des For{\^e}ts}, {Marcowith}, {Commer{\c{c}}on}  \& {Marchand}}{{Guillet} et~al.}{2020}]{Guillet_2020}
{Guillet} V.,  {Hennebelle} P.,  {Pineau des For{\^e}ts} G.,  {Marcowith} A.,  {Commer{\c{c}}on} B.,   {Marchand} P.,  2020, \mn@doi [\aap] {10.1051/0004-6361/201937387}, \href {https://ui.adsabs.harvard.edu/abs/2020A&A...643A..17G} {643, A17}

\bibitem[\protect\citeauthoryear{Hagström, Engvall  \& Pettersson}{Hagström et~al.}{2000}]{Hagstrom_2000}
Hagström M.,  Engvall K.,   Pettersson J. B.~C.,  2000, \mn@doi [J. Phys. Chem. B] {10.1021/jp000311w}, 104, 4457

\bibitem[\protect\citeauthoryear{{Hayashi}}{{Hayashi}}{1981}]{Hayashi_1981}
{Hayashi} C.,  1981, \mn@doi [Prog. Theor. Phys. Suppl.] {10.1143/PTPS.70.35}, \href {https://ui.adsabs.harvard.edu/abs/1981PThPS..70...35H} {70, 35}

\bibitem[\protect\citeauthoryear{{Ilgner} \& {Nelson}}{{Ilgner} \& {Nelson}}{2006}]{Ilgner_Nelson_2006}
{Ilgner} M.,  {Nelson} R.~P.,  2006, \mn@doi [\aap] {10.1051/0004-6361:20053678}, \href {https://ui.adsabs.harvard.edu/abs/2006A&A...445..205I} {445, 205}

\bibitem[\protect\citeauthoryear{{Ivlev}, {Akimkin}  \& {Caselli}}{{Ivlev} et~al.}{2016}]{Ivlev_2016}
{Ivlev} A.~V.,  {Akimkin} V.~V.,   {Caselli} P.,  2016, \mn@doi [\apj] {10.3847/1538-4357/833/1/92}, \href {https://ui.adsabs.harvard.edu/abs/2016ApJ...833...92I} {833, 92}

\bibitem[\protect\citeauthoryear{{Jankovic}, {Owen}, {Mohanty}  \& {Tan}}{{Jankovic} et~al.}{2021}]{Jankovic_2021}
{Jankovic} M.~R.,  {Owen} J.~E.,  {Mohanty} S.,   {Tan} J.~C.,  2021, \mn@doi [\mnras] {10.1093/mnras/stab920}, \href {https://ui.adsabs.harvard.edu/abs/2021MNRAS.504..280J} {504, 280}

\bibitem[\protect\citeauthoryear{{Jin}}{{Jin}}{1996}]{Jin_1996}
{Jin} L.,  1996, \mn@doi [\apj] {10.1086/176774}, \href {https://ui.adsabs.harvard.edu/abs/1996ApJ...457..798J} {457, 798}

\bibitem[\protect\citeauthoryear{{Marchand}, {Guillet}, {Lebreuilly}  \& {Mac Low}}{{Marchand} et~al.}{2021}]{Marchand_2021}
{Marchand} P.,  {Guillet} V.,  {Lebreuilly} U.,   {Mac Low} M.~M.,  2021, \mn@doi [\aap] {10.1051/0004-6361/202040077}, \href {https://ui.adsabs.harvard.edu/abs/2021A&A...649A..50M} {649, A50}

\bibitem[\protect\citeauthoryear{{Marchand}, {Guillet}, {Lebreuilly}  \& {Mac Low}}{{Marchand} et~al.}{2022}]{Marchand_2022_2}
{Marchand} P.,  {Guillet} V.,  {Lebreuilly} U.,   {Mac Low} M.~M.,  2022, \mn@doi [\aap] {10.1051/0004-6361/202142551}, \href {https://ui.adsabs.harvard.edu/abs/2022A&A...666A..27M} {666, A27}

\bibitem[\protect\citeauthoryear{{Mathis}, {Rumpl}  \& {Nordsieck}}{{Mathis} et~al.}{1977}]{MRN_1977}
{Mathis} J.~S.,  {Rumpl} W.,   {Nordsieck} K.~H.,  1977, \mn@doi [\apj] {10.1086/155591}, \href {https://ui.adsabs.harvard.edu/abs/1977ApJ...217..425M} {217, 425}

\bibitem[\protect\citeauthoryear{{Millar}, {Walsh}, {Van de Sande}  \& {Markwick}}{{Millar} et~al.}{2024}]{UMIST_2022}
{Millar} T.~J.,  {Walsh} C.,  {Van de Sande} M.,   {Markwick} A.~J.,  2024, \mn@doi [\aap] {10.1051/0004-6361/202346908}, \href {https://ui.adsabs.harvard.edu/abs/2024A&A...682A.109M} {682, A109}

\bibitem[\protect\citeauthoryear{{Mohanty}, {Ercolano}  \& {Turner}}{{Mohanty} et~al.}{2013}]{Mohanty_2013}
{Mohanty} S.,  {Ercolano} B.,   {Turner} N.~J.,  2013, \mn@doi [\apj] {10.1088/0004-637X/764/1/65}, \href {https://ui.adsabs.harvard.edu/abs/2013ApJ...764...65M} {764, 65}

\bibitem[\protect\citeauthoryear{{Mohanty}, {Jankovic}, {Tan}  \& {Owen}}{{Mohanty} et~al.}{2018}]{Mohanty_2018}
{Mohanty} S.,  {Jankovic} M.~R.,  {Tan} J.~C.,   {Owen} J.~E.,  2018, \mn@doi [\apj] {10.3847/1538-4357/aabcd0}, \href {https://ui.adsabs.harvard.edu/abs/2018ApJ...861..144M} {861, 144}

\bibitem[\protect\citeauthoryear{Mor{\'e}, Sorensen, Hillstrom  \& Garbow}{Mor{\'e} et~al.}{1984}]{more1984minpack}
Mor{\'e} J.~J.,  Sorensen D.~C.,  Hillstrom K.~E.,   Garbow B.~S.,  1984, in Cowell W.~R.,  ed., , Vol.~25, Sources and Development of Mathematical Software.
Prentice-Hall, Inc., USA, pp 88--111

\bibitem[\protect\citeauthoryear{{Mori} \& {Okuzumi}}{{Mori} \& {Okuzumi}}{2016}]{Mori_2016}
{Mori} S.,  {Okuzumi} S.,  2016, \mn@doi [\apj] {10.3847/0004-637X/817/1/52}, \href {https://ui.adsabs.harvard.edu/abs/2016ApJ...817...52M} {817, 52}

\bibitem[\protect\citeauthoryear{{Okuzumi}}{{Okuzumi}}{2009}]{Okuzumi_2009}
{Okuzumi} S.,  2009, \mn@doi [\apj] {10.1088/0004-637X/698/2/1122}, \href {https://ui.adsabs.harvard.edu/abs/2009ApJ...698.1122O} {698, 1122}

\bibitem[\protect\citeauthoryear{{Okuzumi}, {Tanaka}, {Takeuchi}  \& {Sakagami}}{{Okuzumi} et~al.}{2011a}]{Okuzumi_2011}
{Okuzumi} S.,  {Tanaka} H.,  {Takeuchi} T.,   {Sakagami} M.-a.,  2011a, \mn@doi [\apj] {10.1088/0004-637X/731/2/95}, \href {https://ui.adsabs.harvard.edu/abs/2011ApJ...731...95O} {731, 95}

\bibitem[\protect\citeauthoryear{{Okuzumi}, {Tanaka}, {Takeuchi}  \& {Sakagami}}{{Okuzumi} et~al.}{2011b}]{Okuzumi_2011b}
{Okuzumi} S.,  {Tanaka} H.,  {Takeuchi} T.,   {Sakagami} M.-a.,  2011b, \mn@doi [\apj] {10.1088/0004-637X/731/2/96}, \href {https://ui.adsabs.harvard.edu/abs/2011ApJ...731...96O} {731, 96}

\bibitem[\protect\citeauthoryear{{Oppenheimer} \& {Dalgarno}}{{Oppenheimer} \& {Dalgarno}}{1974}]{Oppenheimer_Dalgarno_1974}
{Oppenheimer} M.,  {Dalgarno} A.,  1974, \mn@doi [\apj] {10.1086/153030}, \href {https://ui.adsabs.harvard.edu/abs/1974ApJ...192...29O} {192, 29}

\bibitem[\protect\citeauthoryear{{Ormel} \& {Cuzzi}}{{Ormel} \& {Cuzzi}}{2007}]{OrmelCuzzi_2007}
{Ormel} C.~W.,  {Cuzzi} J.~N.,  2007, \mn@doi [\aap] {10.1051/0004-6361:20066899}, \href {https://ui.adsabs.harvard.edu/abs/2007A&A...466..413O} {466, 413}

\bibitem[\protect\citeauthoryear{Ortega \& Rheinboldt}{Ortega \& Rheinboldt}{2000}]{Ortega_Rheinboldt_2000}
Ortega J.~M.,  Rheinboldt W.~C.,  2000, Iterative Solution of Nonlinear Equations in Several Variables.
SIAM, Philadelphia PA, \mn@doi{10.1137/1.9780898719468}

\bibitem[\protect\citeauthoryear{{Pneuman} \& {Mitchell}}{{Pneuman} \& {Mitchell}}{1965}]{Pneumann_Mitchell_1965}
{Pneuman} G.~W.,  {Mitchell} T.~P.,  1965, \mn@doi [\icarus] {10.1016/0019-1035(65)90026-6}, \href {https://ui.adsabs.harvard.edu/abs/1965Icar....4..494P} {4, 494}

\bibitem[\protect\citeauthoryear{Powell}{Powell}{1970}]{Powell_1970}
Powell M.~J.~D.,  1970, in {Robinowitz} P.,  ed., , Numerical Methods for Nonlinear Algebraic Equations.
Gordon and Breach, London, UK, pp 87--161

\bibitem[\protect\citeauthoryear{{Robinson}, {Booth}  \& {Owen}}{{Robinson} et~al.}{2024}]{Robinson_2024}
{Robinson} A.,  {Booth} R.~A.,   {Owen} J.~E.,  2024, \mn@doi [\mnras] {10.1093/mnras/stae624}, \href {https://ui.adsabs.harvard.edu/abs/2024MNRAS.529.1524R} {529, 1524}

\bibitem[\protect\citeauthoryear{{Salmeron} \& {Wardle}}{{Salmeron} \& {Wardle}}{2008}]{Salmeron_Wardle_2008}
{Salmeron} R.,  {Wardle} M.,  2008, \mn@doi [\mnras] {10.1111/j.1365-2966.2008.13430.x}, \href {https://ui.adsabs.harvard.edu/abs/2008MNRAS.388.1223S} {388, 1223}

\bibitem[\protect\citeauthoryear{{Sano} \& {Stone}}{{Sano} \& {Stone}}{2002}]{Sano_Stone_2002b}
{Sano} T.,  {Stone} J.~M.,  2002, \mn@doi [\apj] {10.1086/342172}, \href {https://ui.adsabs.harvard.edu/abs/2002ApJ...577..534S} {577, 534}

\bibitem[\protect\citeauthoryear{Sano, Miyama, Umebayashi  \& Nakano}{Sano et~al.}{2000}]{Sano_2000}
Sano T.,  Miyama S.~M.,  Umebayashi T.,   Nakano T.,  2000, \mn@doi [\apj] {10.1086/317075}, 543, 486

\bibitem[\protect\citeauthoryear{{Sano}, {Inutsuka}, {Turner}  \& {Stone}}{{Sano} et~al.}{2004}]{Sano_2004}
{Sano} T.,  {Inutsuka} S.-i.,  {Turner} N.~J.,   {Stone} J.~M.,  2004, \mn@doi [\apj] {10.1086/382184}, \href {https://ui.adsabs.harvard.edu/abs/2004ApJ...605..321S} {605, 321}

\bibitem[\protect\citeauthoryear{{Thi}, {Lesur}, {Woitke}, {Kamp}, {Rab}  \& {Carmona}}{{Thi} et~al.}{2019}]{Thi_2019}
{Thi} W.~F.,  {Lesur} G.,  {Woitke} P.,  {Kamp} I.,  {Rab} C.,   {Carmona} A.,  2019, \mn@doi [\aap] {10.1051/0004-6361/201732187}, \href {https://ui.adsabs.harvard.edu/abs/2019A&A...632A..44T} {632, A44}

\bibitem[\protect\citeauthoryear{{Umebayashi} \& {Nakano}}{{Umebayashi} \& {Nakano}}{2009}]{Umebayashi_Nakano_2009}
{Umebayashi} T.,  {Nakano} T.,  2009, \mn@doi [\apj] {10.1088/0004-637X/690/1/69}, \href {https://ui.adsabs.harvard.edu/abs/2009ApJ...690...69U} {690, 69}

\bibitem[\protect\citeauthoryear{{Wardle}}{{Wardle}}{2007}]{Wardle_2007}
{Wardle} M.,  2007, \mn@doi [\apss] {10.1007/s10509-007-9575-8}, \href {https://ui.adsabs.harvard.edu/abs/2007Ap&SS.311...35W} {311, 35}

\bibitem[\protect\citeauthoryear{{Wardle} \& {Ng}}{{Wardle} \& {Ng}}{1999}]{Wardle_Ng_2013}
{Wardle} M.,  {Ng} C.,  1999, \mn@doi [\mnras] {10.1046/j.1365-8711.1999.02211.x}, \href {https://ui.adsabs.harvard.edu/abs/1999MNRAS.303..239W} {303, 239}

\bibitem[\protect\citeauthoryear{{Weidenschilling}}{{Weidenschilling}}{1977}]{Weidenschilling_1977}
{Weidenschilling} S.~J.,  1977, \mn@doi [\mnras] {10.1093/mnras/180.2.57}, \href {https://ui.adsabs.harvard.edu/abs/1977MNRAS.180...57W} {180, 57}

\bibitem[\protect\citeauthoryear{{Weidenschilling}}{{Weidenschilling}}{1980}]{Weidenschilling_1980}
{Weidenschilling} S.~J.,  1980, \mn@doi [\icarus] {10.1016/0019-1035(80)90064-0}, \href {https://ui.adsabs.harvard.edu/abs/1980Icar...44..172W} {44, 172}

\bibitem[\protect\citeauthoryear{{Wurm} \& {Blum}}{{Wurm} \& {Blum}}{1998}]{Wurm_Blum_1998}
{Wurm} G.,  {Blum} J.,  1998, \mn@doi [\icarus] {10.1006/icar.1998.5891}, \href {https://ui.adsabs.harvard.edu/abs/1998Icar..132..125W} {132, 125}

\makeatother
\end{thebibliography}

%\begin{thebibliography}{99}
%\end{thebibliography}

%%%%%%%%%%%%%%%%%%%%%%%%%%%%%%%%%%%%%%%%%%%%%%%%%%

%%%%%%%%%%%%%%%%% APPENDICES %%%%%%%%%%%%%%%%%%%%%

\appendix

\section{MRI activity criteria and adopted alkali}
\label{sec:append-mri-conds}
The alkali we include in our chemical network is guided by the desire to achieve a sufficiently high ionization fraction to sustain the MRI. Specifically, efficient MRI in a Keplerian disc requires that the maximum growth rate of MRI-generated local tangled fields, $\sim$$\Omega$ (the orbital frequency, since the fields are fundamentally amplified by the orbital shear), exceed the Ohmic dissipation rate of the fields, $\sim$$k^2\eta_O$, where $k \sim \Omega/{\rm v}_{\mathcal{A} z}$ is the fastest growing vertical mode, ${\rm v}_{\mathcal{A} z}$ is the vertical component of the local Alfv\'{e}n velocity, and $\eta_O$ is the Ohmic resistivity \cite[see][and references therein]{Mohanty_2018}. This results in the Elssaser number criterion for efficient MRI:     
\begin{equation}
    \Lambda_{\text{O}} \equiv \frac{{\rm v}_{\mathcal{A} z}^2}{\eta_{\text{O}} \Omega} > 1 \,\,\,\,\, .
    \label{eqn:Ohmic_Elsasser}
\end{equation}
We derive a simple scaling relation for the Ohmic Elssaser number, $\Lambda_{\text{O}}$, as follows. Using the expression for Ohmic resistivity $\eta_O$ given by \citet{Wardle_2007}, along with the rate coefficient for collisional momentum transfer between charged species and neutrals from~\cite{Wardle_Ng_2013}, and noting that $\beta_e \gg \beta_i$ (i.e., the Hall parameter for electrons greatly exceeds that for ions), we obtain
\begin{equation}
    \eta_{\text{O}} \sim 230 \,\, T^{\frac{1}{2}} \, {\mathcal{X}}_e^{-1} \,\,\,\,\, ,
    \label{eqn:OhmicResistivity}
\end{equation}
where ${\mathcal{X}}_e \equiv {n_e}/{n_{\text{H}_2}}$ is the ionization fraction, and the $T^{1/2}$ comes from the momentum transfer rate in electron-neutral collisions.  

Concurrently, the vertical Alfv\'{e}n velocity is ${\rm v}_{\mathcal{A} z} \equiv B_z/\sqrt{4\pi\rho}$, where $B_z$ is the strength of the vertical component of the local field, and $\rho$ is the density. We adopt the results of numerical simulations of the MRI, which indicate that $B_z^2 \sim B^2/25$, where $B$ is the strength of the total local r.m.s. field \citep{Sano_2004}. 

We further assume the disc is vertically isothermal \citep[a reasonable approximation for this back-of-the-envelope calculation, at least within a scale-height of the mid-plane in the inner disc; e.g., see][]{Jankovic_2021}. Then, close to the mid-plane, the density is $\rho \sim \Sigma/z_H$, where $\Sigma$ is the surface density and $z_H = c_s/\Omega$ is the disc scale-height, for an isothermal sound speed of $c_s = (kT/\mu m_\text{H})^{1/2}$ (where $\mu$ is the mean molecular weight and $m_H$ is the mass of a hydrogen atom). Using $\mu = 2.34$, we obtain
\begin{equation}
    {\rm v}_{\mathcal{A} z}^2 \sim 19 \,\, B^2 \,T^{\frac{1}{2}}\,\Sigma^{-1}\,\Omega^{-1} \,\,\,\,\, .
    \label{eqn:VerticalAlfvenVelocity}
\end{equation}
Combining equations (\ref{eqn:Ohmic_Elsasser}), (\ref{eqn:OhmicResistivity}) and (\ref{eqn:VerticalAlfvenVelocity}), the Elssaser number condition finally becomes
\begin{equation}
\Lambda_{\text{O}} \sim 8\times10^{-2}\, {\mathcal{X}}_e\, \left(\frac{B^2}{\Sigma\,\Omega^2}\right) \,>\, 1 \,\,\,\,\, .
    \label{eqn:ScaledElssaser}
\end{equation}
Notice  that the explicit dependence on temperature in equations (\ref{eqn:OhmicResistivity}) and (\ref{eqn:VerticalAlfvenVelocity}) has cancelled out in equation (\ref{eqn:ScaledElssaser}): $\Lambda_{\text{O}}$ in this formulation only depends on temperature implicitly through whatever $T$-dependencies ${\mathcal{X}}_e$, $B$ and $\Sigma$ may have.

For a specified set of local disc parameters $B$, $\Sigma$ and $\Omega$, equation (\ref{eqn:ScaledElssaser}) yields a threshold value of ${\mathcal{X}}_e$ that must be exceeded for efficient MRI. Hence, the pertinent alkali(s) to consider will be the one(s) that can surpass this threshold ionization fraction. The local disc parameters will of course depend on the specific disc physics being investigated. As an illustrative example, and to justify the alkali chosen in this paper, we consider: a solar-mass star (implying $\Omega(r)$ = 2$\times$$10^{-7}$ ($r/1$\,au)$^{-3/2}$\,s$^{-1}$), surrounded by a disc with a Minimum Mass Solar Nebula (MMSN) surface density profile \citep[$\Sigma(r)$ = $\Sigma_0$($r/1$\,au)$^{-3/2}$;][]{Hayashi_1981}. Equation (\ref{eqn:ScaledElssaser}) then becomes 
\begin{equation}
    \Lambda_{\text{O}} \sim 10^{11} \, {\mathcal{X}}_e \,\times\, \left( \frac{\Sigma_0}{1700 \, \text{g cm}^{-2}} \right)^{-1} \left( \frac{B}{10 \text{ G}} \right)^2 \left( \frac{r}{1 \, \text{au}} \right)^{\frac{9}{2}} > 1 \,\,\,\,\, 
    \label{eqn:ScaledElssaserMMSN}
\end{equation}
for the Ohmic Elssaser number around a 1\,M$_{\odot}$ star, where we have normalized $\Sigma_0$ (the surface density at 1\,au) by the \citet{Hayashi_1981} MMSN value, and the r.m.s. field strength $B$ by the (plausible) value of $\sim$10\,G found by \citet{Jankovic_2021} for inner disc regions ($\lesssim$1\,au). Equation (\ref{eqn:ScaledElssaserMMSN}) says that, in an MMSN-like disc around a solar-mass star, efficient MRI at $\sim$1\,au requires ${\mathcal{X}}_e > 10^{-11}$ at this location. The radial ($r$) dependence in equation (\ref{eqn:ScaledElssaserMMSN}) also implies that higher ionization fractions are needed closer in (because the Ohmic dissipation rate of the field increases faster than the growth rate with decreasing radius in a vertically isothermal MMSN disc). 

To determine which alkali species can produce the requisite ionization fraction ${\mathcal{X}}_e$, we first invoke the Saha equation
\begin{equation}
    \frac{n_e n_{\text{alk}^+}}{n_{\text{alk}^0}} = \frac{1}{\lambda_e^3} \frac{g_e g_+}{g_0} \exp \left( \frac{-\text{IP}}{kT} \right) \equiv \mathcal{S}(T) \,\,\,\,\, ,
\end{equation}
where $\lambda_e \equiv \sqrt{h^2/(2 \pi m_e k T)}$ is the thermal de Broglie wavelength of electrons, and $g_e (= 2)$, $g_0 (= 2)$ and $g_+ (= 1)$ are the degeneracies of the electron, ground state alkali atom and ground state alkali ion respectively. Assuming $n_e \sim n_{\text{alk}^+}$ (i.e., gas-phase ionization of alkalis is the dominant charge-producing process), and $n_{\text{alk}}^{\text{tot}} (\equiv n_{\text{alk}^0}+n_{\text{alk}^+})\sim n_{\text{alk}^0}$ (weak ionization limit), gives ${\mathcal{X}}_e \sim \sqrt{\mathcal{S}(T) n_{\text{alk}}^{\text{tot}} / n_{\text{H}_2}^2}$. Using this, we calculate the temperature required for either sodium or potassium (the two most abundant alkalis) to produce the threshold ${\mathcal{X}}_e$ $\sim$ $10^{-11}$ needed for MRI at 1\,au. Adopting solar photospheric abundances from \citet{Asplund_2009} ($x_{\text{H}} = 9.21 \times 10^{-1}$, $x_{\text{Na}} = 1.6 \times 10^{-6}$ and $x_{\text{K}} = 9.87 \times 10^{-8}$, where $x_i \equiv n_i/\left(\sum_j n_j\right)$ for any species $i$), together with the respective alkali ionization potentials ($\text{IP}_{\text{Na}} = 5.14$\,eV, $\text{IP}_{\text{K}} = 4.34 \, \text{eV}$), in the same vertically isothermal MMSN disc invoked above, we infer that sodium reaches the requisite ionization threshold at $T$\,$\sim$\,1180\,K, while potassium reaches it at an appreciably lower $T$\,$\sim$\,1060\,K (at this $T$, sodium is a factor of 20 less ionized due to its higher IP, in spite of its higher abundance). Thus, {\it if} gas-phase collisional ionization of alkalis is the primary source of charge, then potassium is the pertinent alkali to consider, at least at $\sim$1\,au in a vertically isothermal MMSN disc (the fiducial conditions we choose to illustrate our results in this paper.\footnote{More precisely, we have adopted $n_{\text{H}_2} = 10^{14}$\,cm$^{-3}$ in much of this paper, to facilitate comparison of our results to the existing literature; this corresponds to the mid-plane number density at $T$$\sim$1000\,K ($\sim$$T$ at which potassium reaches the threshold ${\mathcal{X}}_e$ for MRI), in a vertically isothermal MMSN disc at $\sim$1\,au.})\footnote{\citet{Gammie_1996} also calculate the threshold ionization needed for efficient MRI; they find it requires ${\mathcal{X}}_e > 10^{-13}$, less than our threshold of $10^{-11}$, because they use a slightly different set of assumptions for the onset of MRI than we do.} 

On the other hand, as we show in this work \citep[and also shown by][]{Desch_Turner_2015}, ionic and thermionic emission from grains, following ionization of alkali atoms on grain surfaces, dominate over gas-phase collisional ionization of alkalis as the source of free charges in most of the inner disc. The work function of grains relative to the alkali ionization potential means that, on grain surfaces, ionization of potassium is energetically more favourable than ionization of sodium. Thus, in this case too, potassium is more important than sodium for determining the gas-phase charge abundances (and also for grain charging).

For these reasons, we adopt potassium as the relevant alkali over most of this paper. However, for completeness, we also show the effect of including sodium as an additional alkali, in Appendix\,\ref{sec:sodium} below.

\section{Including another alkali metal}
\label{sec:sodium}
Here we consider the impact of adding another alkali to our network. We choose sodium for its large abundance. Relative to potassium, sodium has a lower activation energy of 3.08 eV, a higher first ionization potential of 5.14 eV and a higher abundance of $1.6 \times 10^{-6}$~\citep{Asplund_2009}. 

\begin{figure}
    \centering
    \includegraphics[width=\columnwidth]{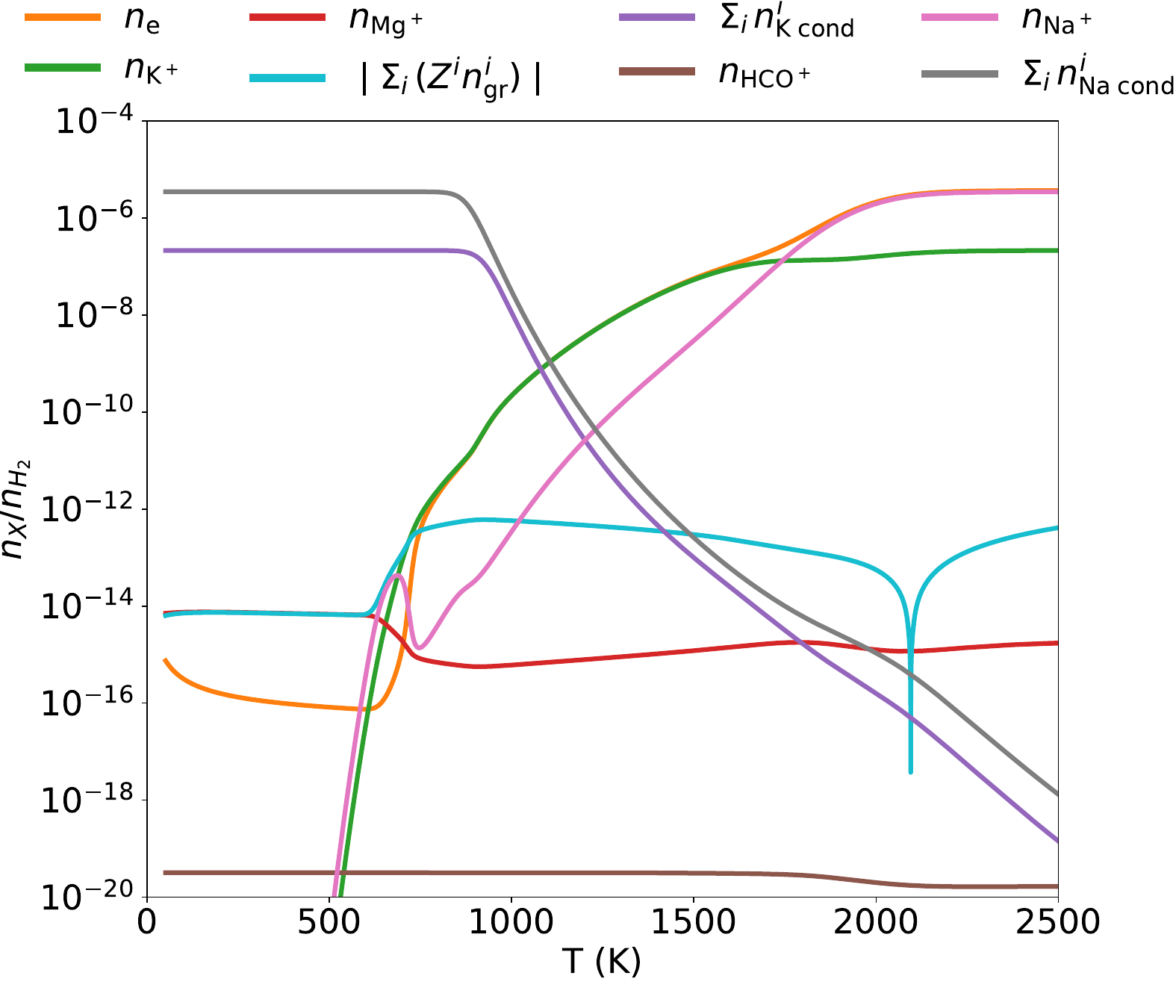}
    \caption{The ionization state of the reaction network including sodium is shown. The distribution is the fiducial MRN distribution with dust-to-gas ratio $f_{\text{dg}} = 0.01$, MRN exponent $q=3.5$, $a_{\text{min}} = 10^{-5} \text{ cm}$, $a_{\text{max}} = 10^{-1} \text{ cm}$ at $n_{\text{H}_2} = 10^{14} \; \mathrm{cm^{-3}}$.} 
    \label{fig:sodium}
\end{figure}

This network is plotted as a function of temperature for our fiducial parameters in Fig.~\ref{fig:sodium}. The effect of the reduced activation energy is to produce a larger number density of Na$^+$ in the gas-phase than that of K$^+$ below around 600\,K. At high enough temperature, the effect of the ionization potential of Na being greater than the ionization potential of K means that the K$^+$ abundance becomes greater than the Na$^+$ abundance. At $T\gtrsim 2000$\,K, the ionization potential of Na$^+$ is surmounted and due to the enhanced abundance of Na relative to K, the abundance of Na$^+$ exceeds the K$^+$ abundance.

The only parts of the temperature range where the inclusion of Na significantly alters the abundances are around 600\,K and above around 2000\,K. In these regions, the inclusion of sodium also affects the grain charge. Above 2000\,K, this effect is irrelevant as grains should be sublimated. However, the region around 600\,K, where Na$^+$ evaporates from the grain surface will affect the grain charge. Therefore, the effect of Na should ideally be included where fragmentation-coagulation calculations are to be co-evolved with the disc chemistry. However, this is a restricted temperature range and is unlikely to affect things significantly across the disc as a whole.

\section{Linear Relationship Between Grain Charge and Grain Size}
\label{sec:append-ds87}

~\cite{DS_87} {(excluding the effects of ion and thermionic emission)} considered a distribution of charges for each grain size, and determined that the { distribution of charges on a single grain size follows a Gaussian, whose mean} $\bar{Z^i} \propto a^i$ {(or equivalently $\tau^i \equiv a^i k T / e^2$)}. {A derivation for this is provided in~\cite{Okuzumi_2009}. We provide a similar derivation here, with ion and thermionic emission included.}

Consider the detailed balance equation between state $Z^i$ and state $Z^i + 1$:

\begin{multline}
  p(Z^i, \tau^i) \biggl[ n_{\mathrm{m^+}} {S}_{\mathrm{m^+}} {\rm{v}}_{\mathrm{m^+}} \Tilde{J}(Z^i, \tau^i) + n_{\mathrm{M^+}} {S}_{\mathrm{M^+}} 
  {\rm{v}}_{\mathrm{M^+}} \Tilde{J}(Z^i, \tau^i)
  \\ + n_{\mathrm{K^+}} {S}_{\mathrm{K^+}} {\rm{v}}_{\mathrm{K^+}} \Tilde{J}(Z^i, \tau^i) 
  \\
  + 4 \lambda_R \frac{4 \pi m_e (kT)^2}{h^3} \exp(-W/kT) \exp(-Z^i/\tau^i) \biggr] 
  \\
   \approx p(Z^i+1, \tau^i) \Bigl[ n_e {S}_e {\rm{v}}_e \Tilde{J}(-Z^i, \tau^i) 
   \\
   + f_+(Z^i/\tau^i)n_{\mathrm{K^+}} {S}_{\mathrm{K^+}} {\rm{v}}_{\mathrm{K^+}}\Tilde{J}(Z^i, \tau^i)
   + f_+(Z^i/\tau^i)n_{\mathrm{K^0}} {S}_{\mathrm{K^0}} {\rm{v}}_{\mathrm{K^0}} \Bigr],
  \label{eqn:detailed-balance}
\end{multline}
{ where $S_{x}$ defines the sticking coefficient, and ${\rm{v}}_x$ the thermal velocity of species $x$. The probability that the grain with size $\tau^i$ will have charge $Z^i$ is denoted $p(Z^i, \tau^i)$}. This is not an exact equality, since the $\Tilde{J}$ and $f_+$ terms on the right-hand side should be functions of $Z^i + 1$; however, provided these functions do not change significantly with small changes in $Z$, we can make this approximation. In the limit of large $\tau$, this approximation is valid. {We shall illustrate why the approximation is valid  for $f_+^i$ . The derivative of ~\cref{eqn:f+i} for $f_+^i$ with respect to $Z^i$ is
\begin{equation}
    \frac{\partial f_+^i}{\partial Z^i} = \frac{1}{\tau^i} \left(1 - f_+^i \right) f_+^i \,\,\,\,,
    \label{eqn:fplus_derivative}
\end{equation}
where $f_+^i<1$, meaning that $\frac{\partial f_+^i}{\partial Z^i} \to 0$ in the limit of large $\tau^i$.}

For $\tau^i \gg 1$ and $\lvert Z^i \rvert \gg 1$, the focusing factor of species $x$ is~\citep{DS_87}

\begin{align}
&\ \Tilde{J}_{x}(Z^ie/q_{x} < 0) \to \left(1-\frac{Z^ie}{q_{x} \tau^i} \right)
\\
&\ \Tilde{J}_{x}(Z^ie/q_{x} > 0) \to \exp\left(-\frac{Z^ie}{q_{x} \tau^i} \right),
\end{align}
where $q_x$ is the charge of the species $x$ interacting with the grain. { Computing the derivatives with respect to $Z^i$, as we do in~\ref{eqn:fplus_derivative}, it is trivial to show that these derivatives also vanish in the limit of large $\tau^i$.}

{
~\cite{Okuzumi_2009} uses the assumption that $\langle (Z^i)^2 \rangle^{-\frac{1}{2}} \sim (\tau^i)^{-\frac{1}{2}} \sim \epsilon \ll 1$, the first relation is justified after deriving the form of $\langle (Z^i)^2 \rangle$. Using the additional assumption that $p(Z^i, \tau^i)$ varies on the scale of $\langle (Z^i)^2 \rangle^{\frac{1}{2}}$
\begin{equation}
    p(Z^i + 1, \tau^i) = p(Z^i,  \tau^i) + \frac{\partial p}{\partial Z^i} (Z^i,  \tau^i) + \mathcal{O}(\epsilon^2) \,\,\,\,\, .
\end{equation}
Substituting this expansion into~\cref{eqn:detailed-balance}, an analogous differential equation to that of~\cite{Okuzumi_2009} is derived with
\begin{equation}
     \frac{\partial p}{\partial Z^i} (Z^i,  \tau^i) + W(Z^i, \tau^i) p(Z^i, \tau^i) \approx 0 \,\,\,\,\, .
\end{equation}
However, in this case $W(Z^i, \tau^i)$ is given by
\begin{equation}
    W(Z^i, \tau^i) = 1 - \frac{\lambda_{\text{ions}} \left(1-\frac{Z^i}{\tau^i} \right) + \alpha_{\text{therm}} \exp\left(-\frac{Z^i}{\tau^i}\right) }{\lambda_{\text{e}} \exp\left(\frac{Z^i}{\tau^i}\right) + \lambda_{\text{K}^+} \left(1-\frac{Z^i}{\tau^i} \right) f_+^i + \lambda_{\text{K}^0} f_+^i} \,\,\,\,\, ,
    \label{eqn:Wfull}
\end{equation}
where we have used $\lambda_{x} = n_x S_x {\rm v}_x$, $\lambda_{\text{ions}} = \lambda_{\text{m}^+} + \lambda_{\text{M}^+} + \lambda_{\text{K}^+}$ and $\alpha_{\text{therm}} = 4 \lambda_R \frac{4 \pi m_e (kT)^2}{h^3} \exp(-W/kT)$ to simplify the notation. In the limit $k T \ll E_a < W$, we have 
\begin{equation}
    W(Z^i, \tau^i) = 1 - \frac{\lambda_{\text{ions}} \left(1-\frac{Z^i}{\tau^i} \right) }{\lambda_{\text{e}} \exp\left(\frac{Z^i}{\tau^i}\right)} \,\,\,\,\, ,
\end{equation}
which is the form in~\cite{Okuzumi_2009}, except allowing for the presence of multiple ions. 

The grain charge $Z_0^i$ with maximum probability, i.e., for which $\frac{\partial p}{\partial Z^i} = 0$, is found by determining the root $W(Z_0^i, \tau^i) = 0$ where $W$ is given by~\cref{eqn:Wfull}. Since $Z_0^i/\tau^i = \psi$ depends only on plasma quantities, all other grain sizes denoted by the superscript $i$ have the same constant $\psi$. Therefore, $Z_0^i = \psi \tau^i$ for all $i$, and the root of only a single equation is required to determine the most probable grain charge for each grain size. For a Gaussian, the most probable grain charge for a single grain size ${Z_0^i}$ is also the mean grain charge for that grain size $\bar{Z^i}$. For the single grain charge $Z_{\text{single}}^i$ we have used for each grain size elsewhere in this work, we also determine the root $W(Z_{\text{single}}^i, \tau^i) = 0$, thus $Z_{\text{single}}^i$ is the most probable grain size, and thus for a Gaussian distribution, the mean grain charge.

$W(Z^i, \tau^i)$ is then expanded about $Z_0^i$ to first order in $\lvert \delta Z^i \rvert / \tau^i$. Since
\begin{multline}
        \frac{\partial W}{\partial Z^i} =  \frac{1}{{\lambda_{\text{e}} \exp\left(\frac{Z^i}{\tau^i}\right) + \lambda_{\text{K}^+} \left(1-\frac{Z^i}{\tau^i} \right) f_+^i + \lambda_{\text{K}^0} f_+^i  }} \times \\
    \left[ \lambda_{\text{ions}} + \alpha_{\text{therm}} \exp\left(-\frac{Z^i}{\tau^i}\right) + \lambda_{e} \exp\left(\frac{Z^i}{\tau^i}\right) - \lambda_{\text{K}^+} f_+^i \right. \\
    - \left. \left(\lambda_{\text{K}^+} \left(1-\frac{Z^i}{\tau^i} \right) + \lambda_{\text{K}^0}\right) f_+^i (f_+^i - 1)    \right] \frac{1   }{\tau^i} \,\,\,\,\ ,
\end{multline}
$\frac{\partial^2 W}{\partial {Z^i}^2} \propto 1/(\tau^i)^2$ and $W(Z_0^i, \tau^i) = 0$, to first order in $\lvert \delta Z^i \rvert / \tau^i$, 
\begin{multline}
        W \approx \frac{1}{{\lambda_{\text{e}} \exp\left(\frac{Z_0^i}{\tau^i}\right) + \lambda_{\text{K}^+} \left(1-\frac{Z_0^i}{\tau^i} \right) f_+^i + \lambda_{\text{K}^0} f_+^i  }} \times \\
    \left[ \lambda_{\text{ions}} + \alpha_{\text{therm}} \exp\left(-\frac{Z_0^i}{\tau^i}\right) + \lambda_{e} \exp\left(\frac{Z_0^i}{\tau^i}\right) - \lambda_{\text{K}^+} f_+^i \right. \\
    - \left. \left(\lambda_{\text{K}^+} \left(1-\frac{Z_0^i}{\tau^i} \right) + \lambda_{\text{K}^0}\right) f_+^i (f_+^i - 1)    \right] \frac{\lvert \delta Z^i \rvert}{\tau^i} \,\,\,\,\, .
\end{multline}
The expansion to first order in $\lvert \delta Z^i \rvert \ll \tau^i$ may be re-written in terms of the variance (which has yet to be determined) $\lvert \delta Z^i \rvert \ll \langle (Z^i)^2 \rangle^{\frac{1}{2}} (\tau^i)^{\frac{1}{2}}$. This implies that the expansion is valid provided $\lvert \delta Z^i \rvert \lesssim \langle (Z^i)^2 \rangle^{\frac{1}{2}}$.

Given this form of $W$, $p(Z^i, \tau^i)$ follows a Gaussian with a mean $Z_0^i$, with a variance $\langle (Z^i)^2 \rangle$, where
\begin{multline}
    \langle (\Delta Z^i)^2  \rangle \approx \left[ {{\lambda_{\text{e}} \exp\left(\frac{Z_0^i}{\tau^i}\right) + \lambda_{\text{K}^+} \left(1-\frac{Z_0^i}{\tau^i} \right) f_+^i + \lambda_{\text{K}^0} f_+^i  }} \right] \times \\
    \left[ \lambda_{\text{ions}} + \alpha_{\text{therm}} \exp\left(-\frac{Z_0^i}{\tau^i}\right) + \lambda_{e} \exp\left(\frac{Z_0^i}{\tau^i}\right) - \lambda_{\text{K}^+} f_+^i \right. \\
    - \left. \left(\lambda_{\text{K}^+} \left(1-\frac{Z_0^i}{\tau^i} \right) + \lambda_{\text{K}^0}\right) f_+^i (f_+^i - 1)    \right]^{-1} {\tau^i} \,\,\,\,\, .
\end{multline}
To show that $\langle (\Delta Z^i)^2  \rangle$ is order $\tau^i$, consider limiting values of $Z_0^i / \tau^i$. In the limit ${Z_0^i}/{\tau^i} \gg 1$, $\langle (\Delta Z^i)^2  \rangle = \tau^i$ and for
${Z_0^i}/{\tau^i} \ll -1$, we also have $\langle (\Delta Z^i)^2  \rangle = \tau^i$. When ${Z_0^i}/{\tau^i} \to 0$, $(\Delta Z^i)^2  \rangle \geq \frac{1}{2} \tau^i$.  

We now plot a full distribution $p(Z^i, \tau^i)$ by using the recurrence relation~\cref{eqn:detailed-balance} and the normalization $\Sigma_{Z^i} p(Z^i, \tau^i) = 1$. We compute $\psi$ and $\langle (\Delta Z^i)^2  \rangle$ according to the method described above and plot the resulting Gaussian alongside the full distribution. To verify the method, we compare the Gaussian derived in this work to the Gaussian found neglecting thermionic and ion emission in~\cite{DS_87} and~\cite{Okuzumi_2009}. For low temperatures $T\lesssim500 \,$K, the Gaussians should be identical as emission processes are not active. We choose $T = 166$\,K, for which $\tau^i = 1$ for our minimum grain size of $10^{-5}$\,cm. In Fig.~\ref{fig:grain_charge_gaussian_ds_comp}, we plot the distributions for $\tau^i = 1$ and $\tau^i = 10$, corresponding to $10^{-5}$\,cm and $10^{-4}$\,cm. We compute the chemical abundances (assuming a single grain charge for each grain size as in the rest of the work) at the fiducial $n_{\text{H}_2} = 10^{14} \, \text{cm}^{-3}$, with the fiducial MRN distribution with parameters $q=3.5$, $a_{\text{min}} = 10^{-5} \text{ cm}$ and $a_{\text{max}} = 10^{-1} \text{ cm}$.
\begin{figure*}
    \centering
    \includegraphics[width=\textwidth]{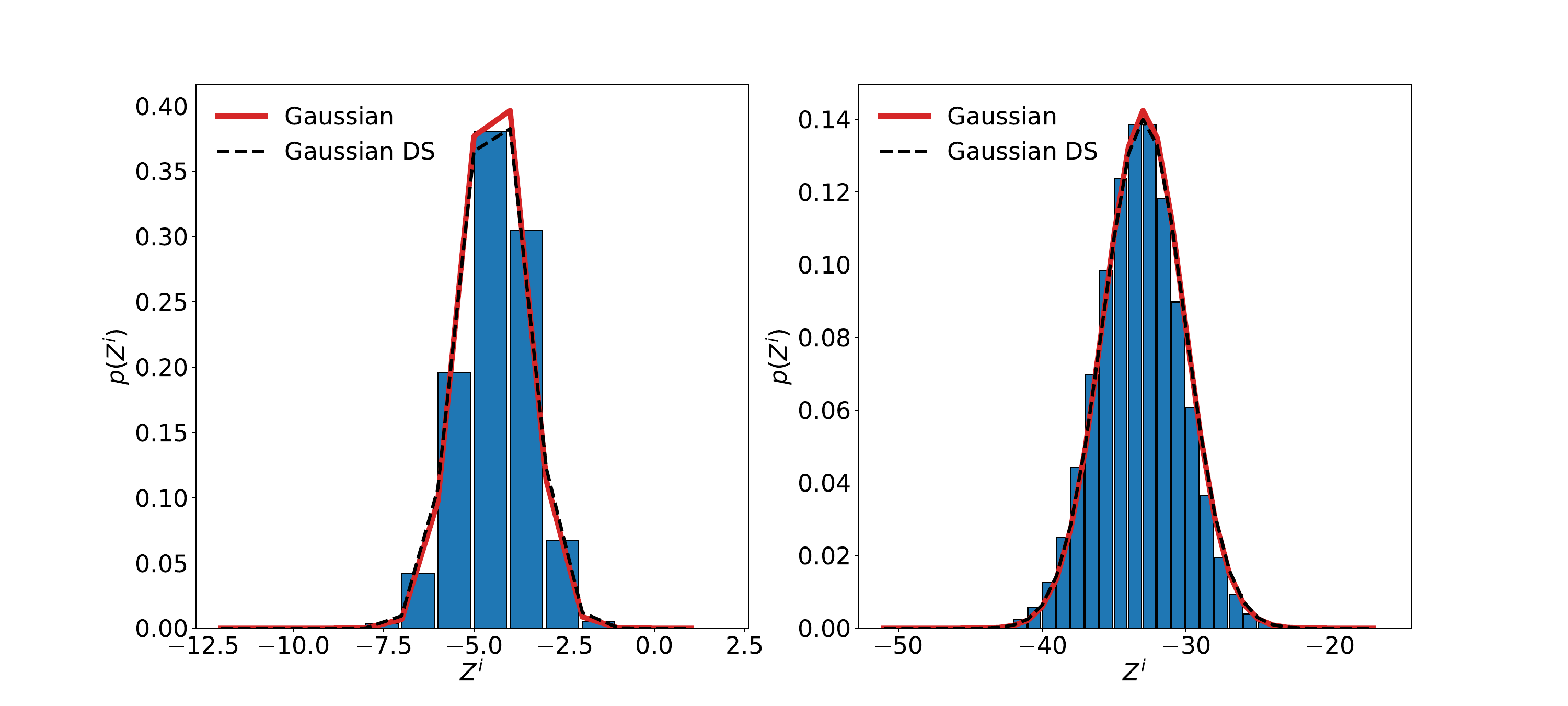}
    \caption{Distributions of $p(Z^i)$ computed at $T=166$\,K. Chemical abundances are computed with our network (using a single charge for each grain size) at the same temperature, with $n_{\text{H}_2} = 10^{14} \, \text{cm}^{-3}$, a fiducial MRN distribution with parameters $q=3.5$, $a_{\text{min}} = 10^{-5} \text{ cm}$ and $a_{\text{max}} = 10^{-1} \text{ cm}$. The distribution $p(Z^i)$ is plotted for two separate grain sizes  $10^{-5} \text{ cm}$ (left) and $10^{-4} \text{ cm}$ (right), corresponding to $\tau = 1$ and $\tau = 10$ respectively. Overplotted are the Gaussians computed including thermionic and ion emission (red; N.B. these effects are insignificant at this temperature), and the Gaussian due to \protect\cite{DS_87} (black, dashed), which does not include these effects. The distribution is well fit by a Gaussian, and agreement can be seen between both Gaussians, as expected.} 
    \label{fig:grain_charge_gaussian_ds_comp}
\end{figure*}
From Fig.~\ref{fig:grain_charge_gaussian_ds_comp}, we see that the distribution $p(Z^i, \tau^i)$ is fit well by both the full Gaussian and Gaussian excluding ion and thermionic emission due to~\cite{DS_87}. 

Finally, we investigate the fit of the Gaussian derived here to the full distribution $p(Z^i, \tau^i)$ at a temperature at which ion and thermionic emission are active. We choose $T = 1000$\,K at the fiducial $n_{\text{H}_2} = 10^{14} \, \text{cm}^{-3}$, with the fiducial MRN distribution with parameters $q=3.5$, $a_{\text{min}} = 10^{-5} \text{ cm}$ and $a_{\text{max}} = 10^{-1} \text{ cm}$. We plot the distributions $p(Z^i, \tau^i)$ along with the Gaussian approximations for $\tau = 6$ and $\tau = 60$, corresponding again to $10^{-5} \text{ cm}$ and $10^{-4} \text{ cm}$, in Fig.~\ref{fig:grain_charge_gaussian}. We see that a Gaussian provides a good approximation to the distributions.

\begin{figure*}
    \centering
    \includegraphics[width=\textwidth]{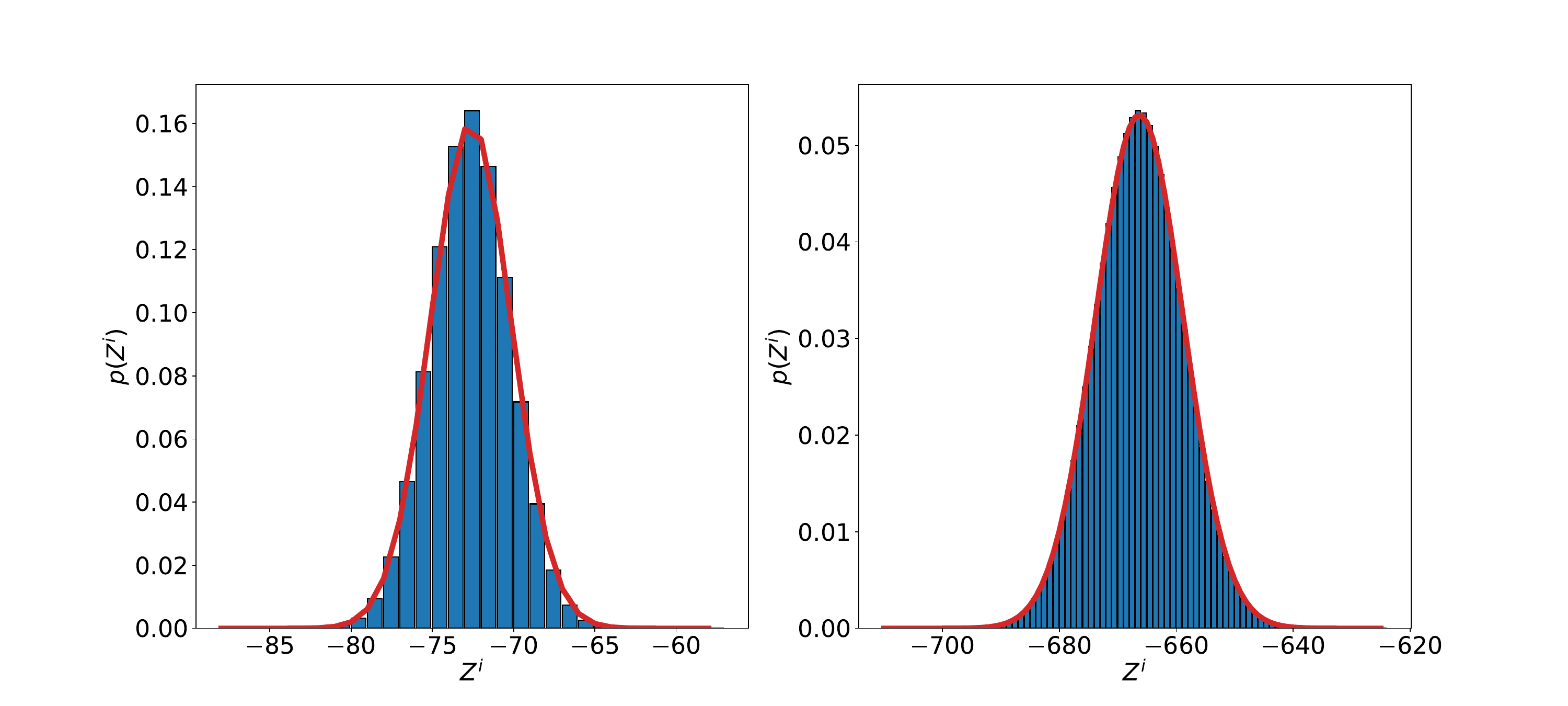}
    \caption{Distributions of $p(Z^i)$ computed at $T=1000$\,K. Chemical abundances are computed with our network (using a single charge for each grain size) at the same temperature, with $n_{\text{H}_2} = 10^{14} \, \text{cm}^{-3}$, a fiducial MRN distribution with parameters $q=3.5$, $a_{\text{min}} = 10^{-5} \text{ cm}$ and $a_{\text{max}} = 10^{-1} \text{ cm}$. The distribution $p(Z^i)$ is plotted for two separate grain sizes  $10^{-5} \text{ cm}$ (left) and $10^{-4} \text{ cm}$ (right), corresponding to $\tau = 6$ and $\tau = 60$ respectively. The Gaussian computed in this work, including thermionic and ion emission, is overplotted (red). The distributions $p(Z^i)$ are well fit by the Gaussian.} 
    \label{fig:grain_charge_gaussian}
\end{figure*}

\section{Comparison with effective dust-to-gas ratio}
\label{sec:appendeffectivedg}
{
Neglecting the effect of grain charge, the total grain surface area determines the chemistry, i.e., any two grain size distributions with the same total surface area produce the same abundances. To explore the general case of charged grains, \cite{Bai_Goodman_2009} used a network specialized to the cooler regions of the disc (i.e., no significant thermal ionization) with two grain sizes. They included charge transfer between grains and found it to be negligible, thus the two grain populations act independently of one another. They found that the abundance of free electrons was controlled by $\sum f_{\text{dg}}^i(a^i) / (a^i)^p$ with a controlling $p$ between 1 and 2. Since the dust-to-gas ratio scales as $a^3$, $p=1$ corresponds to keeping a fixed total grain surface area, the na\"ive expectation.

Adding to this work for the high-temperature network of~\cite{Desch_Turner_2015},~\cite{Jankovic_2021} found that above a certain temperature ($T\gtrsim 1000$\,K), the controlling parameter for the electron number density was $p=1.5$. For a whole population of grain sizes, and an arbitrarily chosen single effective grain size $a_{\text{eff}}$, they thus used an effective dust-to-gas ratio $f_{\text{dg, eff}} = a_{\text{eff}}^p  \int_{a_{\text{min}}}^{a_\text{max}} dn(a) m(a)/ (\rho_{gr} a^p)$ with $p=1.5$, that would produce the same electron abundance as the distribution as a whole.
}

Our model, which includes {\it exactly} the effect of a distribution of grain sizes, is compared with the~\cite{Jankovic_2021} model in Fig.~\ref{fig:mid-plane} for a range of dust-to-gas ratios ($f_{\text{dg}}$) and MRN power-law indices ($q$).

At low temperatures, where non-thermal ionization dominates, an effective dust-to-gas ratio with $p=1.5$ does not yield accurate gas-phase (or grain-phase) abundances. This had been previously demonstrated by~\cite{Jankovic_2021}.

At high temperatures ($T \gtrsim 1000$\,K), the effective dust-to-gas ratio calculation accurately reproduces the dominant gas-phase abundances; therefore, if one only wanted to compute resistivities at these temperatures, an effective dust-to-gas ratio calculation with $p=1.5$ would be valid. 

However, when grains are not one of the dominant charged species, the grain charge can be quite inaccurate. This is particularly evident for the distributions with $q=2.0$, where the fractional error on the effective dust-to-gas ratio calculation is significant. This would considerably alter the collision cross-sections that grains present to one another, as well as the impact velocities of these collisions. Therefore, if one wanted to study the collisional evolution of the grains, an effective dust-to-gas ratio cannot safely be used. 

\begin{figure*}
    \centering
    \includegraphics[width=\textwidth]{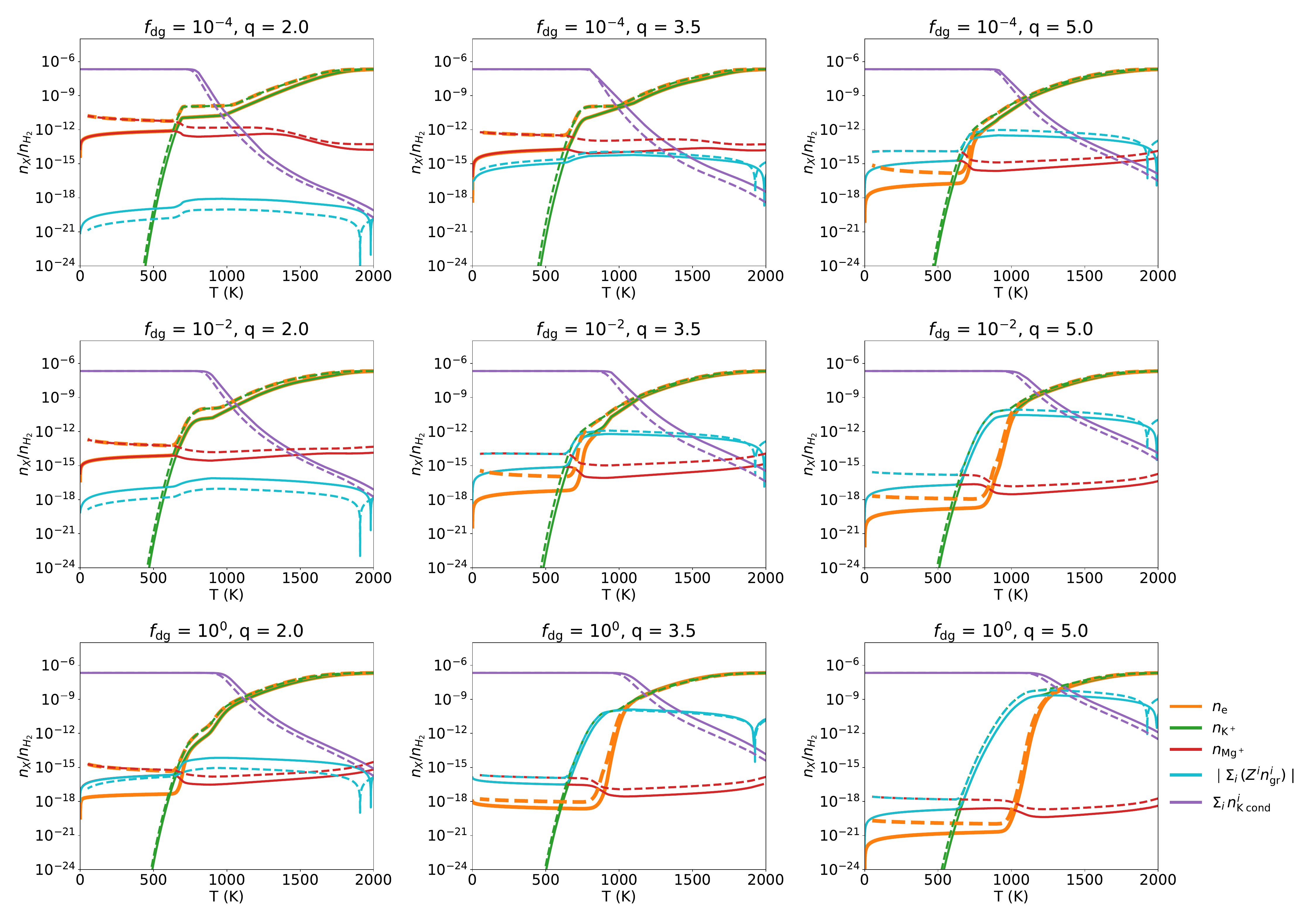}
    \caption{The ionization state of the reaction network is shown for grains following an MRN distribution for a grid of dust-to-gas ratio $f_{\text{dg}}$ and MRN exponent values $q$. The exact model for a distribution of grain sizes is shown in solid lines, while the effective method for modelling a distribution of grain sizes is shown in dashed lines. The remaining parameters used are the fiducial $a_{\text{min}} = 10^{-5} \text{ cm}$, $a_{\text{max}} = 10^{-1} \text{ cm}$ at $n_{\text{H}_2} = 10^{14} \; \mathrm{cm^{-3}}$. The thickness of the lines for $n_e$ have been increased to show where they have gone behind other lines.} 
    \label{fig:mid-plane}
\end{figure*}

%%%%%%%%%%%%%%%%%%%%%%%%%%%%%%%%%%%%%%%%%%%%%%%%%%

% Don't change these lines
\bsp	% typesetting comment
\label{lastpage}
\end{document}